\documentclass[11pt]{article}
\usepackage[utf8]{inputenc}
\usepackage{microtype}\usepackage{amsfonts}
\usepackage{amsmath,amsthm,amssymb, color, tikz, mathtools}
\usepackage{amsfonts}

\usetikzlibrary{positioning}
\usetikzlibrary{shapes}
\usetikzlibrary{decorations.pathreplacing}
\usepackage{enumerate, enumitem} 

\usepackage{thm-restate}
\usepackage[hidelinks]{hyperref}
\usepackage{cleveref}

\usepackage[margin = 2.50cm]{geometry}

\usepackage{graphicx}
\graphicspath{{./image/}}

\hypersetup{colorlinks,linkcolor={blue},citecolor={blue},urlcolor={blue}}

\usepackage{algorithm}
\usepackage[noend]{algpseudocode}
\usepackage{thm-restate}

\usepackage{bbm}

\newcommand{\dist}{\text{dist}}

\renewcommand{\hat}{\widehat}

\newcommand{\R}{\mathbb{R}}
\newcommand{\E}{\mathbb{E}}

\newcommand{\ind}{\mathbbm{1}}

\newcommand{\cL}{\mathcal{L}}

\newtheorem{theorem}{Theorem}[section]
\newtheorem{corollary}[theorem]{Corollary}
\newtheorem{remark}[theorem]{Remark}
\newtheorem{lemma}[theorem]{Lemma}
\newtheorem{claim}[theorem]{Claim}
\newtheorem{defi}[theorem]{Definition}
\newtheorem{question}[theorem]{Question}

\newtheorem{observation}[theorem]{Observation}

\newcommand{\zone}{\{-1, 1\}}
\newcommand{\ftwo}{\mathbb{Z}_2}
\renewcommand{\tilde}{\widetilde}
\renewcommand{\Tilde}{\widetilde}

\newcommand{\dA}{\delta_{\cL}}

\newcommand{\supp}{\mathrm{supp}}

\newcommand{\cS}{\mathcal{S}}
\newcommand{\wh}{\widehat}
\newcommand{\wt}{wt}
\newcommand{\G}{\mathcal{G}}

\newcommand{\Var}{Var}
\newcommand{\Cov}{Cov}
\newcommand{\LCM}{LCM}
\newcommand{\Codim}{Codim}
\newcommand{\Aut}{\text{Aut}}
\newcommand{\Ann}{Ann}

\title{
Testing Isomorphism of Boolean Functions over\\
Finite Abelian Groups
} 

\author{
Swarnalipa Datta\\
Indian Statistical Institute, India\\
\texttt{rimadatta94@gmail.com}
\and
Arijit Ghosh\\
Indian Statistical Institute, India\\
\texttt{arijitiitkgpster@gmail.com}
\and
Chandrima Kayal\\
Universit\'{e} Paris Cit\'{e}, CNRS, IRIF, Paris, France\\
\texttt{chandrimakayal2012@gmail.com}
\and
Manaswi Paraashar\\
University of Copenhagen, Denmark\\
\texttt{manaswi.isi@gmail.com}
\and 
Manmatha Roy
\\ Indian Statistical Institute, India \\
\texttt{reach.manmatha@gmail.com}
}

\begin{document}

\maketitle

\begin{abstract}
Let $f$ and $g$ be Boolean functions over a finite Abelian group $\G$, where $g$ is fully known, and we have {\em query access} to $f$, that is, given any $x \in \G$ we can get the value $f(x)$. 
We study the tolerant isomorphism testing problem: given $\epsilon \geq 0$ and $\tau > 0$, we seek to determine, with minimal queries, whether there exists an automorphism $\sigma$ of $\G$ such that the fractional Hamming distance between $f \circ \sigma$ and $g$ is at most $\epsilon$, or whether for all automorphisms $\sigma$, the distance is at least $\epsilon + \tau$.

We design an efficient tolerant testing algorithm for this problem, with query complexity 
$\mathrm{poly}\left( s, 1/\tau \right)$, where $s$ bounds the spectral norm of $g$. 
Additionally, we present an improved algorithm when $g$ is Fourier sparse.

Our approach uses key concepts from Abelian group theory and Fourier analysis, including the annihilator of a subgroup, Pontryagin duality, and a pseudo inner-product for finite Abelian groups. We believe these techniques will find further applications in property testing.
\end{abstract}

\section{Introduction}

Let $\G$ be a finite Abelian group, and let $\widehat{\G}$ be its dual group, consisting of all characters, that is, homomorphisms from $\G$ to $\mathbb{C}^{\times}$. The Fourier coefficient for a function $f: \G \to \mathbb{C}$ corresponding to a character $\chi \in \G$ is denoted by $\widehat{f}(\chi)$\footnote{Note that Fourier coefficients are complex numbers.}, and the {\em spectral norm} of $f$ is defined as
$$
    \|\widehat{f}\|_{1} := \sum_{\chi \in \widehat{\G}} |\widehat{f}(\chi)|.
$$
For more details refer to Section~\ref{section_prelims}. The spectral norm is a key parameter in the analysis of Boolean functions, with wide-ranging applications in property testing, learning theory, cryptography, pseudorandomness, and quantum computing~\cite{ShpilkaTV17,KushilevitzM93,tsang2013fourier,GirishT021,Tal17, MekaRT19,ChattopadhyayHL19,ReingoldSV13,ForbesK18,RazT22,Tal20,BansalS21}. Moreover, it plays an important role in understanding the coset complexity of subsets of Abelian groups~\cite{Cohen60,GreenS08,GreenSAnnals08, CheungHZZ22}. We will show that, for the isomorphism testing of Boolean functions over finite Abelian groups, the query complexity is essentially controlled by the spectral norm.

The area of property testing is concerned with determining whether a given object satisfies a fixed property or is ``far'' from satisfying it, where ``far'' is typically defined using a suitable distance measure. Over nearly three decades of research, \textit{symmetry} has played a crucial role in property testing, appearing in various forms.

To illustrate the setup and the role of symmetry, we consider property testing of Boolean functions, which are functions of the form $f: \zone^n \to \zone$. The \textit{fractional Hamming distance} between two Boolean functions $f$ and $g$ on $n$-bit inputs is defined as
\[
\delta(f,g) = \Pr_{x \in \{-1,1\}^n}[f(x) \neq g(x)],
\]
that is, the fraction of inputs on which they differ.

Let $S_n$ be the group of all permutations of $[n] : = \{1, \dots, n\}$. For $\sigma \in S_n$, we define the permuted function $f \circ \sigma$ by
\[
(f \circ \sigma)(x_1, \dots, x_n) := f(x_{\sigma(1)}, \dots, x_{\sigma(n)})
\]
for all $(x_1, \dots, x_n) \in \zone^n$. A fundamental question in this setting is the following.

%The area of property testing is concerned with deciding whether a given object satisfies some fixed property or is ``far'' from satisfying that property, where the notion of being ``far'' is usually defined using a suitable notion of distance between the objects. Over the course of nearly three decades of research in this area, \textit{symmetry} has played a crucial role in property testing and shows in various forms.

%We illustrate the setup of property testing and demonstrate the role of symmetry by describing a class of problems in the area of property testing of Boolean functions, which are functions of the form $f: \zone^n \to \zone$. For Boolean functions $f$ and $g$ on $n$-input variables, we denote by $\delta(f,g)$ the fractional Hamming distance between $f,g$, i.e., the fraction of points of $x\in \{-1,1\}^n$ in which $f(x)\neq g(x)$.\footnote{Generally, for $f, g : \mathcal{D} \to \zone$, where $\mathcal{D}$ is some domain, define the Hamming distance between $f$ and $g$ to be $\delta(f,g) = \Pr_{x \in \mathcal{D}}[f(x) \neq g(x)]$, where $x$ is chosen uniformly at random from $\mathcal{D}$.} Let $S_n$ be the group of all permutations of the set $\{1, \dots, n\}$ and $\G$ be a subgroup of $S_n$. For $\sigma \in S_n$, let $f \circ \sigma$ denote the function obtained by permuting the indices of the input of $f$ according to $\sigma$, that is, $f \circ \sigma(x_1, \dots, x_n) = f(x_{\sigma(1)}, \dots, x_{\sigma(n)})$ for all $(x_1, \dots, x_n) \in \zone^n$. A fundamental question is the following.

\begin{question}[Function Isomorphism Testing over $\zone^{n}$]
Given a known Boolean function $f$ and query access to an unknown Boolean function $g$, determine if there exists a permutation $\sigma \in \G$ such that $f \circ \sigma = g$, or for all $\sigma \in \G$, is the Hamming distance between $f \circ \sigma$ and $g$  at least $\epsilon$, under the promise that one of these is the case?
\end{question}
\noindent 
The above question was first investigated by Fischer et al.~\cite{FKRSS04} and was followed by a long 
line of work~\cite{AlonB10, BD10, DBLP:conf/soda/ChakrabortyGM11,ChakrabortyFGM12,AlonBCGM13,BlaisWY15}.

The domain of Boolean functions discussed so far is the set $\zone^n$. It is of great interest, as we 
discuss later, to study Boolean functions whose domain $\mathcal{D}$ has some algebraic structure 
associated with it, like $\mathbb{Z}^{n}_{2}$. In these settings, it is natural to study the isomorphism 
problem for Boolean functions under the {\em group of symmetry} of $\mathcal{D}$ that preserves 
its underlying algebraic structure.

An extremely important example is that of $\mathcal{D} = \ftwo^n$, that is, functions of the form 
$f : \ftwo^n \to \{-1,1\}$ defined over the vector space $\ftwo^{n}$. 
These functions arise in numerous areas of computer science and 
mathematics, see~\cite{ODonnellbook2014} and the references therein for several examples. 
A natural group of symmetry for 
$\ftwo^n$ is the group of all non-singular 
$n \times n$ matrices over $\ftwo$(see~\cite[Chapter~3]{dummit2004abstract}).
We now describe the question of \textit{Linear Isomorphism Testing of Boolean Functions} over
$\mathbb{Z}^{n}_{2}$. For $A \in \ftwo^{n \times n}$, let $f \circ A: \ftwo^n \to \zone$ be the function 
$f \circ A(x) = f(Ax)$ for all $x \in \ftwo^n$. 
The \textit{Linear Isomorphism Distance} between two functions $f: \ftwo^n \to \zone$ and $g: \ftwo^n \to \zone$ is 
defined as
\begin{align}
    \dist_{\ftwo^n}(f,g) = \min_{A \in \ftwo^{n \times n} : A \text{ is non-singular}} \delta(f\circ A,g). 
    \label{eq: defn linear iso dist}
\end{align}
Assume that $f$ and $g$ satisfy the promise that either $\dist_{\ftwo^n}(f,g) = 0$ or 
$\dist_{\ftwo^n}(f,g) \geq \epsilon$, the question of {\em Linear Isomorphism Testing} is that of deciding which 
is the case.

In this work, we study Boolean functions over a finite Abelian group $\G$. The natural symmetries in this setting arise from the automorphism group of $\G$, denoted $\Aut(\G)$ (see Definition~\ref{defn_group_isomorphism}).  
For functions $f,g : \G \to \{0,1\}$, we define their \textit{automorphism distance} as  
\[
    \dist_{\G}(f,g) = \min_{\sigma \in \Aut(\G)} \delta(f \circ \sigma, g),
\]  
where $f \circ \sigma(x) = f(\sigma(x))$ for all $x \in \G$.  
We investigate the following general question.

\begin{question}[Testing Isomorphism of Boolean Functions over Finite Abelian Groups]
\label{question: automorphism testing}
    Let $\G$ be a finite Abelian group, and let $f, g: \G \to \zone$ be Boolean functions such that $f$ is known and given query access to $g$, determine if $f$ and $g$ are isomorphic under the automorphism group $\Aut{(\G)}$ of $\G$ (that is, $\dist_{\G}(f,g) = 0$) or is the automorphism distance between them at least $\epsilon$ (that is, $\dist_{\G}(f,g) \geq  \epsilon$), under the promise that one of these is the case?
\end{question}

We now give an equivalent problem in the setting of finite Abelian groups.
\begin{defi}[Subset Equivalence Problem]\label{def:subset_equivalence_problem}
Given two subsets \( A, B \subseteq \G \) of a finite Abelian group \( \G \), does there exist an automorphism \( \sigma \in \Aut(\G) \) such that \( A = \sigma(B) \)?
\end{defi}

To formulate a testing version of this problem, we introduce the following definitions. For subsets \( A, B \subseteq \G \), the \emph{distance} between \( A \) and \( B \) is defined by
\begin{align}\label{eqn_set_dist}
    \dist(A,B) := \frac{|A \setminus B| + |B \setminus A|}{|\G|}.
\end{align}
We define the \emph{automorphism distance} between \( A \) and \( B \) as
\begin{align}\label{eqn_set_dist_aut}
    \dist_{\G}(A,B) := \min_{\sigma \in \Aut(\G)} \dist(A,\sigma(B)).
\end{align}
We assume \emph{membership query access} to a subset of \( \G \), meaning that we can query whether any given element of \( \G \) belongs to the subset.
The testing version of the Subset Equivalence Problem is formulated below:
\begin{question}[Testing Automorphism Distance Between Subsets of Finite Abelian Groups]\label{question:11}
Let \( \G \) be a finite Abelian group, and let \( A \subseteq \G \) be a known subset. Suppose we have membership query access to an unknown subset \( B \subseteq \G \). Under the promise that either \( \dist_{\G}(A,B) = 0 \) or \( \dist_{\G}(A,B) \geq \epsilon \), determine which is the case; that is, decide whether \( A \) and \( B \) are isomorphic under \( \Aut(\G) \), or they are at least \( \epsilon \)-far from being so.
\end{question}

Question~\ref{question:11} can be viewed as a special case of \emph{Boolean function isomorphism testing} as follows. Define the \emph{indicator function} \( \ind_A : \G \to \{-1,1\} \) of the set \( A \) by
\begin{align*}
    \ind_A(x) =
    \begin{cases}
        -1 & \text{if } x \in A, \\
        1 & \text{otherwise}.
    \end{cases}
\end{align*}
Similarly, define \( \ind_B \) for the set \( B \). Then Question~\ref{question:11} reduces to testing whether the Boolean functions \( \ind_A \) and \( \ind_B \) are isomorphic under the action of \( \Aut(\G) \).

Previous works on property testing of Boolean functions
over $\ftwo^{n}$ critically exploit the Fourier analysis framework, where the vector space structure of the domain is essential to the proofs. However, since we are working with arbitrary finite Abelian groups, these established approaches that rely on the vector space structure do not apply. As a result, we had to develop new ideas, drawing on key concepts from the theory of finite Abelian groups. 
We outline some of these ideas below:

\begin{itemize}
    \item 
        We have used a notion of ``orthonormal" subgroup $H^\perp$ for a subgroup $H$ of a finite Abelian group $\G$ which is not naturally defined in the context of groups. To address this, we used the concept of {\em annihilator} of a subgroup. This concept is described in more detail in Section~\ref{section_H_perp}.
    
    \item 
        For any group automorphism $A: \G \to \G$ and any function $f:\G \to \{-1,+1\}$, we needed a natural way to associate  
        the Fourier coefficients $\left\{\widehat{f\circ A}(\chi) : \chi \in \widehat{\G}\right\}$ for the function $f\circ A$ with the 
        Fourier coefficients $\left\{ \widehat{f}(\chi) : \chi \in \widehat{\G}\right\}$ for $f$. 
        We use \textbf{`Pontryagin duality'} to deal with such scenario, which is of independent interest. More details can be found in Lemma~\ref{lemma_pontryagin} and Lemma~\ref{lemma_aut_perm_char} from Section~\ref{section_prelims}.
    
    \item 
        The vector space $\ftwo^{n}$ can be naturally partitioned into cosets of a subspace. This property is essential in approaches like ``coset hashing", which is used in algorithms for property testing of Boolean functions over such domains (see~\cite{feldman2006new,gopalan2011testing,WY13}).
        However, for a finite Abelian group 
        $\G$, this partitioning is not generally possible, as $\G$ may not form a vector space. To address this, we instead use normal subgroups and partition $\G$ using the cosets of these normal subgroups. For details, see Section~\ref{section_prelims}.

    \item 
        We use a new notion of inner product, called \textbf{pseudo-inner product} (see Definition~\ref{defi:pseudo-inner product}) to analyze our algorithms.

    \item 
        We also introduce a notion of independence for finite Abelian groups, which we call \textbf{pseudo-independence} (see Definition~\ref{defn_independence}), to analyze our algorithms.

    \item 
        Note that the characters of $\ftwo^{n}$ are the 2nd roots of unity, meaning they take the values $-1$ or $+1$ at each point in $\ftwo^{n}$. This property is fundamental to the analysis of Boolean functions over $\ftwo^{n}$, see, for example, Gopalan et al.~\cite{gopalan2011testing}. Since $\G$ is a finite Abelian group, it can be written as $\G = \mathbb{Z}_{p_1^{m_1}} \times \cdots \times \mathbb{Z}_{p_n^{m_n}}$, where $p_{i}$ are primes for all $i \in [n]$, which may not necessarily be distinct. In the case of $\G$, the characters are complex numbers, as they correspond to the $\mathcal{L}$-th roots of unity at each point in $\G$, where $\mathcal{L} = \text{LCM}\left\{ p_1^{m_1}, \ldots, p_n^{m_n} \right\}$. Consequently, there is no natural concept of sign, which we address using a completely different approach. For more details, see Section~\ref{section_implicit seive}, which we believe offers independent interest in the study of Boolean functions on more general domains.

\end{itemize}

\subsection{Related work} 

Fischer et al.~\cite{FKRSS04} were the first to investigate the problem of function isomorphism testing, a line of research that has since been extended by several subsequent works~\cite{AlonB10, BD10, DBLP:conf/soda/ChakrabortyGM11}. The problem of characterizing the {\em testability}\footnote{See the definition of testability from Bhattacharyya et al.~\cite{BFHH13}.} of {\em linearly invariant} properties of functions has been a central focus in property testing. This question has seen significant progress through various works~\cite{GKS08, KS08, BGS15}. Also, Bhattacharyya et al.~\cite{BFHH13} provided a characterization of linearly invariant properties that are constant-query testable with one-sided error. Testing linear isomorphism of Boolean functions over \( \mathbb{Z}_2^n \) was studied by Wimmer and Yoshida~\cite{WY13}, whose algorithm has polynomial query complexity in terms of the spectral norm of the functions. Grigorescu, Wimmer and Xie~\cite{GrigorescuWX13} studied the same problem in the communication setting. Linear isomorphism of Boolean functions has also been explored in the context of combinatorial circuit design~\cite{CV03, BA04, YB12}, error-correcting codes~\cite{ND06, DT01, XH94}, and cryptography~\cite{CC07, CH17, OD12}.

\subsection{Our results}

Our main results provide answers to Question~\ref{question: automorphism testing} for the case when $\G$ is a finite Abelian group. We study Question~\ref{question: automorphism testing} in the following standard setting: the function $g: \G \to \zone$ is known, while the function $f : \G \to \zone$ is unknown. 
The functions are promised to satisfy either $\dist_{\G}(f,g) \leq \epsilon$ or $\dist_{\G}(f,g) \geq \epsilon + \tau$.
The goal is to give a randomized algorithm that determines which case holds, with probability at least $2/3$ over its internal randomness, by making the minimum number of queries to the unknown function $f$. 
A query means that the algorithm obtains the value $f(x)$ for an $x \in \G$ of its choice. Our results are developed in the context of \textit{tolerant testing}, which we refer to as 
$(\epsilon,\tau)$-tolerant testing. This setup is more general than the one considered in Question~\ref{question: automorphism testing}, which corresponds to $(0,\tau)$-tolerant testing.

Our main result presents an algorithm for deciding whether the automorphism distance between 
$f$ and $g$ is at most $\epsilon$ or at least $\epsilon+\tau$. The algorithm's query complexity is polynomial in $(1/\tau)$ (independent of $\epsilon$) and the spectral norm (see Definition~\ref{defi: Fourier spectral norm}) of the known function $g$.

% \vspace{10pt}
% \noindent
% \rule{\textwidth}{1.0pt}
%     {\bf
%     We will be working with the cyclic group of the following form 
%     $$
%         \mathbb{Z}_{p_1^{n_1}}\times \cdots \times \mathbb{Z}_{p_T^{n_T}}
%     $$ 
%     where $p_i$ are distinct primes for all $i \in [T]$. Note that here $p_i^{n_i}$ is considered to be constant and $T$ is growing. So, in the complexity of the algorithms constants will depend on $p_i^{n_i}$ for all $i \in [T]$.  On the other hand, we will also consider Abelian groups of the form 
%     $$
%         \G = \mathbb{Z}^{n_{1}}_{p_1}\times \cdots \times \mathbb{Z}^{n_{T}}_{p_T}
%     $$ 
%     where only $p_i$'s are considered to be constant and $n_i$ can be growing for all $i \in [T]$, so, in this case, the constants in the complexity will depend only on $p_i$'s. 
%     }
    
% \noindent
% \rule{\textwidth}{1.0pt}
% \vspace{5pt}

\begin{restatable}[Main Result: $(\epsilon, \tau)$-Tolerant Isomorphism Testing of Boolean Functions over Finite Abelian Groups]{theorem}{testingalgorithmquery}
     \label{theorem: testing algorithm}
   Let $\G$ be a finite Abelian group, hence $\G =\mathbb{Z}_{p_1^{m_1}}\times \cdots \times \mathbb{Z}_{p_n^{m_n}}$ where $p_i$ are primes for all $i \in [n]$ and not necessarily distinct. Also let $\epsilon, \tau \in (0,1/2]$.
    Given a known function $g: \mathcal{G} \to \zone$ with $\|\widehat{g}\|_1 \leq s$, and query access to an unknown function  $f: \mathcal{G} \to \zone$, the query complexity of deciding whether $\dist_{\G}(f,g) \leq \epsilon$ or $\dist_{\G}(f,g) \geq \epsilon + \tau$ is $\Tilde{O}\left(\frac{s^{8}}{\tau^{8}}\right)$, where $\Tilde{O}(\cdot)$ hides multiplicative factors in $\log s$ and $\log(1/\tau)$ up to constants, and $\mathcal{L}$ is constant. 
\end{restatable}
\noindent
Note that Wimmer and Yoshida~\cite{WY13} proved $\Omega(\|\widehat{g}\|_{1})$ lower bound for the special case of $\G = \mathbb{Z}_2^n$.

The proof of the above theorem generalizes the work of Wimmer and Yoshida~\cite{WY13}, who solved the $(\epsilon/3,2\epsilon/3)$-tolerant linear isomorphism problem for Boolean functions over $\ftwo^{n}$. Their testing algorithm and analysis crucially rely on the fact that $\ftwo^{n}$ is a vector space over the field $\ftwo$.
However, finite Abelian groups do not necessarily have a vector space structure. Due to this, we had to introduce several new ideas and incorporate key concepts from the theory of finite Abelian groups. For more details, see Section~\ref{subsection:proof_sketch} and Section~\ref{subsection:our_difficulties}.

Finite Abelian groups $\G$ which are of the form $G_1 \times G_2 \times \cdots \times G_n$ have been used in various works, namely \cite{grigorescu2006local, amireddy2024local}.
We give the following two corollaries, which immediately follow from Theorem~\ref{theorem: testing algorithm}.

\begin{corollary}
    Let $\G$ be a finite Abelian group such that $\G $ can be expressed as $G \times G \times \dots \times G$ where the product takes place $n$-times and $G$ itself is a finite Abelian group of constant size. Also, let $\epsilon, \tau \in (0,1/2]$.
    Given an known function $g: \mathcal{G} \to \zone$ with $\|\widehat{g}\|_1 \leq s$, and query access to a unknown function  $f: \mathcal{G} \to \zone$, the query complexity of deciding whether $\dist_{\G}(f,g) \leq \epsilon$ or $\dist_{\G}(f,g) \geq \epsilon + \tau$ is $\Tilde{O}(poly(s, 1/\tau))$, where $\Tilde{O}(\cdot)$ hides multiplicative factors in $\log s$ and $\log(1/\tau)$ up to constants, and $\mathcal{L}$ is a constant.
\end{corollary}

More generally we get the following:

\begin{corollary}
    Let $\G$ be a finite Abelian group such that $\G $ can be expressed as $G_1 \times G_2 \times \dots \times G_n$ where $i \in [n]$ and $|G_i| = O(1)$ for all $i \in [n]$. Also let $\epsilon, \tau \in (0,1/2]$.
    Given a known function $g: \mathcal{G} \to \zone$ with $\|\widehat{g}\|_1 \leq s$, and query access to a unknown function  $f: \mathcal{G} \to \zone$, the query complexity of deciding whether $\dist_{\G}(f,g) \leq \epsilon$ or $\dist_{\G}(f,g) \geq \epsilon + \tau$ is $\Tilde{O}(poly(s, 1/\tau))$, where $\Tilde{O}(\cdot)$ hides multiplicative factors in $\log s$ and $\log(1/\tau)$ up to constants, and $\mathcal{L}$ is a constant. 
\end{corollary}

% \textcolor{blue}{Abelian groups have been mentioned in this form, and were being used in certain works, namely, \cite{grigorescu2006local}.}

The main technical result we require is the following algorithm to estimate the values of the {\em large} Fourier coefficients of $f: \G \to \{-1,1\}$, given query access to $f$, without prior knowledge of the coefficients themselves. The set of large Fourier coefficients is defined using a threshold (see Theorem~\ref{lemma: Implicit Sieve}). Note that this set is unknown, as the function 
$f$ is not known.

\begin{restatable}[Generalized Implicit Sieve for Abelian groups]{theorem}{Sieveforabeliangroup}
    \label{lemma: Implicit Sieve}
    Let $\G$ be a finite Abelian group, hence $\G =\mathbb{Z}_{p_1^{m_1}}\times \cdots \times \mathbb{Z}_{p_n^{m_n}}$ where $p_i$ are primes for all $i \in [n]$ and not necessarily distinct.
    Also, let $\theta > 0$ be a threshold and $\mathcal{M} = \left\{x_i\right\}_{i=1}^{\Tilde{m}} \subseteq \mathcal{G}$ be a set of independently and uniformly chosen $\Tilde{m}$ random points from $\mathcal{G}$. 
%    where $\Tilde{m} = \Tilde{\Theta}(\frac{s^2}{\tau^2})$.
    There exists an algorithm, \textsf{Generalized Implicit Sieve Algorithm} (Algorithm~\ref{algo:Implicit_Sieve_generalized}), that takes $\theta$ and $\mathcal{M}$ as input, makes $O\biggl(\frac{\ln(\Tilde{m}/\theta^{16})
    % \frac{\Tilde{m}}{\theta^{16}}
    }{\theta^8}
    + \Tilde{m} \frac{\ln(\mathcal{N} \Tilde{m})}{\theta^2} + \frac{\ln \mathcal{N}}{\theta^8} \biggr)$ queries to the truth table of a Boolean valued function $f: \mathcal{G} \to \zone$, and returns, with probability at least $\frac{94}{100}$, a labeled set of $\Tilde{m}$ examples, and the value of the function $f$ at $x_i$ for each $i \in [\Tilde{m}]$, of the form $\left\{ \chi_{\alpha_1}\left(x_i\right), \dots, \chi_{\alpha_\mathcal{N}}\left(x_i\right), f\left(x_i\right) \mid x_i \in \mathcal{M} \right\}$, where $\mathcal{S} = \left\{\alpha_1, \dots, \alpha_{\mathcal{N}} \right\}$ is some set that satisfies the following two properties:
    \begin{enumerate}
        \item
            $\forall \alpha \in \mathcal{G}$ with $\left|\widehat{f}(\chi_\alpha)\right| \geq \theta$ then $\alpha \in \mathcal{S}$, and
        \item 
            $\forall \alpha \in \mathcal{S}$, $\left| \widehat{f}(\chi_\alpha) \right| \geq {\theta}/{2}$.
    \end{enumerate}
    The output of this algorithm can be seen as a $\Tilde{m} \times (\mathcal{N}+1)$ matrix $Q$ with entries in powers of $\omega_\mathcal{L}$, where $\omega_\mathcal{L}$ is a primitive $\mathcal{L}^{th}$ root of unity, $\mathcal{L}= \LCM\{p_1^{m_1}, \ldots, p_n^{m_n}\}$, and $\mathcal{L}$ is a constant. 
\end{restatable}

\begin{remark}[Comments on the query complexity in Theorem~\ref{lemma: Implicit Sieve}]
    Note that the query complexity of the algorithm in Theorem~\ref{theorem: testing algorithm} satisfies the following:
    \begin{enumerate}
        \item 
            Query complexity is independent of $\mathcal{L}$ and $|\G|$, and 
            
        \item 
            it is polynomial in $\Tilde{m}$, $1/\theta$ and $\ln \mathcal{N}$. 
    \end{enumerate}
\end{remark}

Wimmer and Yoshida~\cite{WY13} proved a special case of the above theorem, which they referred to as the {\em Implicit Sieve Algorithm} for functions over $\ftwo^n$ and they used for testing linear isomorphism testing. Theorem~\ref{lemma: Implicit Sieve} generalizes the Implicit Sieve algorithm of Wimmer and Yoshida to finite Abelian groups. Note that the Implicit Sieve Algorithm of Wimmer and Yoshida~\cite{WY13} is itself a generalization of the celebrated and widely applicable Goldreich-Levin algorithm~\cite{goldreich1989hard}.

Using Theorem~\ref{theorem: testing algorithm} we have the following result for the case of subset isomorphism problem. 
\begin{theorem}
    Let $\G$ be a finite Abelian group and $A, B \subset \G$. Also let $\epsilon, \tau \in (0,1/2]$.
    Given the known set $A \subset \G$ with $\|\widehat{\ind_{A}}\|_1 \leq s$, and membership query access to an unknown set $B \subset G$, the query complexity of deciding whether $\dist_{\G}(A,B) \leq \epsilon$ or $\dist_{\G}(A,B) \geq \epsilon + \tau$ is $\Tilde{O}\left(\frac{s^{8}}{\tau^{8}}\right)$, where $\Tilde{O}(\cdot)$ hides multiplicative factors in $\log s$ and $\log(1/\tau)$ up to constants, and $\mathcal{L}$ is a constant. 
\end{theorem}
The above result is part of a broader research program exploring the interplay between the spectral norm of indicator functions and the structure of subsets of finite Abelian groups. In additive combinatorics and additive number theory, the spectral norm serves as a key tool for measuring the non-uniformity or pseudorandomness of functions defined on finite Abelian groups or their vector spaces. Notable breakthroughs in this area include the quantitative version of Cohen’s Idempotent Theorem due to Green and Sanders~\cite{GreenSAnnals08}, as well as Chang’s Lemma~\cite{Chang02} and its various extensions.

We also prove that for {\em Fourier sparse} functions, testing can be performed more efficiently. The proof follows a similar approach to the theorem on tolerant isomorphism testing for Finite Abelian Groups (Theorem~\ref{theorem: testing algorithm}), with a key observation: when both functions are Fourier $s$-sparse, the number of nonzero Fourier coefficients is at most $s$, which significantly simplifies the learning process in the implicit sieve algorithm (Theorem~\ref{lemma: Implicit Sieve}). In particular, we do not need to estimate $\wt_4$, which is required in the general case.

\begin{restatable}[$(\epsilon, \tau)$-Tolerant Isomorphism Testing of Sparse Boolean Functions over Finite Abelian Groups]{theorem}{}
\label{thm_isomorphism_test_sparse}
    Let $\G$ be a finite Abelian group, hence $\G =\mathbb{Z}_{p_1^{m_1}}\times \cdots \times \mathbb{Z}_{p_n^{m_n}}$ where $p_i$ are primes for all $i \in [n]$ and not necessarily distinct. 
    Given a known function $g: \mathcal{G} \to \zone$ with sparsity $s$ and query access to a unknown $s$-sparse Boolean valued function  $f: \mathcal{G} \to \zone$, the query complexity of deciding whether $\dist_{\G}(f,g) \leq \epsilon$ or $\dist_{\G}(f,g) \geq \epsilon + \tau$ 
    is $\Tilde{O}\left( \frac{s^{4}}{\tau^{4}}\right)$, where $\Tilde{O}(\cdot)$ hides multiplicative factors in $\log s$ and $\log\log(1/\tau)$ up to constants and $\epsilon, \tau \in (0,1/2]$, and $\mathcal{L}$ is a constant. 
    % $c_\mathcal{L}$ is a polynomial function in $\mathcal{L}$, and $\mathcal{L} = \LCM\{p_1^{m_1}, \ldots, p_n^{m_n}\}$.
\end{restatable}

%\textcolor{red}{Add theorem from section 8 and proof ideas }\\
% We also give a new proof of the Goldreich-Levin algorithm for Boolean functions whose domain is a finite Abelian group.

% %\manaswi{State theorem Goldreich-Levin}
% \begin{restatable}[Generalized Goldreich-Levin Algorithm for Abelian groups]{theorem}{Goldreich-Levin}
% \label{th:Goldreich-Levin}
% Let $\G$ be a finite Abelian group, that is, $\G = \mathbb{Z}_{p_1^{n_1}}\times \cdots \times \mathbb{Z}_{p_T^{n_T}}$ where $p_i$ are primes for all $i \in [T]$ and not necessarily distinct. Then there exists a randomized algorithm such that given query access to $f: \mathcal{G} \to \zone$ and $\eta >0 $ it runs in time $\text{poly}(T, 1/\eta)$ and with probability $\geq \frac{2}{3}$ outputs a list of Fourier characters $\mathcal{L}$ such that 

%       \begin{enumerate}
%         \item
%             $\forall \alpha \in \mathcal{G}$ with $\left|\widehat{f}(\chi_\alpha)\right| \geq \eta$ then $\alpha \in \mathcal{L}$.
%         \item 
%             $\forall \alpha \in \mathcal{L}$, $\left| \widehat{f}(\chi_\alpha) \right| \geq {\eta}/{2}$.
%     \end{enumerate}
% \end{restatable}
% Given the wide applicability of the Goldreich-Levin algorithm we belive that it's extension, i.e. Theorem~\ref{th:Goldreich-Levin} and it's implicit version for finite Abelian groups, i.e. Theorem~\ref{lemma: Implicit Sieve}, will find several other applications.

\subsection{Proof sketch}
\label{subsection:proof_sketch}

We start by giving the ideas behind and a sketch of the proof of our main result, Theorem~\ref{theorem: testing algorithm}. Throughout this section $\G$ is assumed to be a finite Abelian group.

\begin{proof}[Proof sketch of Theorem~\ref{theorem: testing algorithm} (Isomorphism Testing for finite Abelian groups)]
%\paragraph{Proof sketch of Theorem~\ref{theorem: testing algorithm} (Isomorphism Testing for finite Abelian groups).}
Given a known function \( g: \G \to \zone \) and an unknown function \( f: \G \to \zone \), where either  
\[
\dist_{\G}(f,g) \leq \epsilon \quad \text{or} \quad \dist_{\G}(f,g) \geq \epsilon + \tau,
\]  
the goal is to determine which case holds with probability at least \( {2}/{3} \).  
The algorithm has query access to \( f \), meaning it can evaluate \( f(x) \) for any \( x \in \G \).

\begin{itemize}
    \item Our algorithm (Algorithm~\ref{algo 1 testing}) uses the \textsf{Generalized Implicit Sieve Algorithm} (see Algorithm~\ref{algo:Implicit_Sieve_generalized} and Theorem~\ref{lemma: Implicit Sieve}) as a subroutine. Given query access to $f$, $\Tilde{m}$ random elements $x_1, \dots, x_{\Tilde{m}} \in \G$ and a parameter $\theta$, the \textsf{Generalized Implicit Sieve Algorithm} returns an $\Tilde{m} \times (\mathcal{N}+1)$ matrix $Q$ which satisfies the following with high probability (ignoring the last column of $Q$ for simplicity):
    \begin{enumerate}
        \item For some unknown set $\cS = \{r_1, \dots, r_\mathcal{N}\}$, the $(i,j)$th entry of $Q$ is $\chi_{r_j}(x_i)$. Thus the $j$th column of $Q$ contains the evaluations of the characters corresponding to $r_j$ on $\Tilde{m}$ random points in $\G$.

        \item If $r \in \cS$ then $|\wh{f}(\chi_r)| \geq \theta/2$ and if $r$ satisfies $|\wh{f}(\chi_r)| \geq \theta$, then $r \in \cS$.
    \end{enumerate}
    
    % Thus the \textsf{Implicit Sieve Algorithm} returns, with high probability, evaluations of $\cS$ (which is a set containing all `large' Fourier coefficients of $f$ and does not contain any `small' Fourier coefficient of $f$) at $m$ random points.

    \item 
        Since
        \begin{align*}
            \widehat{f}(\chi_{r_i}) = \mathbb{E}_{x \in \G} [f(x)\chi_{r_i}(x)]
        \end{align*}
        for all $r_i \in \mathcal{S}$ (see Section~\ref{section_prelims}),
        the output of the \textsf{Generalized Implicit Sieve Algorithm} can be used to estimate the 
        Fourier coefficients $\widehat{f}(\chi_{r_i})$, for $i \in [\mathcal{N}]$, 
        with suitably chosen accuracy and high probability, provided that $\Tilde{m}$ is large enough.
        This follows from an application of the Hoeffding bound. However, note that the set
        $\mathcal{S}$ remains unknown.

    \item 
        We now describe how we use these estimates to construct a function $\tilde{f}: \G \to \R$ such that for some unknown group automorphism $A$, the correlation between $\tilde{f} \circ A$ and $f$ is large. In order to do this, we need the idea of \textit{pseudo-independence} over finite Abelian groups (see Definition~\ref{defn_independence}).

    \item 
    For any set of elements $r_1, \ldots, r_k \in \G$, if there exists $\lambda_1, \ldots, \lambda_k \in \mathbb{N}\cup\{0\}$, such that at least one $\lambda_j$ is invertible in $\mathbb{Z}_\mathcal{L}$, where $\mathcal{L} = \LCM\{p_1^{m_1}, \ldots, p_n^{m_n}\}$,
    and $\sum_{i=1}^k \lambda_i r_i = 0$, then we call $r_1, \ldots, r_k$ to be \textit{pseudo-dependent} (Note that any subset has a pseudo-independent spanning set, see Definition~\ref{defn_independence}). If there does not exist any such $\lambda_i$'s, then $r_1, \ldots, r_k$ are \textit{pseudo-independent}. The idea is, if there exists such $\lambda_1, \ldots, \lambda_k \in \mathbb{Z}_\mathcal{L}$ with $\lambda_j$ invertible in $\mathbb{Z}_\mathcal{L}$, then $r_j$ can be written as a combination of the other elements. That is, 
    \begin{align}
        r_j = -\lambda_j^{-1} \bigl( \sum_{i\in [k], i \neq j} \lambda_i r_i \bigr), 
        \label{eq: intro pseudo dependent} 
    \end{align} 
    where $\lambda_j^{-1}$ is the inverse of $\lambda_{j}$ in $\mathbb{Z}_\mathcal{L}$.
    Observe that the above definition generalized the notion of linear dependence and independence over vector spaces.
    Furthermore, whenever we define a group automorphism on the elements $r_1, \ldots, r_{j-1}, r_{j+1}, \ldots, r_k$, by the property of homomorphism it automatically gets defined on $r_j$. That is,
    \begin{align*}
        \psi(r_j) = -\lambda_j^{-1} \bigl( \sum_{i\in [k], i \neq j} \lambda_i \psi(r_i) \bigr).
    \end{align*}
    Recall that $\lambda_j$ is invertible in $\mathbb{Z}_\mathcal{L}$ and $\lambda_j^{-1}$ is the inverse of $\lambda_{j}$ in $\mathbb{Z}_\mathcal{L}$.

    \item To construct $\Tilde{f}$ we first identify a $R \subseteq \cS$ (without knowing $R$) such that the elements in $R$ are {pseudo-independent} and $\cS$ is a subset of the subgroup generated by $R$ (which can be checked by a brute-force algorithm, see Algorithm~\ref{algo 1 testing}). We prove that such an $R$ can be found with high probability. 
    This will allow us to relabel the elements in $R$ as $e_1, \dots, e_{|R|} \in \widehat{\G}$ and the elements in $\cS \setminus R$ as a combination of $e_1, \dots, e_{|R|}$, using Equation~\eqref{eq: intro pseudo dependent}, for some automorphism $A$ of $\G$. 
    Relabel the columns of $Q$ by $\{S_1, \dots ,S_{\mathcal{N}}\}$ where $S_1 = e_1, \dots S_{|\Tilde{B}|} = e_{|\Tilde{B}|}$ such that for all $j \in \{{|\Tilde{B}|}+1, \dots, \mathcal{N}\}$, $S_j$ can be written as 
    $$
        \lambda^{-1} \biggl( \sum_{i\in [|\Tilde{B}|]} \lambda_i S_i \biggr),
    $$ 
    where 
    $\lambda, \lambda_1, \ldots, \lambda_{|\Tilde{B}|} \in \mathbb{N}\cup\{0\}$, and $\lambda$ is invertible 
    in $\mathbb{Z}_\mathcal{L}$. Note that the columns $S_1, \dots, S_{|\Tilde{B}|}$ correspond to those in $\Tilde{B}$.
    
    \item 
        We will create $\Tilde{f}$ such that for all $j \in [\mathcal{N}]$, $\widehat{\Tilde{f}}(\chi_{S_j}) = r_j$
        and $\widehat{\Tilde{f}}(\chi_S) = 0$ if $S \notin \{S_1, \dots, S_{\mathcal{N}}\}$ where  
        \begin{align*}
            r_j = \frac{\sum_{i=1}^{\Tilde{m}} f(x^{(i)})Q[i,j]}{\Tilde{m}}.
        \end{align*}

    \item 
        Thus at this point we have constructed $\Tilde{f}$ that has high correlation with $f$ under some unknown group automorphism $A$, with high probability. We show that to decide whether $ \dist_{\G}(f,g) \leq \epsilon$ or $ \dist_{\G}(f,g) \geq \epsilon + \tau$, it is sufficient to 
        check the correlation between $\tilde{f}$ and $f$ under all possible group automorphisms on $\G$ (see Observation~\ref{obs: distance vs. Fourier coefficient}) and this can be done without making queries to $f$.

\end{itemize}
We refer the reader to Section~\ref{sec: query algo} for the details.
\end{proof}

As evident, the \textsf{Generalized Implicit Sieve Algorithm} (Algorithm~\ref{algo:Implicit_Sieve_generalized}) plays an important role in the proof of our main result.

\begin{proof}[Proof idea of Theorem~\ref{lemma: Implicit Sieve}. ({Generalized Implicit Sieve Algorithm})]
%\paragraph{Proof idea of Theorem~\ref{lemma: Implicit Sieve}. ({Generalized Implicit Sieve Algorithm})}
Before we outline the proof of Theorem~\ref{lemma: Implicit Sieve}, let us first explain the goal of
Algorithm~\ref{algo:Implicit_Sieve_generalized}. 
\begin{itemize}
    \item The algorithm takes a Boolean valued function $f: \G \to \zone$ as input, where $\G = \mathbb{Z}_{p_1^{m_1}} \times \cdots \times \mathbb{Z}_{p_n^{m_n}}$, and $p_i, i \in [n]$ are primes, and not necessarily distinct. It is also given a threshold $\theta$ and a set $\mathcal{M} = \{x_1, \ldots, x_{\Tilde{m}}\}$ of uniformly random points from $\G$, where $\Tilde{m}$ is sufficiently large. The algorithm considers a randomly permuted coset structure $(H, \mathcal{C})$ such that the subgroup $H$ has codimension $\leq t$ in $\G$ (see Definition~\ref{defn_codimension}), where $t\geq \log_{\mathcal{L}} \bigl( \frac{100^4 \Tilde{m}^4}{\theta^{32}} \bigr)$, where $\mathcal{L} = \LCM\{p_1^{m_1}, \ldots, p_n^{m_n}\}$. Then $\wt_2$ (see Definition~\ref{defn_wt_2}) is estimated for each bucket with error $\pm \frac{\theta^2}{4}$ and confidence $1- \frac{1}{100\mathcal{L}^t}$. 
    If the estimated weight $\Tilde{\wt_2}$ is $\geq \frac{3\theta^2}{4}$, then the algorithm keeps the bucket. Otherwise, it discards the bucket. For each surviving bucket $C$, $\wt_4$ (see Definition~\ref{defn_wt_4}) is estimated with error $\pm \frac{\theta^4}{8} \Tilde{\wt_2}(C)$ and confidence $1- \frac{1}{100\mathcal{L}^t}$. If the estimated weight $\Tilde{\wt_4}$ is $\geq \frac{3\theta^4}{4}$, then the algorithm keeps the bucket. Otherwise, it discards the bucket. Then it randomly picks $|\mathcal{M}|=\Tilde{m}$ points $\{y_1, \ldots, y_{\Tilde{m}}\}$ from $\G$, and calculates $P_Cf(y_i)\overline{P_Cf(y_i-x_i)}$ for all $i \in [\Tilde{m}]$.
    Note that $P_C$ is the projection operator on functions \( f :\G \to \mathbb{R} \), which is defined as follows $(C \subseteq \G)$
    $$ 
        {P}_{C}f(x) := \sum_{\beta \in C} \hat{f}(\beta) \chi_\beta(x).
    $$
    The Generalized Implicit Sieve Algorithm makes $O\biggl(\frac{\ln \frac{\Tilde{m}}{\theta^{16}}}{\theta^8} + \Tilde{m} \frac{\ln(\mathcal{N} \Tilde{m})}{\theta^2} + \frac{\ln \mathcal{N}}{\theta^8} \biggr)$ many queries.
    % $c_\mathcal{L}$ is a function of $\mathcal{L}$.
\end{itemize}

We now outline the proof of Theorem~\ref{lemma: Implicit Sieve}.
\begin{itemize}
%    \item 
%        Let first consider the case when the unknown function $f$ is Fourier $s$-sparse (see Definition~\ref{}).
%        Observe that by taking $s^2$ many buckets ensures us that the characters fall into separate buckets with high probability. Then separating out the buckets with big Fourier coefficients, we construct a matrix $Q$ whose $(i,j)^{th}$ entry is given by $\frac{P_{\alpha(C_i)}f(y_j) \overline{P_{\alpha(C_i)}f(y_j-x_j)}}{|P_{\alpha(C_i)}f(y_j) \overline{P_{\alpha(C_i)}f(y_j-x_j)}|} = \chi_{\alpha(C_i)}(x_j)$.

    \item
        The first step in the proof of the theorem is to partition $\mathcal{G}$ into cosets of a random subgroup. Since $\mathcal{G}$ is Abelian, all its subgroups are normal.  
        Define a random subgroup $H$ by sampling $\beta_1, \dots, \beta_k$ independently and uniformly from $\mathcal{G}$ and using the pseudo-inner product operator on $\mathcal{G}$ to obtain  
        \[
            H = \left\{\alpha \in \mathcal{G} : \alpha * \beta_i = 0 \ \forall i \in [k]\right\}.
        \]  
        By Definition~\ref{defi:pseudo-inner product} and Definition~\ref{defn_cosets_groups}, $H$ forms a subgroup of $\mathcal{G}$. Its cosets have the form  
        \[
            C(b) = \{\alpha \in \mathcal{G} : \alpha * \beta_i = b_i \ \forall i \in [k]\},
        \]  
        where $b = (b_1, \dots, b_k) \in \mathbb{Z}_{\mathcal{L}}^k$. We refer to the cosets of $H$ as ``buckets".

    \item 
        We show that with high probability, all the big Fourier coefficients, that is, the coefficients with weight $\geq \frac{\theta^2}{4}$ belong to different buckets (see Lemma~\ref{lemma_large_char_diff_buckets}).
    
    \item 
        Moreover, 
        we show that the weight of a bucket $C_\alpha$ is determined by the big coefficient $\chi_{\alpha(C)}$ contained in $C_\alpha$ with high probability, assuming that the big coefficients are in different buckets with high probability. We use the second moment method to show this. 
    Let for each $\gamma \in \G$,
    \begin{align*}
        Y_{\gamma,\alpha} =
        \begin{cases}
            1, &\text{if } \gamma\in C_\alpha \\
            0, &\text{otherwise}.
        \end{cases}
    \end{align*}
    And, $$Y_\alpha = \sum_{\gamma\in S} Y_{\gamma, \alpha}.$$
    Also, let    
    \begin{align*}
        X_{\gamma,C_\alpha} = 
        \begin{cases}
            |\widehat{f}(\chi_\gamma)|^2 &\text{if } \gamma \in C_\alpha \\
            0 &\text{otherwise.}
        \end{cases}
    \end{align*}
    And, $$X_{C_\alpha} = \sum_{\gamma \in S} X_{\gamma,C_\alpha}.$$ 
    
    The $b=0$ case (see Definition~\ref{defn_random_coset_structure}) is problematic, as the $\Cov[Y_{\gamma_1, \alpha} Y_{\gamma_2,\alpha}] >0$ in this case and hence finding an upper bound for $\Pr[|X_{C_\alpha} - \mathbb{E}[X_{C_\alpha}]| \geq \epsilon]$ becomes difficult using Chebyshev. We use \textbf{random shift} (see Definition~\ref{def:random-shift}) to avoid this, and study case by case to show that $\Cov[Y_{\gamma_1, \alpha} Y_{\gamma_2,\alpha}] = 0$. This gives us an upper bound on the variance of $X_{C_\alpha}$, which in turn helps us to show that the weight of a bucket $C_\alpha$ is determined by the big coefficient $\chi_{\alpha(C)}$ contained in $C_\alpha$ with high probability. This implies that with high probability, the Fourier coefficients other than the big coefficient $\chi_{\alpha(C)}$ does not contribute much to the bucket $C_\alpha$.

    \item 
        We then show that the algorithm does not discard any bucket $C$ with $|\widehat{f}(\chi_\gamma)| \geq \theta$, but it discards any bucket $C$ with $|\widehat{f}(\chi_\gamma)| < \frac{\theta^2}{4}$, for all $\gamma \in C$. Furthermore, we establish that $|\widehat{f}(\chi_{\alpha(C)})| \geq \frac{\theta}{2}$ for every bucket $C=C_\alpha$ (where $\alpha(C)$ is the dominating element of $C_\alpha$) among those that the algorithm has not discarded.

\item 
    Additionally, we show that the small weight coefficients do not contribute much to the weight of a bucket. That is, we show that $\Pr_x[|\sum_{r \in C_\alpha \setminus \{\chi_{\alpha(C)}\}} \widehat{f}(\chi_r) \chi_r(x)| \geq \frac{\theta^2}{4}]$ is very small, where $\chi_{\alpha(C)}$ is the dominating Fourier coefficient of the bucket $C_\alpha$.

\item Finally, our target is to output a matrix $Q$ (Equation~\eqref{eqn_matrix_Q}) with a small error, whose entries are characters evaluated at the points $x_i, i \in [\Tilde{m}]$, where $x_i$ are picked independently and uniformly at random. To show that, we pick $y_i$ independently and uniformly at random, and then prove that the distance between $P_{C_\alpha}f(y_i) \overline{P_{C_\alpha}f(y_i-x_i)}$ and $|\widehat{f}(\chi_{\alpha(C)})|^2 \chi_{\alpha(C)}(x_i)$ is $\leq \frac{9\theta^4}{16}$ with high probability. Then we construct a matrix $Q'$ (see Equation~\eqref{eqn_matrix_Q'}), and estimate the Fourier coefficients for all the large buckets. We then show that, up to some small error, the algorithm outputs the matrix $Q$ (see Equation~\eqref{eqn_matrix_Q}) whose entries are given by $\chi_{\alpha(C_i)}(x_j)$, $i \in [\mathcal{N}], j \in [\Tilde{m}]$, where $\mathcal{N}$ denotes the number of survived buckets.
\end{itemize}
\end{proof}

\bigskip

\subsection{Challenges with finite Abelian Groups}
\label{subsection:our_difficulties}

We outline the main challenges encountered in proving our results for finite Abelian groups $\mathcal{G}$.
Recall that $\G$ can be written as $\G = \mathbb{Z}_{p_1^{m_1}} \times \cdots \times \mathbb{Z}_{p_n^{m_n}}$, where $p_i, i \in [n]$ are primes, and not necessarily distinct.

\begin{itemize}
    \item While estimating $\wt_2$ and $\wt_4$, it is required to find $H^\perp$ for a subgroup $H$ of $\G$. But due to the lack of vector space structure, $H^\perp$ cannot be defined in terms of `basis', which is possible when $\G=\mathbb{Z}_2^n$, as $\mathbb{Z}_2^n$ is a vector space. This problem had to be taken care of in a different way. Given a subgroup $H$ of $\G$, we define $H^\perp$ by 
    \begin{align*}
        H^\perp := \{ x \in \G : x * y =0 \ \forall y \in H \},
    \end{align*}
    where $*$ is a pseudo inner product defined by $$x * y := \biggl( \sum_{i=1}^{T} \frac{\mathcal{L}}{p_i^{m_i}} x^{(i)} \cdot y^{(i)} \biggr) \pmod{\mathcal{L}},$$ and $\mathcal{L} = \LCM \{ p_1^{m_1}, \ldots, p_n^{m_n} \}$. 
    It is to be noted that $H^\perp$ is also a subgroup of $\G$.

    % Also, since any subgroup of $\G$ is of the form $\langle p_1^{i_1} \rangle \times \cdots \times \langle p_T^{i_T} \rangle$, where $i_j \in \{0, \ldots n_j-1\}$ for all $j \in [T]$, let $H = \langle p_1^{i_1} \rangle \times \cdots \times \langle p_T^{i_T} \rangle$. Here $\langle x \rangle$ denotes the subgroup generated by $x$. Then, by Lemma~\ref{lemma_H_perp_equivalence}, it follows that $H^\perp$ is the subgroup $\langle p_1^{n_1-i_1} \rangle \times \cdots \times \langle p_T^{n_T-i_T} \rangle$. 
    
    This is extremely useful in proving the fact that, given a bucket $C_\alpha$, $\wt_2(C_\alpha)$ and $\wt_4(C_\alpha)$ can be written as an average of some quantity, which in turn helps to reduce the query complexity of $\wt_2$ and $\wt_4$ estimation. More precisely, by Lemma~\ref{lemma_wt_of_bucket_new} and Claim~\ref{claim_wt_4_avg}, it follows that, 
    $$\wt_2\bigl(\mathcal{C}_{\alpha}\bigr) = \mathbb{E}_{x \in \G, z \in H^\perp} \biggl[ f(x) f(x+z) \chi_r(z) \biggr]$$ 
    and
    $$\wt_4(C_\alpha) = \mathbb{E}_{x,z_1,y_1\in \G, z,y \in H^\perp} \biggl[ f(z_1) f(x-z_1-z) f(y_1) f(x-y_1-y) \chi_r(z-y) \biggr]$$
    respectively. Applying a concentration inequality the query complexity becomes dependent on $\theta$ and $\Tilde{m}$ only, where $\theta$ and $\Tilde{m}$ are as defined in Theorem~\ref{lemma: Implicit Sieve}, and it is not dependent on the order of the group $\G$.

    \item To prove our main theorem, we also need to show that, for any automorphism $A: \G \to \G$ and any function $f: \G\to \{-1,+1\}$, the Fourier coefficients $\widehat{f\circ A}(\chi_r), \ r \in \G$ of $f\circ A$ are same as the Fourier coefficients of $f$ upto some permutation. In case of $\mathbb{Z}_2^n$, it was quite easy, as $\mathbb{Z}_2^n$ is a vector space. As an Abelian group is not a vector space in general, we needed to use Pontryagin duality and dual automorphisms. For more details, see Lemma~\ref{lemma_pontryagin} and Lemma~\ref{lemma_aut_perm_char} in Section~\ref{section_prelims}.

    \item Due to the lack of vector space structure, we cannot partition $\G$ into subspaces. Instead, we define a normal subgroup $V_{0,r_1,\ldots,r_k}$ (see Lemma~\ref{lemma_cosets_of_V_0}), and partition $\G$, and hence $\widehat{\G}$ into its cosets. For details see Section~\ref{section_prelims}.

    \item While testing isomorphism, the notion of linear independence is required. But when there is no vector space structure, linear independence does not make any sense. A different notion of `independence' needed to be defined in order to tackle this problem. 
    
    For any set of elements $r_1, \ldots, r_k \in \G$, if there exists $\lambda_1, \ldots, \lambda_k \in \mathbb{N}\cup\{0\}$, with at least one $\lambda_j$ an unit in $\mathbb{Z}_\mathcal{L}$ (that is, $\lambda_j$ is invertible in $\mathbb{Z}_\mathcal{L}$), 
    $\mathcal{L} = \LCM\{p_1^{m_1}, \ldots, p_n^{m_n}\}$, 
    such that $\sum_{i=1}^k \lambda_i r_i = 0$, then we call $r_1, \ldots, r_k$ to be dependent. If there does not exist any such $\lambda_i$'s, then $r_1, \ldots, r_k$ are independent. The idea is, if there exists such $\lambda_1, \ldots, \lambda_k \in \mathbb{Z}_\mathcal{L}$ with $\lambda_j$ an unit, then $r_j$ can be written as a combination of the other elements. That is, 
    $$r_j = -\lambda_j^{-1} \bigl( \sum_{i\in [k], i \neq j} \lambda_i r_i \bigr).$$ So, whenever we define a group automorphism on the elements $r_1, \ldots, r_{j-1}, r_{j+1}, \ldots, r_k$, by the property of homomorphism it automatically gets defined on $r_j$.
    This plays an important role in Algorithm~\ref{algo 1 testing}.

    \item We know that the characters of a Boolean function from $\mathbb{Z}_2^n$ to $\{0,1\}$ takes values $-1$ or $+1$ at each point of $\mathbb{Z}_2^n$. So, to determine the dominating character in a bucket at a point $x_i \in M$ with high probability, determining the sign of $P_{C_\alpha}f(y_i) P_{C_\alpha}f(y_i+x_i)$ is sufficient, where $\chi_{\alpha(C)}$ being the dominating character of the bucket $C_\alpha$.

    But in our case, the characters of a finite Abelian group $\G$ are complex-valued at each point of $\G$. So sign does not make any sense here. The problem needed to be tackled differently. We first calculate $P_{C_\alpha}f(y_i) \overline{P_{C_\alpha}f(y_i-x_i)}$, and then we show that $P_{C_\alpha}f(y_i) \overline{P_{C_\alpha}f(y_i-x_i)} - |\widehat{f}(\chi_{\alpha(C)})|^2 \chi_{\alpha(C)}(x_i)$ is small with high probability, which helps us to understand the dominating character in the bucket $C_\alpha$ at the point $x_i\in M$. For details see Section~\ref{section_implicit seive}.
\end{itemize}

\section{Background on Finite Abelian Groups}\label{section_prelims}
%\manaswi{General Comment: If something is not proved it might be nice to provide a reference.}\textcolor{blue}{(SD: Okay!)}
Throughout this paper, $\G$ denotes a finite Abelian group $\mathbb{Z}_{p_1^{m_1}}\times \cdots \times \mathbb{Z}_{p_n^{m_n}}$, where $p_i$ are primes for all $i \in [n]$ and may not be distinct, and $\omega_\mathcal{L}$ denotes a primitive $\mathcal{L}^{th}$ root of unity, where $\mathcal{L} = \LCM \{p_1^{m_1}, \ldots, p_n^{m_n}\}$. 

Let us define the characters of $\G$ first.

\begin{defi}\label{defn_character_1_groups}
    \textbf{(Character)} For $\G = \mathbb{Z}_{p_1^{m_1}} \times \cdots \times \mathbb{Z}_{p_n^{m_n}}$, where $p_i, i \in [n]$ are primes, a character of the group $\G$ is a homomorphism $\chi: \G \to \mathbb{C}^\times$ of $\G$, that is, $\chi$ satisfies the following: for all $x,y \in \G$ we have $\chi(x+y)=\chi(x) \chi(y)$.

    \noindent Equivalently, a character of $\G$ is of the form $$\chi(x^{(1)},\dots, x^{(n)}) = \chi_{r^{(1)}}(x^{(1)}) \cdots \chi_{r^{(n)}}(x^{(n)}),$$ where $r^{(i)} \in \mathbb{Z}_{p_i^{m_i}}$ and $\chi_{r^{(i)}}$ is a character of $\mathbb{Z}_{p_i^{m_i}}$ and is defined by $$\chi_{r^{(i)}}(x^{(i)}) = \omega_{p_i^{m_i}}^{r^{(i)} x^{(i)}}.$$ 
    Thus for any $(r^{(1)}, \dots, r^{(n)})\in \G$ we can define a corresponding character of $\G$ as $\chi_{r^{(1)}, \dots, r^{(n)}}$ that on input  $x = (x^{(1)}, \dots, x^{(n)})$ is defined as
    \begin{align}\label{eq:character}
    \chi_{(r^{(1)}, \ldots, r^{(n)})}(x^{(1)}, \ldots, x^{(n)}) 
    &= \omega_{p_1^{m_1}}^{r^{(1)}\cdot x^{(1)}} \cdots \omega_{p_n^{m_n}}^{r^{(n)}\cdot x^{(n)}} \\
    &= \omega_\mathcal{L}^{\sum_{i=1}^T r^{(i)}\cdot x^{(i)} \frac{n}{p_i^{m_i}} \pmod{\mathcal{L}}},
\end{align}
where $\mathcal{L} = \LCM \{ p_1^{m_1}, \ldots, p_n^{m_n} \}$.
\end{defi}

Now let us look at some properties of characters.
%\textcolor{blue}{should add a clean note that now characters are complex and the complex inner product as well?}
\begin{lemma}\label{lemma_character_groups}
Let $\chi$ be a character of $\G$. Then,
\begin{enumerate}
    \item $\chi_0(x)=1$ for all $x\in \G$. 
    \item $\chi(-x)= \chi(x)^{-1}=\overline{\chi(x)}$ for all $x \in \G$. 
    \item For any character $\chi$ of $\G$, where $\chi\neq \chi_0$, $\sum_{x\in \G} \chi(x)=0$.
    \item $|\chi(x)|=1$ for all $x \in \G$.
\end{enumerate}
\end{lemma}

Now let us define the dual group of $\G$.

\begin{defi}
    \textbf{(Dual group)} The set of characters of $\G$ forms a group under the operation $(\chi\psi) (x) = \chi(x) \psi(x)$ and is denoted by $\widehat{\G}$, where $\chi$ and $\psi$ are characters of $\G$. $\widehat{\G}$ is called the dual group of $\G$.
\end{defi}

The following theorem states that $\G$ is isomorphic to its dual group.

\begin{theorem}
    $\widehat{\G} \cong \G$, that is $\G$ and $\widehat{\G}$ are isomorphic to each other.
    % \manaswi{More detailed statement along with a reference?}
\end{theorem}

Let us look at the definition of Fourier transform for functions on $\G$.

\begin{defi} \label{def:foruier-transform_groups}
    \textbf{(Fourier transform)} For any $\G = \mathbb{Z}_{p_1^{m_1}} \times \cdots \times\mathbb{Z}_{p_n^{m_n}}$, with $p_i, i \in [n]$ primes, and any function $f: \G \to \mathbb{C}$, the Fourier transform $\widehat{f}: \widehat{\G}\to \mathbb{C}$ is 
    $$\widehat{f}(\chi_{r^{(1)},\ldots,r^{(T)}}) =\frac{1}{|\G|} \sum_{x\in \G} f(x) \omega_{p_1^{m_1}}^{-r^{(1)}\cdot x^{(1)}} \cdots \omega_{p_n^{m_n}}^{-r^{(n)}\cdot x^{(n)}},$$ where $x = (x^{(1)}, \dots, x^{(n)})$. 

    % The Fourier transform of a function $f:\G \to \mathbb{C}$ is defined as $\widehat{f}(\chi)= \frac{1}{|\G|} \sum_{x\in \G} f(x) \overline{\chi(x)}$, where $\overline{\chi(x)}$ is the conjugate of $\chi(x)$. \manaswi{Repeated below.}
\end{defi}

\begin{remark}
    The Fourier transform of a function $f:\G \to \mathbb{C}$ is defined by $$\widehat{f}(\chi)= \frac{1}{|\G|} \sum_{x\in \G} f(x) \overline{\chi(x)},$$ where $\overline{\chi(x)}$ is the conjugate of $\chi(x)$. The Definition~\ref{def:foruier-transform_groups} follows from this, as $\chi=\chi_{r^{(1)},\ldots,r^{(n)}}$ for some $r^{(i)} \in \mathbb{Z}_{p_i^{m_i}}, \ i\in \{1,\ldots,n\}$.
\end{remark}

The following theorem states that any function from $\G$ to $\mathbb{C}$ can be written as a linear combination of characters of $\G$.

\begin{theorem}\label{Fourier_inversion_groups}
    \textbf{(Fourier inversion formula)} Any function $f: \G \to \mathbb{C}$ can be uniquely written as a linear combination of characters of $\G$, that is,
    \begin{equation} \label{eq:fourierex}
    f(x)= \sum_{\chi_{r^{(1)},\ldots,r^{(n)}} \in \widehat{\G}} \widehat{f}(\chi_{r^{(1)},\ldots,r^{(n)}}) \chi_{r^{(1)}, \dots, r^{(n)}}(x),\end{equation} 
    where $x = (x^{(1)}, \dots, x^{(n)})$. 
\end{theorem}

\begin{theorem}\label{theorem_Parseval_groups}
    \textbf{(Parseval)} For any two functions $f,g: \G \to \mathbb{C}$, $$\mathbb{E}_{x\in \G} [f(x)\overline{g(x)}] =\sum_{\chi\in \widehat{\G}} \widehat{f}(\chi) \overline{\widehat{g}(\chi)}.$$
    More specifically, if $f:\G \to \{-1,1\}$ is a Boolean-valued function then
    $$
        \sum_{\chi \in \widehat{\G}} |\widehat{f}(\chi)|^2=1.
    $$
\end{theorem}

Now let us define the Fourier sparsity $s_f$ of a function $f$ on $\G$. 

\begin{defi}
\label{defi: Fourier spectral norm}
\textbf{(Fourier Spectral Norm)}
    The Fourier Spectral Norm or the Spectral norm of $f : \G \to \mathbb{C}$, denoted by $\|f\|_1$ is defined as the sum of the absolute value of its Fourier coefficients. That is,
    \begin{align*}
        \left\| \widehat{f} \right\|_1 = \sum_{r \in \G} \left| \widehat{f}(\chi_r) \right|.
    \end{align*}
\end{defi}

\begin{defi}
\textbf{(Sparsity and Fourier Support)} 
\begin{itemize}
\item 
        Fourier support $\supp(\widehat{f})$ of a function $f: \G \to \mathbb{C}$ denotes the set $\left\{ \chi \,\mid\, \widehat{f}(\chi) \neq 0\right\}$. 
    \item
        The Fourier sparsity $s_f$ of a function $f: \G \to \mathbb{C}$ is defined to be the number of non-zero Fourier coefficients in the Fourier expansion of $f$ (Theorem~\ref{Fourier_inversion_groups}).  Alternately, Fourier sparsity can be defined as the size of the Fourier support. In this paper, by sparsity of a function, we mean the Fourier sparsity of the function. Moreover, by $s$-sparse function we mean functions with Fourier Sparsity $s$. 
\end{itemize}
\end{defi}

We know that since $\mathbb{Z}_2^n$ is also a vector space over the field $\ftwo$, it has a linear algebraic structure. The characters are in one-to-one correspondence with $\ftwo$-linear forms. However, since a general Abelian group is not a vector space, these do not hold here. But one can define something along similar lines. 
% \manaswi{A line about what is going to defined?}\sdc{I have mentioned it below.}

An element $r$ of $\G$ is 
of the form $(r^{(1)}, \dots, r^{(n)})$, where $r^{(i)}$ is an element of $\mathbb{Z}_{p_i^{m_i}}$.

\begin{defi}
\label{defi:pseudo-inner product}
    \textbf{(Pseudo inner product)} For $x, r \in \G$ we denote by $*$ the following pseudo inner product. $$r * x := \biggl( \sum_{i=1}^{n} \frac{\mathcal{L}}{p_i^{m_i}} r^{(i)} \cdot x^{(i)} \biggr) \pmod{\mathcal{L}},$$ where $\mathcal{L} = \LCM \{ p_1^{m_1}, \ldots, p_n^{m_n} \}$.
\end{defi}

\begin{observation}\label{obs_chi_r_x}
    $r* x \neq 0 \Rightarrow \chi_r(x)\neq 1$. 
\end{observation}

Now, we partition $\G$ with the help of a normal subgroup.

\begin{defi}\label{defn_normal_subgroup}
\label{defn_cosets_groups}
    Let $r_1,\ldots,r_k \in \G$ such that $r_j= (r_j^{(1)},\ldots, r_j^{(n)})$, where $r_{j}^{(i)} \in \mathbb{Z}_{p_i^{m_i}}$ for all $i \in [n]$ and $j \in [k]$. Let $b= (b_1, \ldots, b_k) \in \mathbb{Z}_\mathcal{L}^k$, where $\mathcal{L} = \LCM\{ p_1^{m_1}, \ldots, p_n^{m_n}\}$. 
    We define the set $V_{b,r_1,\ldots,r_k}$ by
    \begin{align}\label{eqn_cosets_groups}
        V_{b,r_1,\ldots,r_k} = \{ x\in \G : r_j * x = b_j \pmod{\mathcal{L}} \ \forall j \in [k]\}.
    \end{align}
     That is, 
    \begin{align*}
        V_{b,r_1,\ldots,r_k} 
        &= \biggl\{x \in \G : \biggl( \sum_{i=1}^{T} \frac{\mathcal{L}}{p_i^{m_i}} r_j^{(i)} \cdot x^{(i)} \biggr) = b_j \pmod{\mathcal{L}} \ \forall j \in [k] \biggr\}.
    \end{align*}

    If $b_1 = b_2 = \dots = 0$ then we use $V_{0,r_1,\ldots,r_k}$ to denote $V_{b,r_1,\ldots,r_k}$.
\end{defi}

\begin{remark}\label{remark_cosets_groups}
    When $k=1$, then $b \in \mathbb{Z}_\mathcal{L}$, and we have only one element $r \in \G$, so this set becomes $V_{b,r} = \{ r * x =b\pmod{\mathcal{L}}\}$.
\end{remark}

\begin{lemma}\label{lemma_cosets_of_V_0}
    $V_{0,r_1,\ldots,r_k}$ is a normal subgroup of $\G$ and for any $b, r_1, \dots, r_k$ either $V_{b,r_1,\ldots,r_k}$ is a coset of $V_{0,r_1,\ldots,r_k}$ or $V_{b,r_1,\ldots,r_k} = \emptyset$.
\end{lemma}

\begin{proof}
    Clearly, $0\in V_{0,r_1,\ldots,r_k}$, where $0$ is the identity element of $\G$. Let $x,y \in V_{0,r_1,\ldots,r_k}$. Then, since $r_j * (x-y) = \sum_{i=1}^{T} \frac{\mathcal{L}}{p_i^{m_i}} r_j^{(i)} \cdot x^{(i)} - \sum_{i=1}^{T} \frac{\mathcal{L}}{p_i^{m_i}} r_j^{(i)} \cdot y^{(i)} =0 \pmod{\mathcal{L}}$ for all $j\in [k]$, so $x-y \in V_{0,r_1,\ldots,r_k}$. So $V_{0,r_1,\ldots,r_k}$ is a subgroup of $\G$, and hence a normal subgroup of $\G$ since $\G$ is Abelian.

    Since $V_{0,r_1,\ldots,r_k}$ is a normal subgroup of $\G$ we can now consider its cosets. 
    For each coset $B$ of $V_{0,r_1, \ldots,r_k}$, let $a(B)$ be a fixed coset representative (we choose a coset representative and fix it all through the proof). Then, if $y \in B$, then $y = a(B)+x$, where $x \in V_{0,r_1,\ldots,r_k}$. Now, for each $r_j$, 
    $$r_j * y = r_j * (a(B)+x) = r_j * a(B),$$ 
    since $r_j * x =0$ for all $r_j, \ j \in [k]$. 
    So, $r_j * y$ is fixed for each coset $B$. If we assume $r_j * a(B) = b_j \pmod{\mathcal{L}}$ for each $j \in [k]$, then we have $V_{b,r_1,\ldots,r_k}$ as a coset of $V_{0,r_1,\ldots,r_k}$, where $b=(b_1, \ldots ,b_k)$.

    If for a $b = (b_1, \ldots,b_k) \in \mathbb{Z}_\mathcal{L}^k$, there does not exist any $a(B)$ such that $r_j * a(B) = b_j \ \forall j \in [k]$, $V_{b, r_1, \ldots, r_k} = \emptyset$. 
\end{proof}

\begin{remark}\label{remark_subgroup_implies_normal}
    Note that since $\G$ is an Abelian group, any subgroup of $\G$ is a normal subgroup of $\G$.
\end{remark}

\begin{defi}\label{defn_codimension}
    \textbf{(Codimension)} Let $V_{b,r_1,\ldots,r_k}$ be as defined in Definition~\ref{defn_cosets_groups}. Then the codimension of $V_{b,r_1,\ldots,r_k}$ is given by 
    \begin{align*}
        \Codim(&V_{b,r_1,\ldots,r_k}) =  \min_{k'} \biggl\{k: \exists r_1',\ldots, r_{k'}' \in \{r_1, \ldots, r_k\} \text{ such that } \\
        &V_{b,r_1,\ldots,r_k} = \{x\in \mathcal{G}: r_j' * x =b_j \pmod{\mathcal{L}} \ \forall j \in [k]\} \biggr\},
    \end{align*}
    where $b =(b_1, \ldots, b_k) \in \mathbb{Z}_\mathcal{L}^k$, and $\mathcal{L} = \LCM\{p_1^{m_1}, \ldots, p_n^{m_n}\}$. 
\end{defi}

\begin{defi}\label{defn_random_coset_structure}
    \textbf{(Random coset structure)} Let us consider $\beta_1, \ldots, \beta_k \in \mathcal{G}$, which are chosen independently and uniformly at random from $\mathcal{G}$. Also let $H= \{\alpha \in \mathcal{G} :  \alpha * \beta_i = 0 \ \forall i \in [k]\}$. So $H$ is a subgroup, hence a normal subgroup of $\mathcal{G}$. Then, we define the sets $C(b)$ by $C(b)= \{\alpha \in \mathcal{G} : \alpha * \beta_i = b_i \ \forall i \in [k]\}$, for all $b=(b_1,\ldots, b_k) \in \mathbb{Z}_{\mathcal{L}}^k$. 
\end{defi}

\begin{remark}\label{remark_random_coset_structure}
    Observe that the nonempty $C(b)$ are cosets of $H$. We will often refer to them as buckets.
\end{remark}

\begin{defi}\label{defn_group_isomorphism}
    Let $G$ be a group, with group operation $\odot$. A group automorphism on $G$ is a bijective function $\psi: G \to G$ which satisfies the following.
    \begin{align*}
        \psi(x \odot y) = \psi(x) \odot \psi(y), \ \ \forall x,y \in G.
    \end{align*}
    The set of all automorphisms of $G$ forms a group under the composition of functions and is denoted by $\Aut(G)$.
\end{defi}

\begin{defi}\label{defn_independence}
    Let $r_1, \ldots, r_k$ be elements in $\G$. We call $r_1, \ldots, r_k$ to be \textbf{dependent} if there exists $\lambda_1,\ldots, \lambda_k \in \mathbb{Z}_\mathcal{L}$ with at least one $\lambda_j$ invertible in $\mathbb{Z}_\mathcal{L}$ such that $\sum_{i=1}^k \lambda_i r_i = 0$.

    If there does not exist any such $\lambda_1,\ldots, \lambda_k \in \mathbb{Z}_\mathcal{L}$ with at least one $\lambda_j$ invertible in $\mathbb{Z}_\mathcal{L}$ such that $\sum_{i=1}^k \lambda_i r_i = 0$, then we call $r_1, \ldots, r_k$ to be \textbf{independent}. 
    % \manaswi{Examples of dependent and independent would be nice. It is not clear where they exist so examples will be convincing.}\sdc{I have given an example below.}
\end{defi}

For example, consider the group $\mathbb{Z}_4$. The element $3$ is dependent on the element $1$, but it is independent of the element $2$. Whereas, $2$ is dependent on $3$, as $2 = 2 \cdot 3 (\mod 4)$. So the set $\{2, 3\}$ is dependent.

% \manaswi{Not sure if there should be cube-root in the following bound.}
\begin{lemma}[Hoeffding's Inequality] \label{Hoeffding}
Let $X_1, \cdots, X_k$ be real independent random variables, each taking value in [-1, 1]. Then 
\[\Pr \left[\left|\sum_{i=1}^k X_i - \mathbb{E}\sum_{i=1}^k X_i\right| \geq \epsilon\right] \le 2\exp\left(-\frac{\epsilon^2}{2k}\right). 
\]
\end{lemma}

The following structural result about an Abelian group and its dual will play a key role in our analysis. 
For a first reading, the proof of this lemma may be skipped without affecting the readability of the rest 
of the paper.

\begin{lemma}\label{lemma_pontryagin}
    Let $A$ be a map from $\G$ to $\G$. Let us define a map $\widehat{A}: \widehat{\G} \to \widehat{\G}$ by $\widehat{A}(\chi_r) = \chi_r \circ A$. Also, let us define another map $\widehat{\widehat{A}}: \G \to \G$ by $\chi_{\widehat{\widehat{A}} (r)} = \widehat{A}(\chi_r)$. Then $A$ is an automorphism if and only if $\widehat{A}$ is an automorphism if and only if $\widehat{\widehat{A}}$ is an automorphism.
\end{lemma}

% The proof of this lemma is given in \textcolor{red}{move the proof in appendix and refer here.}

\begin{proof}
\begin{description}
\item[Proof of the part [$A$ is an automorphism if and only if $\widehat{A}$ is an automorphism]] 
\begin{description}
    \item[When $A$ is an automorphism.] 
    First let us show that $\widehat{A}$ is also a homomorphism. Observe that,
    \begin{align*}
        \widehat{A}(\chi_{r_1} \chi_{r_2})(x) &= \widehat{A}(\chi_{r_1+r_2})(x) \\
        &= \chi_{r_1+r_2}\circ A(x) \\
        &= \chi_{r_1}(A(x)) \chi_{r_2}(A(x)) \\
        &= \widehat{A}(\chi_{r_1})(x) \widehat{A}(\chi_{r_2})(x),
    \end{align*}
which shows that $\widehat{A}$ is indeed a homomorphism.

To show injectivity, let $\widehat{A}(\chi_r) = 1$. Then, 
\begin{align*}
    \widehat{A}(\chi_r)(x) =1 \ \forall x \in \G 
    &\Rightarrow \chi_r(A(x)) =1 \ \forall x \in \G \\
    &\Rightarrow \chi_r(y) =1 \ \forall y \in \G, & \text{where } y = A(x) \\
    &\Rightarrow \chi_r = \chi_0,
\end{align*}
where the equality in the second last line holds because $A$ is an automorphism.
Since $\chi_0$ is the identity element of $\widehat{G}$, so $\widehat{A}$ is injective.

Since the domain and the range of $\widehat{A}$ are same, so $\widehat{A}$ is bijective, and hence, it is an automorphism on $\widehat{G}$.

\item[When $\widehat{A}$ is an automorphism.] To show that $A$ is a homomorphism, we observe that, for all $r \in \G$,
\begin{align*}
    &\widehat{A}(\chi_r)(x+y) = \widehat{A}(\chi_r)(x) \widehat{A}(\chi_r)(y) = \chi_r(A(x)) \chi_r(A(y)) = \chi_r(A(x)+A(y)) \\
    &\Rightarrow \chi_r(A(x+y)) = \chi_r(A(x)+A(y)) \\
    &\Rightarrow \chi_r (A(x+y) - A(x) -A(y)) =0.
\end{align*}
Therefore, $A(x+y) = A(x) + A(y)$, hence $A$ is a homomorphism.

Now let us show that $A$ is injective. To do that, let $A(x) =0$. Then,
\begin{align*}
    \widehat{A}(\chi_r)(x) = \chi_r(A(x)) = \chi_r(0) = 1 \ \forall r \in \G,
\end{align*}
which is only possible if $x=0$, since $\widehat{A}$ is an automorphism and maps characters to characters. So, $A$ is injective.

Since the domain and the range of $A$ are same, so $A$ is bijective, and hence, $A$ is an automorphism.
\end{description}

\item[Proof of the part [$\widehat{A}$ is an automorphism if and only if $\widehat{\widehat{A}}$ is an automorphism]]
\begin{description}
    \item[When $\widehat{A}$ is an automorphism.] Then, for each $x \in \G$,
    \begin{align*}
    &\chi_{\widehat{\widehat{A}}(r_1+r_2)}(x) 
    = \widehat{A}(\chi_{r_1+r_2})(x)
    = \chi_{r_1+r_2} (A(x)) 
    = \chi_{r_1}(A(x)) \chi_{r_2}(A(x)) = \widehat{A}(\chi_{r_1})(x) \widehat{A}(\chi_{r_2})(x) \\
    &\Rightarrow \chi_{\widehat{\widehat{A}}(r_1+r_2)}(x) = \chi_{\widehat{\widehat{A}}(r_1)}(x) \chi_{\widehat{\widehat{A}}(r_2)}(x) \\
    &\Rightarrow \chi_{\{\widehat{\widehat{A}}(r_1+ r_2) - \widehat{\widehat{A}}(r_1) - \widehat{\widehat{A}}(r_2)\}}(x) =0,
    \end{align*}
which implies that $\widehat{\widehat{A}}(r_1+ r_2) = \widehat{\widehat{A}}(r_1) + \widehat{\widehat{A}} (r_2)$ for all $r_1, r_2\in \G$. So $\widehat{\widehat{A}}$ is a homomorphism.

Now, let $\widehat{\widehat{A}}(r) =0$. Then, for all $x \in \G$,
\begin{align*}
    \chi_{\widehat{\widehat{A}}(r)}(x) = 1 \Rightarrow \widehat{A}(\chi_r)(x) = 1 \Rightarrow \chi_r(A(x)) =1.
\end{align*}
Since $\widehat{A}$ is an automorphism, so $A$ is an automorphism by the first part of the proof, therefore $\chi_r(y) = 1$ for all $y \in \G$, where $y = A(x)$. So $r =0$, hence $\widehat{\widehat{A}}$ is injective.

Since the domain and the range of $\widehat{\widehat{A}}$ is same, so $\widehat{\widehat{A}}$ is bijective, and hence, it is an automorphism.

\item[When $\widehat{\widehat{A}}$ is an automorphism.] Then, for each $x \in \G$,
\begin{align*}
    &\widehat{A}(\chi_{r_1+r_2})(x) = \chi_{\widehat{\widehat{A}}(r_1+r_2)}(x) = \chi_{\widehat{\widehat{A}}(r_1) + \widehat{\widehat{A}} (r_2)}(x) = \chi_{\widehat{\widehat{A}}(r_1)}(x) \chi_{\widehat{\widehat{A}}(r_2)}(x) \\
    &\Rightarrow \widehat{A}(\chi_{r_1}\chi_{r_2})(x) = \widehat{A}(\chi_{r_1})(x) \widehat{A}(\chi_{r_2})(x),
\end{align*}
which shows that $\widehat{A}$ is a homomorphism.

Now, let $\widehat{A}(\chi_r)(x) =1$ for all $x \in \G$. Then, for all $x \in \G$,
\begin{align*}
    \chi_{\widehat{\widehat{A}}(r)}(x) =1 \Rightarrow \widehat{\widehat{A}}(r) = 0 \Rightarrow r =0,
\end{align*}
since $\widehat{\widehat{A}}$ is an automorphism. Therefore, $\widehat{A}$ is injective.

Since the domain and the range of $\widehat{A}$ is same, so $\widehat{A}$ is bijective. Hence, $\widehat{A}$ is an automorphism.
\end{description}
\end{description}
\end{proof}

\begin{defi}\label{defn_func_iso}
    \textbf{(Boolean function isomorphism)} Let $f,g:\G \to \{-1,+1\}$ be Boolean valued functions. Then $f$ is isomorphic to $g$ if there exists an automorphism $A$ on $\G$ such that $f = g\circ A$.
\end{defi}

\begin{defi}\label{defn_iso_dist}
    \textbf{(Automorphism distance)} Let $f,g:\G \to \{-1,+1\}$ be Boolean valued functions. The automorphism distance between $f$ and $g$ is defined by
    \begin{align*}
        \dist_{\G}(f,g) = \min_{A \in \Aut(\G)} \delta(f,g\circ A),
    \end{align*}
    where $\delta(f,g\circ A)$ is the fractional Hamming distance between $f$ and $g \circ A$.
\end{defi}

\begin{defi}
    \textbf{($\epsilon$-close and $\epsilon$-far from isomorphic)} Let $f,g : \G \to \{-1,+1\}$ be two Boolean valued functions. Then $f$ is said to be $\epsilon$-close ($\epsilon$-far) from being isomorphic to $g$ if $\dist_{\G}(f,g) \leq \epsilon$ ($\dist_{\G}(f,g) \geq \epsilon$). That is, $f$ is $\epsilon$-close to being isomorphic to $g$ if there exists an automorphism $A$ on $\G$ such that $\delta(f, g\circ A) \leq \epsilon$; and $f$ is $\epsilon$-far from being isomorphic to $g$ if for all automorphism $A$ on $\G$, $\delta(f, g\circ A) \geq \epsilon$.
\end{defi}

\begin{lemma}\label{lemma_aut_perm_char}
    For any automorphism $A: \G \to \G$, 
    \begin{align*}
        \widehat{f\circ A}(\chi_r) = \widehat{f} (\chi_{\widehat{\widehat{A^{-1}}}(r)}).
    \end{align*}
\end{lemma}

\color{black}

\begin{proof}
\begin{align*}
    \widehat{f\circ A}(\chi_r) &= \frac{1}{|\G|} \sum_{x \in \G} f \circ A(x) \chi_r(x) \\
    &= \frac{1}{|\G|} \sum_{x \in \G} f(A(x)) \chi_r(x) \\
    &= \frac{1}{|\G|} \sum_{y \in \G} f(y) \chi_r(A^{-1}(y)) &\text{where } y =A(x) \\
    &= \frac{1}{|\G|} \sum_{y \in \G} f(y) \widehat{A^{-1}}(\chi_r)(y) &\text{by Lemma~\ref{lemma_pontryagin}} \\
    &= \frac{1}{|\G|} \sum_{y \in \G} f(y) \chi_{\widehat{\widehat{A^{-1}}}(r)}(y) &\text{by Lemma~\ref{lemma_pontryagin}} \\
    &= \widehat{f}(\chi_{\widehat{\widehat{A^{-1}}}(r)}).
\end{align*}
\end{proof}

% \begin{observation}
%     \label{obs: distance vs. Fourier coefficient}
%    For any two Boolean function  $f:\ftwo^n \to \{0,1\}$ and $g:\ftwo^n \to \{0,1\}$,  $\delta(f , g) \leq \epsilon$ if and only if $\sum_{S \in \ftwo^n} \widehat{f}(S) \widehat{g}(S) \geq (1 -2\epsilon)$.
% \end{observation}
% \begin{proof}
% For any  $f:\ftwo^n \to \R$ and $g:\ftwo^n \to \R$ we have, 
% \begin{align*}
%     \E_x[f(x) g(x)] = \sum_{S \in \ftwo^n  } \widehat{f}(S) \widehat{g}(S). \tag*{see~\cite[Section 1.4]{ODonnellbook2014}}
% \end{align*}
% Now for $f:\ftwo^n \to \{0,1\}$ and $g:\ftwo^n \to \{0,1\}$ we have, 
% \begin{align*}
%     \E_x[f(x) g(x)] =  \Pr_{x} [f(x) = g(x)] - \Pr_{x} [f(x) \neq g(x)] = 1 - 2\delta(f,g).
% \end{align*}
% Hence, $\delta(f , g) \leq \epsilon$ if and only if $\sum_{S \in \ftwo^n} \widehat{f}(S) \widehat{g}(S) \geq (1 -2\epsilon)$.
% \end{proof}

\section{The subgroup $H^\perp$}\label{section_H_perp}

Throughout this section, $\G$ denotes the finite Abelian group $\mathbb{Z}_{p_1^{m_1}}\times \cdots \times \mathbb{Z}_{p_n^{m_n}}$, where $p_i$ are primes for all $i \in [n]$ and not necessarily distinct, $\langle x \rangle$ denotes the subgroup generated by $x$, and $\omega_\mathcal{L}$ denotes a primitive $\mathcal{L}^{th}$ root of unity and $\mathcal{L} = \LCM\{p_1^{m_1},\cdots,p_n^{m_n}\}$. Let $H$ be a subgroup of $\G$.

Let us look at the definition of $H^\perp$.

\begin{defi}\label{defn_1_new}
    \textbf{(Definition 1)} Let $H$ be a subgroup of $\G$. Then $H^\perp$ is the subgroup given by
    \begin{align*}
        H^\perp := \{ x \in \G : x * y =0 \ \forall y \in H \},
    \end{align*}
    where $$x * y := \biggl( \sum_{i=1}^{T} \frac{\mathcal{L}}{p_i^{m_i}} x^{(i)} \cdot y^{(i)} \biggr) \pmod{\mathcal{L}},$$ and $\mathcal{L} = \LCM \{p_1^{m_1} \cdots p_n^{m_n}\}$.
    % $n = \LCM \{ p_1^{m_1}, \ldots, p_n^{m_n} \}$.
\end{defi}

\begin{remark}
    Observe that $H^\perp$ is also a subgroup of $\G$.
\end{remark}

\begin{claim}\label{claim_H_perp_new}
    For $z \in H^\perp$, 
    \begin{align*}
        \sum_{\beta \in H} \omega_\mathcal{L}^{\beta * z} =|H|,
    \end{align*}
    where $\omega_\mathcal{L}$ is a primitive $\mathcal{L}^{th}$ root of unity of order $\mathcal{L}$, and $|H|$ denotes the order of the subgroup $H$. 
    
    Also, for $z \notin H^\perp$, 
    \begin{align*}
        \sum_{\beta \in H} \omega_\mathcal{L}^{\beta * z} =0.
    \end{align*}
\end{claim}

\begin{proof}
    \begin{description}
        \item [Case 1: When $z\in H^\perp$.] Then $z * x =0$ for all $x \in H$. Hence 
    \begin{align*}
        \sum_{\beta \in H} \omega_\mathcal{L}^{\beta * z} =|H|.
    \end{align*}

        \item [Case 2: When $z\notin H^\perp$.] Let $\sum_{\beta \in H} \chi_\beta (z) =A$. Since $z \notin H^\perp$, so, by Definition~\ref{defn_1_new}, there exists $\gamma \in H$ such that $\gamma * z \neq 0$, that is, $\chi_\gamma(z) \neq 1$. (see Observation~\ref{obs_chi_r_x})
        % \manaswi{Last property seems quite important. Can we prove it in a lemma and refer the lemma here?}\sdc{This follows from the definition of $H^\perp$, I have referred it here.} 
        Then, 
        \begin{align*}
            \chi_\gamma(z) \times A &= \chi_\gamma(z) \sum_{\beta \in H} \chi_\beta(z) \\
            &= \sum_{\beta \in H} \chi_{\beta+\gamma}(z) \\
            &= \sum_{\gamma' \in H} \chi_{\gamma'}(z), &\gamma' = \beta+\gamma \\
            &= A,
        \end{align*}
        which implies that 
        \begin{align*}
            A (\chi_\gamma(z)-1) =0 \Rightarrow A=0,
        \end{align*}
        since $\chi_\gamma(z) \neq 1$.
    \end{description}
\end{proof}

We need the following isomorphism result:

\begin{lemma}\label{lemma_iso_H_perp_G/H}
    $H^\perp$ is isomorphic to the quotient group $\G/H$.
\end{lemma}

\color{black}
\begin{proof}
    We will show that $\widehat{H^\perp}$ is isomorphic to $\widehat{\G/H}$, which implies that $H^\perp$ is isomorphic to the quotient group $\G/H$, as $G \equiv \widehat{G}$ for any group $G$.

    The set of characters of $H^\perp$ is given by $$\Ann_{\G}(H) = \{\chi \in \widehat{\G} : \chi(y) = 1 \ \forall y \in H\},$$ where $\Ann_{\G}(H)$ is known as the annihilator of the subgroup $H$ in $\G$. From Definition~\ref{defn_1_new}, observe that for $x \in \G$, $\chi_x(y) = \omega_\mathcal{L}^{x*y} = 1$  if and only if $x* y =0 \Leftrightarrow x \in H^\perp$.

    Let us define a group homomorphism $\mathcal{F} : \widehat{\G/H} \to \Ann_{\G}(H)$ by $\mathcal{F}(\zeta) = \zeta \circ q$, where $q: \G \to \G/H$ is the quotient is the quotient group homomorphism defined by $q(r)=r+H$. 
    \begin{itemize}
        \item \textbf{($\mathcal{F}$ is injective)} Let $\zeta \in \widehat{\G/H}$ such that $\zeta \circ q = \Tilde{0}$, where $\Tilde{0}=(0, \ldots, 0) [T \text{ times}]$ is the identity element of $\G$. So, $\zeta(r + H) = \zeta \circ q(r) = 1$ for all $r \in \G$, which implies $\zeta$ is the identity element of $\G/H$. Therefore, $\mathcal{F}$ is injective.

        \item \textbf{($\mathcal{F}$ is surjective)} Let $\psi \in \Ann_{\G}(H)$. Let $\zeta : \G/H \to \mathbb{C}$ by $\zeta(r+H) = \psi(r)$ for all $r \in \G$. Since $\chi_{r+H}(x) = \omega_\mathcal{L}^{r *x + H*x}$, so any character $\chi_{r+H}$ of $\G$ is a character of $\G/H$ if $H * x=0$ for all $x\in \G$ (as then, the value of $\omega_\mathcal{L}^{r *x + H*x}$ will be determined only by the coset representatives). Since $\psi(r+H) = \psi(r)\psi(H) = \psi(r)$ and $H$ is the identity element of $G/H$, so $\zeta$ is a character of $\G/H$. Also, $\psi = \zeta \circ q$. Therefore, $\mathcal{F}(\zeta) = \zeta \circ q =\psi$. Hence, $\mathcal{F}$ is surjective.
    \end{itemize}

    Therefore, $\mathcal{F}$ is an isomorphism, which implies that $H^\perp$ is isomorphic to the quotient group $\G/H$.
\end{proof}

\begin{corollary}\label{lemma_H_perp_new}
    $|H| \times |H^\perp| = |\G|$.
\end{corollary}

\begin{proof}
    Follows from the fact that $|H^\perp|= \frac{|\G|}{|H|}$, since $H^\perp$ is isomorphic to the quotient group $\G/H$ by Lemma~\ref{lemma_iso_H_perp_G/H}, and $|\G/H| = \frac{|\G|}{|H|}$ by Lagrange's theorem.
\end{proof}

\section{$\wt_2$ and $\wt_4$ estimation}

\subsection{$\wt_2$ estimation}

In this section, we estimate $\wt_2$ of a bucket (see Definition~\ref{defn_random_coset_structure} and Remark~\ref{remark_random_coset_structure}).

\begin{defi}\label{defn_wt_2}
    Let $f:\G \to \{-1,+1\}$ be a Boolean valued function. Then $\wt_2$ of a bucket $C$ is defined in the following way.
    \begin{align*}
        \wt_2(C) = \sum_{\beta\in \mathcal{C}} |\widehat{f}(\chi_\beta)|^2.
    \end{align*}
\end{defi}

Now let us define the projection operator 
${P}_{r+H}$, $r \in \G$ on a coset $r+H$ of $H$ by
\begin{align*}
    \widehat{P_{r+H}}(f) =
    \begin{cases}
        \widehat{f}(\chi_\gamma) & \gamma \in r+H \\
        0 & \text{otherwise.}
    \end{cases}
\end{align*}

\begin{claim}\label{claim_projection_expectation_new}
\begin{align*}
     P_{r+H} f(x) = \mathbb{E}_{z \in H^\perp} [f(x-z)\chi_r(z)].
\end{align*}
\end{claim}

\begin{proof}
\begin{align*}
    {P}_{r+H} f(x)
    &= \sum_{\gamma \in r+H} \hat{f}(\gamma) \chi_\gamma(x) & \\
    &= \sum_{\beta \in H} \hat{f}(r+\beta) \chi_{r+\beta} (x) &\text{ where } \gamma = r+\beta, \beta \in H \\
    &= \sum_{\beta \in H} \hat{f}(r+\beta) \chi_r(x) \chi_\beta(x) \\
    &= \chi_r(x) \sum_{\beta \in H} \hat{f}(r+\beta) \chi_\beta(x) \\
    &= \chi_r(x) \sum_{\beta\in H} \chi_\beta (x) \frac{1}{|\G|} \sum_{y\in \G} f(y) \overline{\chi_{r+\beta} (y)} \\
    &= \frac{1}{|\G|} \chi_r(x) \sum_{y\in \G} f(y) \sum_{\beta \in H} \chi_\beta(x) \chi_{-r-\beta} (y) \\
    &= \frac{1}{|\G|} \chi_r(x) \sum_{y\in \G} f(y) \sum_{\beta \in H} \chi_\beta(x) \chi_{r+\beta} (-y) \\
    &= \frac{1}{|\G|} \chi_r(x) \sum_{y\in \G} f(y) \chi_r(-y) \sum_{\beta \in H} \chi_\beta (x-y) \\
    &= \frac{|H|}{|\G|} \sum_{y \in H^\perp} f(y) \chi_r (x-y) &\text{by Claim} ~\ref{claim_H_perp_new} \\
    &= \frac{1}{|H^\perp|} \sum_{z \in H^\perp} f(x-z) \chi_r (z) \\
    &= \mathbb{E}_{z \in H^\perp} [f(x-z)\chi_r(z)],
\end{align*}
where $z = x-y$.
\end{proof}

Now let us estimate the weights of the buckets at the $k^{th}$ step.

\begin{lemma}\label{lemma_wt_of_bucket_new}
    The weight $\wt_2$ of a bucket $\mathcal{C}_{\alpha}$, where $\alpha$ is the dominating character of the bucket, and $\mathcal{C}_{\alpha}$ corresponds to the coset $r+H$, is given by
    \begin{align*}
        \wt_2 \bigl(\mathcal{C}_{\alpha}\bigr) = \mathbb{E}_{x \in \G, z \in H^\perp} \biggl[ f(x) f(x+z) \chi_r(z) \biggr].
    \end{align*}
\end{lemma}

\begin{proof}
    \begin{align*}\wt_2 \bigl(\mathcal{C}_{\alpha}\bigr) 
        &= \sum_{\gamma\in r+H} |\widehat{f}(\chi_\gamma)|^2 \\
        &= \sum_{\gamma\in \G} |\widehat{{P}_{r+H}}(\chi_\gamma)|^2 \\
        &= \mathbb{E}_{x' \in \G} \biggl[ {P}_{r+H}f(x') \overline{{P}_{r+H}f(x')} \biggr] &\text{by Parseval} \\
        &= \mathbb{E}_{x' \in \G, z_1,z_2 \in H^\perp} [ \chi_r(z_1)f(x'-z_1) \overline{\chi_r(z_2)} f(x'-z_2)] &\text{by Claim} ~\ref{claim_projection_expectation_new} \\
        &= \mathbb{E}_{x' \in \G, z_1,z_2 \in H^\perp} \biggl[ f(x'-z_1) f(x'-z_2) \chi_r(z_1-z_2) \biggr] \\
        &= \mathbb{E}_{x \in \G, z \in H^\perp} \biggl[ f(x) f(x+z) \chi_r(z) \biggr],
    \end{align*}
    taking $z=z_1-z_2$ and $x=x'-z_1$.
\end{proof}

\begin{lemma}\label{lemma_wt_estimate_new}
    Given query access to a Boolean valued function $f:\mathcal{G} \to \{-1,+1\}$, one can estimate the weight $\wt_2$ of each bucket $\mathcal{C}_{\alpha}$ with accuracy $\pm \epsilon$ except with probability $\delta$ using $O(4 \ln(\frac{4}{\delta}) /\epsilon^2)$ many samples.
\end{lemma}

\begin{proof}
    By Lemma~\ref{lemma_wt_of_bucket_new}, 
    \begin{align*}
        \wt_2 \bigl(\mathcal{C}_{\alpha}\bigr) = \mathbb{E}_{x \in \G, z \in H^\perp} \biggl[ f(x) f(x+z) \chi_r(z) \biggr],
    \end{align*}
    where $r+H$ is the coset corresponding to the bucket $\mathcal{C}_{\alpha}$.
    
    Here, $f(x), \ f(x+z)$ are $\pm 1$ valued and $\chi_r(z)$ is a complex number with absolute value equal to $1$. So, $\Re(\chi_r(z)), \ \Im(\chi_r(z))$ both are in $[-1,+1]$, where $\Re(\chi_r(z))$ and $\Im(\chi_r(z))$ are the real and imaginary parts of $\chi_r(z)$ respectively. Let $E_i= f(x) f(x+z) \chi_r(z) = \Re(E_i) + \Im(E_i)$, where $\Re(E_i) = \Re(\chi_r(z))$ and $\Im(E_i) = \Im(\chi_r(z))$. Then $\Re(E_i)$ and $\Im(E_i)$ take values in $[-1,+1]$. Let the total number of samples required be $N$. Also let $F_1=\frac{\sum_{i=1}^N \Re(E_i)}{N}$, $F_2=\frac{\sum_{i=1}^N \Im(E_i)}{N}$ and $F=F_1+F_2$. Clearly, $\Re(E_i)$ are independent for all $i\in [N]$, since the samples $x_i,z_i$ are drawn uniformly and independently at random (and the same holds for $\Im(E_i)$, $i \in [N]$). Now let us apply Hoeffding's inequality (Lemma~\ref{Hoeffding}) on $\Pr[|F_1 - \mathbb{E}(F_1)|\geq \frac{\epsilon}{\sqrt{2}}]$ and $\Pr[|F_2 - \mathbb{E}(F_2)|\geq \frac{\epsilon}{\sqrt{2}}]$ and upper bound both the bounds by $\dfrac{\delta}{2}$, as then
    \begin{align*}
        \Pr[|F - \mathbb{E}(F)|\geq \epsilon] &= \Pr[|F_1 - \mathbb{E}(F_1)|^2 + |F_2 - \mathbb{E}(F_2)|^2 \geq \epsilon^2] \\
        &\leq \Pr\left[|F_1 - \mathbb{E}(F_1)|\geq \frac{\epsilon}{\sqrt{2}}\right] + \Pr\left[|F_2 - \mathbb{E}(F_2)|\geq \frac{\epsilon}{\sqrt{2}}\right] \\
        &\leq \frac{\delta}{2} + \frac{\delta}{2} \\
        &= \delta.
    \end{align*}
    So,
    \begin{align*}
        \Pr\left[\left|F_1 - \mathbb{E}(F_1)\right|\geq \frac{\epsilon}{\sqrt{2}}\right] &= \Pr \left[  \left| \frac{\sum_{i=1}^N \Re(E_i)}{N} - \frac{\mathbb{E} \left( \sum_{i=1}^N \Re(E_i) \right) }{N} \right| \geq \frac{\epsilon}{\sqrt{2}} \right] \\
        &= \Pr \left[ \left| \sum_{i=1}^N \Re(E_i) - \mathbb{E} \left( \sum_{i=1}^N \Re(E_i) \right) \right| \geq \frac{N\epsilon}{\sqrt{2}} \right] \\
        &\leq 2 e^{-\frac{N^2\epsilon^2}{2\times 2N}} \\
        &= 2 e^{-\frac{N\epsilon^2}{4}} \\
        &\leq \frac{\delta}{2},
    \end{align*}
    The last inequality follows from the fact that 
    which implies that 
    \begin{align}\label{eqn_wt_2_estimate}
        N \geq \frac{4 \ln \left( {4}/{\delta} \right)}{\epsilon^{2}}.
    \end{align}
\end{proof}

\subsection{$\wt_4$ estimation}

Let us first define the $\wt_4$ of a bucket (see Definition~\ref{defn_random_coset_structure} and Remark~\ref{remark_random_coset_structure}).

\begin{defi}\label{defn_wt_4}
    Let $f:\G \to \{-1,+1\}$ be a Boolean valued function. Then $\wt_4$ of a bucket $C$ is defined in the following way.
    \begin{align*}
        \wt_4(C) = \sum_{\beta\in \mathcal{C}} |\widehat{f}(\chi_\beta)|^4.
    \end{align*}
\end{defi}

Let $H$ be a subgroup of $\G$. Let us define a squared projection operator $F_{r+H}f$ on a coset $r+H$ of $H$, whose Fourier coefficients are given by
\begin{align*}
    \widehat{F_{r+H}f}(\chi_\gamma) =
    \begin{cases}
        \widehat{f}(\chi_\gamma)^2 & \text{if } \gamma \in r+H \\
        0 & \text{otherwise.}
    \end{cases}
\end{align*}

\begin{claim}\label{claim_F_r+H}
    $$F_{r+H}(x) = \mathbb{E}_{z_1 \in \G, z \in H^\perp} \biggl[ f(z_1) f(x-z_1-z) \chi_r(z) \biggr].$$
\end{claim}

\begin{proof}[Proof of Claim~\ref{claim_F_r+H}]
\begin{align*}
    F_{r+H}f(x) &= \sum_{\gamma \in r+H} \widehat{f}(\chi_\gamma)^2 \chi_\gamma(x) \\
    &= \sum_{\beta\in H} \widehat{f}(\chi_{r+\beta})^2 \chi_{r+\beta} (x) & \gamma=r+\beta \\
    &= \sum_{\beta \in H} \frac{1}{|\G|^2} \biggl( \sum_{z \in \G} f(z) \overline{\chi_{r +\beta}(z)} \biggr)^2 \chi_{r+\beta}(x) \\
    &= \frac{1}{|\G|^2} \sum_{z_1,z_2 \in \G} \sum_{\beta \in H} f(z_1)f(z_2) \overline{\chi_{r+H}(z_1)} \overline{\chi_{r+H}(z_2)} \chi_{r+H}(x) \\
    &= \frac{1}{|\G|^2} \sum_{z_1,z_2 \in \G} f(z_1)f(z_2) \chi_r(-z_1-z_2+x) \sum_{\beta \in H} \chi_\beta(-z_1-z_2+x) \\
    &= \frac{|H|}{|\G|^2} \sum_{z_1 \in G} f(z_1) \sum_{z_2 \in (x-z_1)+H^\perp} f(z_2) \chi_r(x-z_1-z_2) \\
    &= \frac{1}{|\G|} \sum_{z_1 \in \G} f(z_1) \biggl[ \frac{1}{|H^\perp|} \sum_{z_2 \in (x-z_1)+H^\perp} f(z_2) \chi_r(x-z_1-z_2) \biggr] \\
    &= \frac{1}{|\G| |H^\perp|} \sum_{z_1 \in \G, z_2 \in (x-z_1) +H^\perp} f(z_1)f(z_2) \chi_r(x-z_1-z_2) \\
    &= \mathbb{E}_{z_1 \in \G, z_2 \in (x-z_1) + H^\perp} \biggl[ f(z_1) f(z_2) \chi_r(x-z_1-z_2) \biggr] \\
    &= \mathbb{E}_{z_1 \in \G, z \in H^\perp} \biggl[ f(z_1) f(x-z_1-z) \chi_r(z) \biggr] & x-z_1-z_2=z.
\end{align*}
\end{proof}

\begin{claim}\label{claim_wt_4_avg}
    $$\wt_4(C_\alpha) = \mathbb{E}_{x,z_1,y_1\in \G, z,y \in H^\perp} \biggl[ f(z_1) f(x-z_1-z) f(y_1) f(x-y_1-y) \chi_r(z-y) \biggr].$$
\end{claim}

\begin{proof}[Proof of Claim~\ref{claim_wt_4_avg}]
Let $C_\alpha = r+H$ for some $r \in \G$. Then, 

\begin{align*}
    \wt_4(C_\alpha) &= \sum_{\gamma\in r+H} \widehat{f}(\chi_\gamma)^4 \\
    &= \mathbb{E}_{x \in \G} \biggl[ F_{r+H}f(x) \overline{F_{r+H}f(x)} \biggr] &\text{by Parseval} \\
    &= \mathbb{E}_{x,z_1,y_1\in \G, z,y \in H^\perp} \biggl[ f(z_1) f(x-z_1-z) f(y_1) f(x-y_1-y) \chi_r(z-y) \biggr].
\end{align*}
\end{proof}

\section{Generalized Implicit Sieve}\label{section_implicit seive}
% \textcolor{red}{For now I am keeping the error and the threshold the same as done in the Implicit sieve for $\mathbb{Z}_2^n$. We may need to change it later. I have taken the error while estimating $\wt_4$ to be $= \pm \frac{\theta^4}{8} \Tilde{\wt_2}(C)$ instead of $= \pm \frac{\theta^4}{4} \Tilde{\wt_2}(C)$, as their calculation seems wrong.}

Throughout this section, $\mathcal{G}$ denotes a finite Abelian group, hence $\G = \mathbb{Z}_{{p_1}^{m_1}} \times \cdots \times \mathbb{Z}_{{p_n}^{m_n}}$, $p_i, i \in [n]$ are primes and not necessarily distinct. Also, let $\mathcal{L}= \LCM\{p_1^{m_1}, \ldots, p_n^{m_n}\}$, and $ x * y$ denotes the following. 

\begin{align*}
   x*y = \sum_{i=1}^n \frac{\mathcal{L}}{p_i^{m_i}} x^{(i)} y^{(i)} \pmod{\mathcal{L}},
\end{align*}
where $x^{(i)},y^{(i)}$ are the $i^{th}$ component of $x$ and $y$ respectively.

Let us start by restating our theorem which we will be proving in this section. 

\Sieveforabeliangroup*

\begin{algorithm}[H]
\caption{Generalized Implicit Sieve}\label{algo:Implicit_Sieve_generalized}
\begin{algorithmic}[1]
    \Statex \textbf{Input:} Let $\G = \mathbb{Z}_{p_1^{m_1}}\times \cdots \times \mathbb{Z}_{p_n^{m_n}}$ be a finite Abelian group where $p_i$ are primes for all $i \in [n]$ and not necessarily distinct, its set of all characters, a threshold $\theta$ and a set $M = \{x_1, \ldots, x_{\Tilde{m}}\}$ of uniformly and independently chosen characters, which is sufficiently large.
    \Statex \textbf{Output:} A matrix $Q$, whose columns are given by $\chi_\alpha(x_i), \ i \in [\Tilde{m}]$ for some character $\chi_\alpha$. Also output $\mathcal{F}$, the $\Tilde{m} \times 1$ column matrix whose entries are $f(x_1), \ldots, f(x_{\Tilde{m}})$.

    \bigskip

    \State \textbf{Initialize} $t\geq \log_{\mathcal{L}} \bigl( \frac{100^4 \Tilde{m}^4}{\theta^{32}} \bigr)$. 
    % \manaswi{What is the parameter $n$?}

    \State Consider a randomly permuted coset structure $(H,\mathcal{C})$ such that the subgroup $H$ has codimension $\leq t$ (see Definition~\ref{defn_random_coset_structure}).

    \For{for each $C \in \mathcal{C}$}
    
    \State Estimate $\wt_2$ (see Definition~\ref{defn_wt_2}) for each bucket with error $= \pm \frac{\theta^2}{4}$ and with confidence $1- \frac{1}{100 \mathcal{L}^t}$. Let $\Tilde{\wt_2}(C)$ be the estimated weight of the bucket $C$.

    \If{$\Tilde{\wt_2}(C) \geq \frac{3\theta^2}{4}$} 
    
    \State Keep the bucket $C$.
    \Else { discard.}

    \EndIf

    \EndFor

    \For{for each $C \in \mathcal{C}$ which have survived}
    
    \State Estimate $\wt_4$ (see Definition~\ref{defn_wt_4}) for each bucket with error $= \pm \frac{\theta^4}{8} \Tilde{\wt_2}(C)$ and with confidence $1-\frac{1}{100 \mathcal{L}^t}$. Let $\Tilde{\wt_4}(C)$ be the estimated weight of the bucket $C$.

    \If{$\Tilde{\wt_4}(C) \geq \frac{3\theta^4}{4}$} 
    
    \State Keep the bucket $C$.
    \Else{ discard.}

    \EndIf

    \EndFor

    \State Pick $\{y_1, \ldots,y_{\Tilde{m}}\}$ randomly from $\mathcal{G}$.

    \For{for each $i \in [\Tilde{m}]$}

    \State Estimate $P_Cf(y_i)$ and $P_Cf(y_i-x_i)$ with error $\leq \frac{\theta}{2}$ and confidence $1-\frac{1}{100\Tilde{m}\mathcal{N}}$, for each coset $C$, where $\mathcal{N}$ is the number of survived buckets. 

    \State Estimate $\widehat{f}(\chi_{\alpha(C_j)})$ with error $\leq \frac{\theta^4}{32}$ and with confidence $1- \frac{1}{100\mathcal{N}}$. Let us denote the estimates by $\widehat{f}(\chi_{\alpha(C_j)})'$.
    
    \State $Q_{ij} = \frac{P_{C_j}f(y_i) \overline{P_{C_j}f(y_i-x_i)}}{|\widehat{f}(\chi_{\alpha(C_j)})'|^2}$, where $C_j, \ j \in [\mathcal{N}]$ are the survived buckets, and $Q_{ij}$ is the element in the $i^{th}$ row and $j^{th}$ column of the matrix $Q$.
    % \textcolor{red}{two more steps for creating the output Q matrix}
    \EndFor
\end{algorithmic}
\end{algorithm}

% To start with, we will estimate $\wt_2$ and $\wt_4$ of a bucket.

% \input{Wt_2_estimation}

% \input{Wt_4_estimation}

\subsection{Proof of Theorem~\ref{lemma: Implicit Sieve}}

In this section, we will prove Theorem~\ref{lemma: Implicit Sieve}. Recall the definition of random coset structure (Definition~\ref{defn_random_coset_structure} and Remark~\ref{remark_random_coset_structure}).

\begin{defi}[Random shift]\label{def:random-shift}
    Let us pick a non-zero element $u \in \mathbb{Z}_\mathcal{L}^t$. For each $b$, let us define $C(b+u)$ as $C_u(b)$. This is known as a random shift. It is easy to see that $C_u(b)$ is also a coset and is isomorphic to $C(b)$. 
\end{defi}

% \textcolor{red}{Here we need random shift because $b=0$ is a problematic case.}

\begin{defi}
    Let $\mathcal{C}$ be the set of all $C(b)$'s. We call $(H, \mathcal{C})$ a random permuted coset structure.
\end{defi}

\begin{lemma}\label{lemma_big_coeff_diff_buckets}
    For all $\alpha \in \mathcal{G}$ and $b \in \mathbb{Z}_{\mathcal{L}}^t$, $\Pr[ \alpha \in C_u(b)]= \frac{1}{\mathcal{L}^t}$.
\end{lemma}

\begin{proof}
    It is clear that the event $\alpha \in C_u(b)$ is equivalent to the event $\forall i \in [t], \langle \chi, \beta_i \rangle = b_i + u_i$. So, $\Pr[\langle \alpha, \beta_i \rangle = b_i + u_i]=\Pr[u_i=\langle \alpha, \beta_i \rangle - b_i ]=\frac{1}{\mathcal{L}}$ for each $i \in [t]$. Since, $u_1, u_2,\ldots, u_t$ and $b_1, \ldots, b_t$ are all independent, hence $\Pr[ \alpha \in C_u(b)]= \frac{1}{\mathcal{L}^{t}}$.
\end{proof}

\begin{lemma}\label{lemma_large_char_diff_buckets}
    Let $B$ be the set of all characters $\alpha$ with $|\widehat{f}(\chi_\alpha)| \geq \frac{\theta^2}{4}$. Then, if $t \geq 2\log_{\mathcal{L}} \bigl( \frac{16}{\theta^4} \bigr) + \log_{\mathcal{L}} (\frac{1}{\delta})$, then each $\alpha \in B$ belong to different buckets $C_u(b)$ except with probability $\leq \delta$.
\end{lemma}

\begin{proof}
    Since $\widehat{f}(\chi_\alpha) \geq \frac{\theta^2}{4}$ for all $\alpha \in B$, so $|B| \leq \frac{16}{\theta^4}$ by Parseval. Now, consider two distinct characters $\alpha_1$ and $\alpha_2$ in $\mathcal{G}$. Then,
    \begin{align*}
    \Pr\left[ \alpha_1, \alpha_2 \text{ belong to the same bucket}\right]
    =\Pr\left[\forall i \in [t], \langle \alpha_1-\alpha_2, \beta_i\rangle=0\right] 
    =\frac{1}{\mathcal{L}^t},
    \end{align*}
    where the last equality holds because $\alpha_1$ and $\alpha_2$ are distinct, and $\beta_1, \ldots, \beta_t$ are picked independently.
    
    Observe that the number of ways in which two distinct $\alpha_1, \alpha_2$ can be chosen from $B$ is $\binom{\frac{16}{\theta^4}}{2} \leq (\frac{16}{\theta^4})^2 = \frac{256}{\theta^{8}}$, since $|B|\leq \frac{16}{\theta^4}$. Therefore, the probability that all the characters in $B$ belong to different buckets is given by
        \begin{align*}
            \Pr[\text{all } \alpha\in B \text{ belong to different buckets}] &\geq 1- \binom{\frac{16}{\theta^4}}{2} \frac{1}{\mathcal{L}^{t}} \\
            &\geq 1- \frac{256}{\theta^{8}} \frac{1}{\mathcal{L}^{t}} \\
            &\geq 1- \frac{256}{\theta^{8}} \frac{1}{\mathcal{L}^{(2\log_{\mathcal{L}} \frac{16}{\theta^4} + \log_{\mathcal{L}} (\frac{1}{\delta}))}} \\
            &= 1- \frac{256}{\theta^{8}} \frac{1}{\frac{\frac{256}{\theta^{8}}}{\delta}} \\
            &= 1- \delta.
        \end{align*}
\end{proof}

From Lemma~\ref{lemma_big_coeff_diff_buckets}, we know that the big coefficients, that is the coefficients corresponding to the characters in $B= \left\{\alpha\in \mathcal{G} : |\widehat{f}(\chi_\alpha)| \geq \frac{\theta^2}{4} \right\}$ are in different buckets with high probability, when $t$ is large enough. Let $S= \left\{\alpha\in \mathcal{G} : |\widehat{f}(\chi_\alpha)| < \frac{\theta^2}{4} \right\}$. We also know that at most $\frac{16}{\theta^4}$ elements are present in $B$. We will first show that the coefficients in $S$ are uniformly distributed in the buckets containing the big coefficients. For an $\alpha \in B$, let us denote the bucket containing $\alpha$ by $C_\alpha$, taking into consideration the random shift.

\begin{lemma}\label{lemma_small_char_uniform_dist}
   For each bucket $C_u(b)$, $$\wt_2(C_u(b))\leq |\widehat{f}(\chi_\alpha)|^2 + \frac{\theta^{16}}{100^2 \Tilde{m}^2}$$ with probability $\geq 1-\frac{2}{100}$, assuming that the big characters are in different buckets with high probability, and setting $t\geq \log_{\mathcal{L}} \bigl( \frac{100^4 \Tilde{m}^4}{\theta^{32}} \bigr)$. 
\end{lemma}

\begin{proof}
    For each $\gamma \in S$, let us define the following indicator variables:
    \begin{align*}
        Y_{\gamma,\alpha} =
        \begin{cases}
            1, &\text{if } \gamma\in C_\alpha \\
            0, &\text{otherwise}.
        \end{cases}
    \end{align*}
    Let $Y_\alpha = \sum_{\gamma\in S} Y_{\gamma, \alpha}$.

    Also, let us define the random variables $X_{\gamma,C_\alpha}$ in the following way:    
    \begin{align*}
        X_{\gamma,C_\alpha} = 
        \begin{cases}
            |\widehat{f}(\chi_\gamma)|^2 &\text{if } \gamma \in C_\alpha \\
            0 &\text{otherwise.}
        \end{cases}
    \end{align*}
    Let $X_{C_\alpha} = \sum_{\gamma \in S} X_{\gamma,C_\alpha}$. 
    Therefore,
    \begin{align}\label{eqn_exp_X_alpha}
        \mathbb{E}[X_{C_\alpha}] \leq \frac{1}{\mathcal{L}^{t}},
    \end{align}
        since the number of buckets equals $\mathcal{L}^{t}$ and the sum of the square of the absolute value of all the Fourier coefficients is equal to $1$.
        
        \begin{claim}
            $\Cov[Y_{\gamma_1, \alpha} Y_{\gamma_2,\alpha}] =0$.
        \end{claim}
        \begin{description}
        \item[Case 1: When $\gamma_2= \lambda_1\gamma_1 + \lambda_2\alpha$ for some $\lambda_1, \lambda_2 \in \mathbb{N}\cup\{0\}$, and both $\gamma_1, \gamma_2$ are nonzero.] Then we have,
        \begin{align*}
            &\langle \gamma_2, \beta_i \rangle = \lambda_1 \langle \gamma_1, \beta_i \rangle + \lambda_2 \langle \alpha, \beta_i \rangle \\
            &\Rightarrow (b_i+u_i)(\lambda_1+\lambda_2-1)=0 \ \forall i\in [t] \\
            &\Rightarrow (b+u)(\lambda_1+\lambda_2-1)=0.
        \end{align*}

        \begin{description}
            \item[Subcase 1: When $\lambda_1+\lambda_2-1$ is invertible in $\mathbb{Z}_\mathcal{L}$.] Then we have $b+u=0$. Observe that $Y_{\gamma_1, \alpha} Y_{\gamma_2,\alpha}=1$ if and only if 
       \begin{enumerate}
       \item $b+u=0^t$, i.e., $u=-b$, and
       \item $\forall i \in [t], \langle \gamma_1, \beta_i \rangle=b_i+u_i$.
       \end{enumerate}
       Clearly, $\Pr[u=-b]= \frac{1}{\mathcal{L}^{t}}$. Since the $\beta_1,\ldots, \beta_t$ and $u$ are all independent random variables, $\Pr[\forall i \in [t], \langle \gamma_1, \beta_i \rangle=b_i+u_i | u=-b] = \frac{1}{\mathcal{L}^{t}}$. Therefore, $\mathbb{E}[Y_{\gamma_1, \alpha} Y_{\gamma_2,\alpha}]=\frac{1}{\mathcal{L}^{t}}\cdot \frac{1}{\mathcal{L}^{t}}=\frac{1}{\mathcal{L}^{2t}}$.

       Now, $\mathbb{E}[Y_{\gamma_1,\alpha}]=\Pr[\gamma_1 \in C_\alpha]=\frac{1}{\mathcal{L}^{t}}$. Similarly, it follows that $\mathbb{E}[Y_{\gamma_2,\alpha}]=\frac{1}{\mathcal{L}^{t}}$.
       We thus have
        $\Cov[Y_{\gamma_1, \alpha} Y_{\gamma_2,\alpha}] = \frac{1}{\mathcal{L}^{2t}}-\frac{1}{\mathcal{L}^{2t}} = 0$.

        \item[Subcase 2: When $\lambda_1+\lambda_2-1$ is not invertible in $\mathbb{Z}_\mathcal{L}$.] Then $\lambda_1+\lambda_2-1$ is a zero divisor in $\mathbb{Z}_\mathcal{L}$. It follows that $b_i+u_i= n_i'$, for some $n_i'$ such that $n_i'$ divides $\mathcal{L}$, for all $i \in [t]$.
        Observe that $Y_{\gamma_1, \alpha} Y_{\gamma_2,\alpha}=1$ if and only if 
       \begin{enumerate}
       \item $b_i+u_i= n_i'$ for all $i \in [t]$, i.e., $u_i=n_i'-b_i$ for all $i \in [t]$, and
       \item $\forall i \in [t], \langle \gamma_1, \beta_i \rangle=b_i+u_i$.
       \end{enumerate}
       Clearly, $\Pr[u_i=n_i'-b_i \ \forall i \in [t]] = \frac{1}{\mathcal{L}^{t}}$, since the values of $\lambda_1$ and $\lambda_2$ determine that of $n_i'$'s. Since the $\beta_1,\ldots, \beta_t$ and $u$ are all independent random variables, $\Pr[\forall i \in [t], \langle \gamma_1, \beta_i \rangle=b_i+u_i | u_i=n_i'-b_i] = \frac{1}{\mathcal{L}^{t}}$. Therefore, $\mathbb{E}[Y_{\gamma_1, \alpha} Y_{\gamma_2,\alpha}]=\frac{1}{\mathcal{L}^{t}}\cdot \frac{1}{\mathcal{L}^{t}}=\frac{1}{\mathcal{L}^{2t}}$.

       Now, $\mathbb{E}[Y_{\gamma_1,\alpha}]=\Pr[\gamma_1 \in C_\alpha]=\frac{1}{\mathcal{L}^{t}}$. Similarly, it follows that $\mathbb{E}[Y_{\gamma_2,\alpha}]=\frac{1}{\mathcal{L}^{t}}$.
       We thus have
        $\Cov[Y_{\gamma_1, \alpha} Y_{\gamma_2,\alpha}] = \frac{1}{\mathcal{L}^{2t}}-\frac{1}{\mathcal{L}^{2t}} = 0$.
        \end{description}

        \item [Case 2: When $\gamma_1, \gamma_2$ are both nonzero, distinct and $\gamma_2 \neq \lambda_1 \gamma_1 + \lambda_2\alpha$ for any $\lambda_1, \lambda_2 \in \mathbb{N}\cup \{0\}$.]
        Here we have, $Y_{\gamma_1, \alpha} Y_{\gamma_2,\alpha}=1$ if and only if $\forall i \in [t], \langle \gamma_1, \beta_i \rangle=b_i+u_i$, $\langle \gamma_2, \beta_i \rangle=b_i+u_i$, and $\langle \alpha, \beta_i \rangle=b_i+u_i$. Let us fix $i\in [t]$ and $u_i$. Since $\gamma_2 \neq \lambda_1 \gamma_1 + \lambda_2\alpha$ for any $\lambda_1, \lambda_2 \in \mathbb{N}\cup \{0\}$, the bucket $\gamma_2$ belongs to is not determined by that of $\gamma_1$ and $\alpha$. This fact, along with the fact that $u_1,\ldots,u_t,\beta_1,\ldots,\beta_t$ are all independent, implies that
        \begin{align*}
            \mathbb{E}[Y_{\gamma_1, \alpha} Y_{\gamma_2,\alpha}]= (\frac{1}{(\mathcal{L})^2})^t =\frac{1}{\mathcal{L}^{2t}}.
        \end{align*}
            
        Similarly, it follows that $\mathbb{E}[Y_{\gamma_1,\alpha}]=\mathbb{E}[Y_{\gamma_2,\alpha}]=\frac{1}{\mathcal{L}^{t}}$. We thus have that $\Cov[Y_{\gamma_1,\alpha}, Y_{\gamma_2,\alpha}]=0$.
        \end{description}

        Therefore,
        \begin{align}\label{cov_0}
            \Cov[Y_{\gamma_1,\alpha}, Y_{\gamma_2,\alpha}]=0.
        \end{align}
        The case $\alpha=0$ follows similarly, putting $\lambda_2=0$ in the above explanation we can get $\Cov[Y_{\gamma_1,0}, Y_{\gamma_2,0}]=0$.

        So, $\Cov[X_{\gamma_1,C_\alpha}, X_{\gamma_2,C_\alpha}]=0$, which follows directly from the proof of Lemma~\ref{lemma_small_char_uniform_dist} and the fact that $X_{\gamma,C_\alpha} = |\widehat{f}(\chi_\gamma)|^2 Y_{\gamma,\alpha}$.

        Therefore,
    \begin{align*}
        \Var[X_{C_\alpha}] &= \mathbb{E}[(\sum_{\gamma\in S} X_{\gamma, C_\alpha})^2] - (\mathbb{E}[\sum_{\gamma\in S} X_{\gamma, C_\alpha}])^2 \\
        &= \sum_{\gamma\in S} \mathbb{E}[(X_{\gamma, C_\alpha})^2] - \sum_{\gamma\in S} (\mathbb{E}[X_{\gamma, C_\alpha}])^2 +2\sum_{\gamma_1, \gamma_2 \in S, \gamma_1 \neq \gamma_2} \Cov[X_{\gamma_1,C_\alpha}, X_{\gamma_2,C_\alpha}] \\
        &\leq \mathbb{E}[(X_{C_\alpha})^2] \\
        &\leq \frac{\theta^4}{16} \cdot \frac{1}{\mathcal{L}^{t}},
    \end{align*}
    where the last inequality follows from the fact that $\mathbb{E}[X_{C_\alpha}] = \sum_{\gamma\in S} X_{\gamma, C_\alpha} \leq \frac{1}{\mathcal{L}^{t}}$ by \cref{eqn_exp_X_alpha}, and from the fact that $|X_{\gamma,C_\alpha}| = |\widehat{f}(\chi_\gamma)|^2 \leq \frac{\theta^4}{16}$.

        % Now, we have,
        % \begin{align*}
        %     \Var[Y_\alpha] &= \mathbb{E}[(\sum_{\gamma\in \overline{S}} Y_{\gamma, \alpha})^2] - (\mathbb{E}[\sum_{\gamma\in \overline{S}} Y_{\gamma, \alpha}])^2 \\
        %     &= \sum_{\gamma\in \overline{S}} \mathbb{E}[(Y_{\gamma, \alpha})^2] + 2\sum_{\gamma_1, \gamma_2 \in \overline{S}, \gamma_1 \neq \gamma_2} \mathbb{E}[Y_{\gamma_1, \alpha} Y_{\gamma_2,\alpha}] - \sum_{\gamma\in \overline{S}} (\mathbb{E}[Y_{\gamma, \alpha}])^2 - 2\sum_{\gamma_1, \gamma_2 \in \overline{S}, \gamma_1 \neq \gamma_2} \mathbb{E}[Y_{\gamma_1, \alpha}] \mathbb{E}[Y_{\gamma_2,\alpha}] \\
        %     &= \sum_{\gamma\in \overline{S}} \mathbb{E}[(Y_{\gamma, \alpha})^2] - \sum_{\gamma\in \overline{S}} (\mathbb{E}[Y_{\gamma, \alpha}])^2 +2\sum_{\gamma_1, \gamma_2 \in \overline{S}, \gamma_1 \neq \gamma_2} \Cov[Y_{\gamma_1,\alpha}, Y_{\gamma_2,\alpha}] \\
        %     &\leq \mathbb{E}[Y_\alpha] =\frac{|\overline{S}|}{p^{mt}},
        % \end{align*}
        % where the last inequality follows from the fact that $(Y_{\gamma, \alpha})^2 = Y_{\gamma, \alpha}$ and from \cref{cov_0}. 
        
        % Therefore, 
        
        % \begin{align*}
        %     \Pr[|Y_\alpha - \mathbb{E}[Y_\alpha]| \geq \epsilon] &\leq \frac{\Var[Y_\alpha]}{\epsilon^2} \\
        %     &\leq \frac{|\overline{S}|}{p^{mt}\epsilon^2}.
        % \end{align*}

        So, by Chebyshev,
    \begin{align*}
        \Pr\left[ \left| X_{C_\alpha} - \mathbb{E}[X_{C_\alpha}] \right| \geq \epsilon \right] \leq \frac{\Var[X_{C_\alpha}]}{\epsilon^2}
        \leq \frac{\theta^4}{16 \mathcal{L}^{t}\epsilon^2}.
    \end{align*}
    
    Therefore, by union bound,
    \begin{align*}
        \Pr\left[ \bigcap_{\alpha \in B}|X_{C_\alpha} - \mathbb{E}[X_{C_\alpha}]| \leq \epsilon \right] 
        &= 1 - \Pr[\bigcup_{\alpha \in B}|X_{C_\alpha} - \mathbb{E}[X_{C_\alpha}]| \geq \epsilon] \\
        &\geq 1 - \sum_{\alpha \in B} \Pr[|X_{C_\alpha} - \mathbb{E}[X_{C_\alpha}]| \geq \epsilon] \\
        &\geq 1- \frac{16}{\theta^4} \frac{\theta^4}{16 \mathcal{L}^{t}\epsilon^2} \\
        &=1- \frac{1}{\mathcal{L}^{t}\epsilon^2}.
    \end{align*}
        
        Taking $t\geq \log_{\mathcal{L}} \bigl( \frac{100^4 \Tilde{m}^4}{\theta^{64}} \bigr)$, and $\epsilon= \frac{128^2\times \theta^{32}}{100^2 \Tilde{m}^2}$, we have, 
        \begin{align*}
            \Pr[\bigcap_{\alpha \in B}|X_{C_\alpha} - \mathbb{E}[X_{C_\alpha}]| \leq \frac{\theta^{32}}{128^2\times 100^2 \Tilde{m}^2}] \geq 1-\frac{1}{100} = \frac{99}{100}.
        \end{align*}

    Also, taking $\delta=\frac{1}{100}$ in Lemma~\ref{lemma_large_char_diff_buckets}, we have that the characters in $B$ are in different buckets with high probability. So, except with probability $\frac{2}{100}$, 
    \begin{align*}
        \wt_2(C_\alpha) \leq |\widehat{f}(\chi_{\alpha})|^2 + \frac{\theta^{32}}{128^2\times 100^2 \Tilde{m}^2}.
    \end{align*}
\end{proof}

\begin{lemma}
Algorithm~\ref{algo:Implicit_Sieve_generalized} does the following with probability $\geq 1-\frac{1}{100}$:
    \begin{enumerate}
        \item It does not discard any bucket $C$ with $|\widehat{f}(\chi_\gamma)| \geq \theta$.
        \item It discards any bucket $C$ with $|\widehat{f}(\chi_\gamma)| < \frac{\theta^2}{4}$, for all $\gamma \in C$.
        \item $|\widehat{f}(\chi_{\alpha(C)})| \geq \frac{\theta}{2}$ for every bucket $C=C_\alpha$ (that is, $\alpha(C)$ is the dominating element of $C_\alpha$) in the set of buckets that the algorithm has not discarded.
    \end{enumerate}
\end{lemma}

\begin{proof}
\begin{description}
    \item[\textbf{Proof of part (1).}] From Algorithm~\ref{algo:Implicit_Sieve_generalized} we know that $\Tilde{\wt_2}(C) \geq \wt_2(C) - \frac{\theta^2}{4} \geq \theta^2 -\frac{\theta^2}{4} = \frac{3\theta^2}{4}$. 
    
    Also, $\Tilde{\wt_4}(C) \geq \wt_4(C) - \frac{\theta^4}{8} \Tilde{\wt_2}(C) \geq \theta^4 -\frac{\theta^4}{8} \Tilde{\wt_2}(C) \geq \theta^4 -\frac{\theta^4}{8} (1+ \frac{\theta^2}{4}) = \frac{7}{8}\theta^4 - \frac{1}{32} \theta^6 \geq \frac{3\theta^4}{4}$, since $\frac{7}{8}\theta^4 - \frac{1}{32} \theta^6 - \frac{3\theta^4}{4} = \frac{\theta^4}{8} - \frac{\theta^6}{32} = \frac{\theta^4}{8} (1- \frac{\theta^2}{4})>0$ as $\frac{\theta^2}{4}<1$. Hence the bucket $C$ is not discarded.

    \item[\textbf{Proof of part (2).}] From Algorithm~\ref{algo:Implicit_Sieve_generalized} we know that the estimated weight $\Tilde{\wt_2}(C)$ of a bucket $C$ is $\geq \frac{3\theta^2}{4}$. Since the error $=\pm \frac{\theta^2}{4}$ while estimating $\wt_2(C)$, we have $\wt_2(C) \geq \frac{3\theta^2}{4} \pm \frac{\theta^2}{4} = \frac{\theta^2}{2} \text{ or } \theta^2$, hence $\wt_2(C) \leq \frac{4}{3} \Tilde{\wt_2}(C)$. Therefore, since the error $=\pm \frac{\theta^4}{8}\Tilde{\wt_2}(C)$ while estimating $\wt_4(C)$, we have,
    \begin{align*}
        \Tilde{\wt_4}(C) &\leq \wt_4(C) + \frac{\theta^4}{8}\Tilde{\wt_2}(C) \\
        &= \sum_{\gamma \in C} |\widehat{f}(\chi_\gamma)|^4 + \frac{\theta^4}{8}\Tilde{\wt_2}(C) \\
        &\leq \max_{\gamma \in C} \{|\widehat{f}(\chi_\gamma)|^2\} \times \wt_2(C) + \frac{\theta^4}{8}\Tilde{\wt_2}(C) \\
        &\leq \frac{\theta^4}{16} \times \wt_2(C) + \frac{\theta^4}{8}\Tilde{\wt_2}(C) &\text{since } |\widehat{f}(\chi_\gamma)| \leq \frac{\theta^2}{4} \\
        &\leq \frac{\theta^4}{12} \Tilde{\wt_2}(C) + \frac{\theta^4}{8}\Tilde{\wt_2}(C) &\text{since } \wt_2(C) \leq \frac{4}{3} \Tilde{\wt_2}(C) \\
        &= \frac{5\theta^4}{24} \Tilde{\wt_2}(C) \\
        &\leq \frac{5\theta^4}{24}(1 + \frac{\theta^2}{4}) < \frac{3\theta^4}{4}  &\text{since } \Tilde{\wt_2}(C) \leq 1 + \frac{\theta^2}{4}.
    \end{align*}
    Hence, the bucket $C$ will be discarded.

    \item[\textbf{Proof of part (3).}] Let there exists a bucket $C=C_\alpha$ such that $|\widehat{f}(\chi_\alpha)|<\frac{\theta}{2}$, where $\alpha \in B$. We know that $\wt_2(C_\alpha) \leq |\widehat{f}(\chi_{\alpha})|^2 + \frac{\theta^{32}}{128^2\times 100^2 \Tilde{m}^2}$ from Lemma~\ref{lemma_small_char_uniform_dist}. So, $\wt_2(C_\alpha) \leq \frac{\theta^2}{4} + \frac{\theta^{32}}{128^2\times 100^2 \Tilde{m}^2} < \frac{\theta^2}{4} + \frac{\theta^2}{4} = \frac{\theta^2}{2}$. Therefore, $\Tilde{\wt_2}(C_\alpha) \leq \wt_2(C_\alpha) + \frac{\theta}{4} < \frac{\theta^2}{2} + \frac{\theta^2}{4} = \frac{3\theta^2}{4}$, hence it will be discarded.
\end{description}
    
\end{proof}

\begin{lemma}\label{lemma_small_char_bound}
    The small weight coefficients do not contribute much to the weight of a bucket. That is,
    \begin{align*}
        \Pr_{x} \left[ \left| \sum_{r \in C_\alpha \setminus \{\alpha\}} \widehat{f}(\chi_r) \chi_r(x) \right| \geq \frac{\theta^4}{4} \right] \leq \frac{4\theta^4}{128\times 100 \Tilde{m}}.
    \end{align*}
\end{lemma}

\begin{proof}
    From Lemma~\ref{lemma_small_char_uniform_dist} we know that $\mathbb{E}[|\sum_{r \in C_\alpha \setminus \{\alpha\}} \widehat{f}(\chi_r) \chi_r(x)|^2] \leq \frac{\theta^{32}}{128^2\times100^2 \Tilde{m}^2}$. Therefore,
    \begin{align*}
        \mathbb{E} \left[ \left| \sum_{r \in C_\alpha \setminus \{\alpha\}} \widehat{f}(\chi_r) \chi_r(x) \right| \right] \leq \sqrt{\mathbb{E} \left[ \left| \sum_{r \in C_\alpha \setminus \{\alpha\}} \widehat{f}(\chi_r) \chi_r(x) \right|^{2} \right]} \leq \frac{\theta^{16}}{128\times 100\Tilde{m}}.
    \end{align*}
    
    Hence, 
    \begin{align*}
        \Pr_{x} \left[\left|\sum_{r \in C_\alpha \setminus \{\alpha\}} \widehat{f}(\chi_r) \chi_r(x)\right| \geq \frac{\theta^4}{4}\right] \leq \frac{4\theta^4}{128\times 100\Tilde{m}},
    \end{align*}
    by Markov's inequality, which is very small.
\end{proof}

\begin{lemma}
Let $y_i, \ i \in [\Tilde{m}]$ are chosen uniformly and independently at random. Since $x_i, i \in [\Tilde{m}]$ are also chosen uniformly and independently at random, so $(y_i-x_i)$ are also chosen uniformly and independently at random for all $i \in [\Tilde{m}]$. Then,
    \begin{align*}
        |P_{C_\alpha}f(y_i) \overline{P_{C_\alpha}f(y_i-x_i)}- |\widehat{f}(\chi_{\alpha(C)})|^2 \chi_{\alpha(C)}(x_i)| \leq \frac{\theta^2}{2} + \frac{\theta^4}{16}
    \end{align*}
with probability $\geq 1- \frac{4\theta^4}{128\times 100\Tilde{m}}$, for each $i \in [\Tilde{m}]$.
\end{lemma}

\begin{proof}
For each bucket $C_\alpha$, let $\alpha(C)$ be the dominating element, $\alpha(C) \in B$. By Lemma~\ref{lemma_big_coeff_diff_buckets} the elements in $B$ are present in different buckets with high probability. Observe that
    \begin{align*}
        &P_{C_\alpha}(y_i) \overline{P_{C_\alpha}(y_i-x_i)} \\
        &= \biggl( \widehat{f}(\chi_{\alpha(C)}) \chi_{\alpha(C)}(y_i) + \sum_{r \in C_\alpha \setminus \{ \alpha(C) \}} \widehat{f}(\chi_r) \chi_r(y_i) \biggr) \\
        &\times \overline{\biggl( \widehat{f}(\chi_{\alpha(C)}) \chi_{\alpha(C)}(y_i-x_i) + \sum_{r \in C_\alpha\setminus \{ \alpha(C) \}} \widehat{f}(\chi_r) \chi_r(y_i-x_i) \biggr)} \\
        &= \biggl( \widehat{f}(\chi_{\alpha(C)}) \chi_{\alpha(C)}(y_i) + \sum_{r \in C_\alpha \setminus \{ \alpha(C) \}} \widehat{f}(\chi_r) \chi_r(y_i) \biggr) \\
        &\times \biggl( \overline{\widehat{f}(\chi_{\alpha(C)})} \chi_{\alpha(C)}(x_i-y_i) + \sum_{r \in C_\alpha \setminus \{ \alpha(C) \}} \overline{\widehat{f}(\chi_r)} \chi_r(x_i-y_i) \biggr) \\
        &= |\widehat{f}(\chi_{\alpha(C)})|^2 \chi_{\alpha(C)}(x_i) + \biggl(\sum_{r \in C_\alpha \setminus \{ \alpha(C) \}} \widehat{f}(\chi_r) \chi_r(y_i) \biggr) \overline{\widehat{f}(\chi_{\alpha(C)})} \chi_{\alpha(C)}(x_i-y_i) \\
        &+ \biggl( \sum_{r \in C_\alpha \setminus \{ \alpha(C) \}} \overline{\widehat{f}(\chi_r)} \chi_r(x_i-y_i) \biggr) \widehat{f}(\chi_{\alpha(C)}) \chi_{\alpha(C)}(y_i) \\
        &+ \biggl( \sum_{r \in C_\alpha \setminus \{ \alpha(C) \}} \widehat{f}(\chi_r) \chi_r(y_i) \biggr) \biggl( \overline{\sum_{r \in C_\alpha \setminus \{ \alpha(C) \}} \widehat{f}(\chi_r) \chi_r(y_i-x_i)} \biggr)
    \end{align*}

    Now, 
    $$
        \left| \sum_{r \in C_\alpha \setminus \{ \alpha(C) \}} \widehat{f}(\chi_r) \chi_r(y_i) \right| \leq \frac{\theta^4}{4}
    $$ 
    and 
    $$
        \left| \sum_{r \in C_\alpha \setminus \{ \alpha(C) \}} \widehat{f}(\chi_r) \chi_r(x_i - y_i) \right| \leq \dfrac{\theta^4}{4}
    $$ 
    with probability $\geq 1- \dfrac{4\theta^4}{128\times 100\Tilde{m}}$ by Lemma~\ref{lemma_small_char_bound}. 
    % \manaswi{The statement of the lemma should say that $y_i$ are chosen at random. Also mention why $(x_i - y_i)$ is random (and where). Finally, union bound is applied with Lemma~\ref{lemma_small_char_bound}, which should show up in the constant (I think 4 should go up to 12, but needs to be checked).}\sdc{Done.}

    Therefore, by triangle inequality of $\ell_1$ norm, with probability $\geq 1- 2\times\dfrac{4\theta^4}{128 \times 100\Tilde{m}}$,
    \begin{align*}
        \bigg| P_{C_\alpha}(y_i) \overline{P_{C_\alpha}(y_i-x_i)} - |\widehat{f}(\chi_{\alpha(C)})|^2 \chi_{\alpha(C)}(x_i) \bigg|
        &\leq \frac{\theta^4}{4} + \frac{\theta^4}{4} + \frac{\theta^4}{16} = \frac{9\theta^4}{16}.
    \end{align*}
\end{proof}

So, by taking union bound over $\Tilde{m}$-elements, the above result holds for all $x_i$'s and $y_i$'s, $i \in [\Tilde{m}]$ with probability $\geq 1- \dfrac{8\theta^4}{128\times 100}$. Therefore, by taking union bound over $\mathcal{N} \leq \dfrac{16}{\theta^4}$ buckets, the above holds with probability $\geq 1- \dfrac{1}{100}$.

Hence, Algorithm~\ref{algo:Implicit_Sieve_generalized} works with probability $\geq 1 - (\frac{1}{100} + \frac{2}{100} + \frac{1}{100} + \frac{1}{100} + \frac{1}{100}) = \frac{94}{100}$.

Let $\mathcal{N}$ denotes the number of survived buckets. Let us construct a matrix $Q'$ upto an error $\frac{9\theta^4}{16}+ \frac{\theta^4}{16}$, which is given by 
% \manaswi{We do not output this matrix.}
\begin{align}\label{eqn_matrix_Q'}
Q'=
\begin{bmatrix}
    |\widehat{f}(\chi_{\alpha(C_1)})|^2\chi_{\alpha(C_1)}(x_1) & |\widehat{f}(\chi_{\alpha(C_2)})|^2\chi_{\alpha(C_2)}(x_1) & \cdots & |\widehat{f}(\chi_{\alpha(C_{\mathcal{N}})})|^2\chi_{\alpha(C_\mathcal{N})}(x_1) & \mid & f(x_1) \\
    |\widehat{f}(\chi_{\alpha(C_1)})|^2\chi_{\alpha(C_1)}(x_2) & |\widehat{f}(\chi_{\alpha(C_2)})|^2\chi_{\alpha(C_2)}(x_2) & \cdots & |\widehat{f}(\chi_{\alpha(C_{\mathcal{N}})})|^2\chi_{\alpha(C_\mathcal{N})}(x_2) & \mid & f(x_2) \\
    . & . & . & . & \mid & . \\
    . & . & . & . & \mid & . \\
    . & . & . & . & \mid & . \\
    |\widehat{f}(\chi_{\alpha(C_1)})|^2\chi_{\alpha(C_1)}(x_{\Tilde{m}}) & |\widehat{f}(\chi_{\alpha(C_2)})|^2\chi_{\alpha(C_2)}(x_{\Tilde{m}}) & \cdots & |\widehat{f}(\chi_{\alpha(C_{\mathcal{N}})})|^2\chi_{\alpha(C_\mathcal{N})}(x_{\Tilde{m}}) & \mid &  f(x_{\Tilde{m}})
\end{bmatrix}
\end{align}

We estimate the Fourier coefficients with tolerance $\leq \frac{\theta^4}{32}$ and confidence $\frac{1}{100\mathcal{N}}$. Then,
\begin{align*}
    ||\widehat{f}(\chi_{\alpha(C_i)})|^2 - |\widehat{f}(\chi_{\alpha(C_i)})'|^2|
    &= \biggl||\widehat{f}(\chi_{\alpha(C_i)})| + |\widehat{f}(\chi_{\alpha(C_i)})'|\biggr| \times \biggl||\widehat{f}(\chi_{\alpha(C_i)})| - |\widehat{f}(\chi_{\alpha(C_i)})'|\biggr| \\
    &\leq \left(1+1+\frac{\theta^4}{32}\right) \biggl|\widehat{f}(\chi_{\alpha(C_i)}) - \widehat{f}(\chi_{\alpha(C_i)})'\biggr| \\
    &\leq \left( 2 +\frac{\theta^4}{32} \right) \frac{\theta^4}{32} \leq \frac{3\theta^4}{32}.
\end{align*}

Therefore, we get the following matrix $Q''$, upto error $\leq \frac{9\theta^4}{16} + \frac{\theta^4}{16} +  \frac{3\theta^4}{32} = \frac{23\theta^4}{32}$.

\begin{align*}
Q''=
\begin{bmatrix}
    |\widehat{f}(\chi_{\alpha(C_1)})'|^2 \chi_{\alpha(C_1)}(x_1) & |\widehat{f}(\chi_{\alpha(C_2)})'|^2 \chi_{\alpha(C_2)}(x_1) & \cdots & |\widehat{f}(\chi_{\alpha(C_{\mathcal{N}})})'|^2\chi_{\alpha(C_\mathcal{N})}(x_1) & \mid & f(x_1) \\
    |\widehat{f}(\chi_{\alpha(C_1)})'|^2 \chi_{\alpha(C_1)}(x_2) & |\widehat{f}(\chi_{\alpha(C_2)})'|^2\chi_{\alpha(C_2)}(x_2) & \cdots & |\widehat{f}(\chi_{\alpha(C_{\mathcal{N}})})'|^2 \chi_{\alpha(C_\mathcal{N})}(x_2) & \mid & f(x_2) \\
    . & . & . & . & \mid & . \\
    . & . & . & . & \mid & . \\
    . & . & . & . & \mid & . \\
    |\widehat{f}(\chi_{\alpha(C_1)})'|^2 \chi_{\alpha(C_1)}(x_{\Tilde{m}}) & |\widehat{f}(\chi_{\alpha(C_2)})'|^2 \chi_{\alpha(C_2)}(x_{\Tilde{m}}) & \cdots & |\widehat{f}(\chi_{\alpha(C_{\mathcal{N}})})'|^2 \chi_{\alpha(C_\mathcal{N})}(x_{\Tilde{m}}) & \mid &  f(x_{\Tilde{m}})
\end{bmatrix}
\end{align*}

For each $i \in [\Tilde{m}]$ and $j \in [\mathcal{N}]$, $Q_{ij}''- Q_{ij}' \leq \frac{3\theta^4}{32}$ with probability $\geq 1- \frac{1}{100\mathcal{N}}$.
% \manaswi{``upto error'' should be specified.}

We divide $i^{th}$ column, $i \in [\Tilde{m}]$ of $Q''$ by the estimated values $|\widehat{f}(\chi_{\alpha(C_i)})'|^2, \ i \in [\Tilde{m}]$. Hence, upto error $\frac{23\theta^4}{32}$, the matrix $Q$ is given by
\begin{align}\label{eqn_matrix_Q}
Q=
\begin{bmatrix}
    \chi_{\alpha(C_1)}(x_1) & \chi_{\alpha(C_2)}(x_1) & \cdots & \chi_{\alpha(C_\mathcal{N})}(x_1) & \mid & f(x_1) \\
    \chi_{\alpha(C_1)}(x_2) & \chi_{\alpha(C_2)}(x_2) & \cdots & \chi_{\alpha(C_\mathcal{N})}(x_2) & \mid & f(x_2) \\
    . & . & . & . & \mid & . \\
    . & . & . & . & \mid & . \\
    . & . & . & . & \mid & . \\
    \chi_{\alpha(C_1)}(x_{\Tilde{m}}) & \chi_{\alpha(C_2)}(x_{\Tilde{m}}) & \cdots & \chi_{\alpha(C_\mathcal{N})}(x_{\Tilde{m}}) & \mid &  f(x_{\Tilde{m}})
\end{bmatrix}
\end{align}

\color{black}

\begin{lemma}\label{lemma_wt__2_estimate}
    Given query access to a Boolean valued function $f:\mathcal{G} \to \{-1,+1\}$, one can estimate the weight $\wt_2$ of each bucket $\mathcal{C}_{\alpha}$ with accuracy $\pm \epsilon$ except with probability $\delta$ using $O\left(\frac{1}{\theta^{4}} \ln \left(\frac{\Tilde{m}}{\theta^8}\right)\right)$ many samples.
\end{lemma}

\begin{proof}
    The proof follows exactly along the lines of the proof of Lemma~\ref{lemma_wt_estimate_new}. Putting $\epsilon = \frac{\theta^2}{4}$, $\delta = \frac{1}{100 \mathcal{L}^t}$ in Lemma~\ref{eqn_wt_2_estimate}, and since $t \geq \log_{\mathcal{L}} \bigl( \frac{100^4 \Tilde{m}^4}{\theta^{32}} \bigr)$ by \cref{lemma_small_char_uniform_dist}, we have, the number of samples 
    $$
        N \geq 4 \ln (400 \mathcal{L}^t) / (\theta^4/16) = 4 \ln (400 \frac{100^4 \Tilde{m}^4}{\theta^{32}}) / (\theta^4/16). 
    $$
    So taking $N= O(\frac{\ln \frac{\Tilde{m}}{\theta^8}}{\theta^4})$ many samples would work.
\end{proof}

\begin{lemma}\label{lemma_wt_4_estimation}
    Given query access to a Boolean valued function $f:\mathcal{G} \to \{-1,+1\}$, one can estimate the weight $\wt_4$ of each bucket $\mathcal{C}_{\alpha}$ with accuracy $\pm \epsilon$ except with probability $\delta$ using $O \biggl( \frac{\ln \frac{\Tilde{m}}{\theta^8}}{\theta^{8}} \biggr)$ many samples.
\end{lemma}

\begin{proof}
    The proof follows exactly along the lines of the proof of Lemma~\ref{lemma_wt_estimate_new}. Putting $\epsilon = \frac{\theta^4}{8} \Tilde{\wt_2}(C_\alpha)$, $\delta = \frac{1}{100 \mathcal{L}^t}$ in \cref{eqn_wt_2_estimate}, and since $t \geq \log_{\mathcal{L}} \bigl( \frac{100^4 \Tilde{m}^4}{\theta^{32}} \bigr)$, we have, the number of samples 
    $$
    N \geq 4 \ln(\frac{4}{\delta}) /\epsilon^2 = 4 \ln(400 \mathcal{L}^t) \frac{64}{\theta^8 (\Tilde{\wt_2}(C_\alpha))^2} \geq 4 \ln(400 \frac{100^4 \Tilde{m}^4}{\theta^{32}}) \frac{64}{\theta^8 (1+\frac{\theta^2}{4})^2},
    $$ 
    since $\Tilde{\wt_2}(C_\alpha) \leq 1+\frac{\theta^2}{4}$. Therefore, taking the number of samples $N = 4 \ln(400 \frac{100^4 \Tilde{m}^4}{\theta^{32}}) \frac{64}{\theta^8 (1+\frac{\theta^2}{4})^2} = O \biggl( \frac{\ln \frac{\Tilde{m}}{\theta^{8}}}{\theta^{8}} \biggr)$ serves our purpose.
\end{proof}

\begin{lemma}\label{lemma_queries_proj}
    $P_Cf(y_i)$ and $P_Cf(y_i-x_i)$ can be estimated using $O(\frac{\ln(\mathcal{N} \Tilde{m})}{\theta^2})$ many queries.
\end{lemma}

\begin{proof}
    The proof follows exactly along the lines of Lemma~\ref{lemma_wt_estimate_new}. Putting $\epsilon = \frac{\theta}{2}$ and $\delta = \frac{1}{100\mathcal{N}\Tilde{m}}$, we have the number of queries to be equal to $O(\frac{\ln(\mathcal{N} \Tilde{m})}{\theta^2})$.
\end{proof}

\begin{theorem}
    The query complexity of the Generalized Implicit Sieve Algorithm (Algorithm~\ref{algo:Implicit_Sieve_generalized}) is 
    $$
        O\biggl( \frac{\ln \frac{\Tilde{m}}{\theta^{16}}}{\theta^8} + \Tilde{m} \frac{\ln(\mathcal{N} \Tilde{m})}{\theta^2} + \frac{\ln \mathcal{N}}{\theta^8} \biggr).
    $$
\end{theorem}

\begin{proof}
    Since taking $t = \log_{\mathcal{L}} \left( \frac{100^4 \Tilde{m}^4}{\theta^{64}} \right)$ is sufficient (Lemma~\ref{lemma_small_char_uniform_dist}), so the number of buckets $= \mathcal{L}^t = \frac{100^4 \Tilde{m}^4}{\theta^{64}}$. For each bucket, the $\wt_2$ and $\wt_4$ estimates make $O(\frac{\ln \frac{\Tilde{m}}{\theta^{16}}}{\theta^4})$ (from Lemma~\ref{lemma_wt__2_estimate}) and $O \biggl( \frac{\ln \frac{\Tilde{m}}{\theta^{16}}}{\theta^8} \biggr)$ (from Lemma~\ref{lemma_wt_4_estimation}) queries respectively. Also, each $P_{C_\alpha}$ needs to be estimated for $y_i, y_i-x_i, \ i \in [\Tilde{m}]$ and for each bucket, which makes $O\left(\frac{\ln(\mathcal{N} \Tilde{m})}{\theta^2}\right)$ queries by Lemma~\ref{lemma_queries_proj}. 
    We have also estimated $\widehat{f}(\chi_{\alpha(C_i)}), \ i \in [\Tilde{m}]$, with tolerance $\leq \frac{\theta^4}{32}$ and confidence $\frac{1}{100\mathcal{N}}$. The number of queries here is equal to $O\left( \frac{\ln(\mathcal{N})}{\theta^8} \right)$.
    The total number of queries made is equal to $O\left( \frac{\ln \frac{\Tilde{m}}{\theta^{16}}}{\theta^8} + \Tilde{m} \frac{\ln(\mathcal{N} \Tilde{m})}{\theta^2} + \frac{\ln \mathcal{N}}{\theta^8} \right)$.
    % Therefore, the total number of queries made by the Generalized Implicit Sieve algorithm is $O\biggl( \frac{\Tilde{m}^5}{\theta^{64}} + \frac{\Tilde{m}^4}{\theta^{64}} \times \frac{\ln \frac{\Tilde{m}}{\theta^{16}}}{\theta^8}\biggr) = O\biggl( \frac{\Tilde{m}^5}{\theta^{64}} + \frac{\Tilde{m}^4 \ln \frac{\Tilde{m}}{\theta^{16}}}{\theta^{72}}\biggr)$.
\end{proof}

\section{Query algorithm for isomorphism testing of $\G$}\label{sec: query algo}

In this section, we give a query algorithm for testing if a Boolean valued function $g:\mathcal{G} \to \zone$ is $\epsilon$-close or $(\epsilon + \tau)$-far from being isomorphic to an unknown function $f: \mathcal{G} \to \zone$, where $\mathcal{G}$ is a finite Abelian group, hence $\G = \mathbb{Z}_{{p_1}^{m_1}} \times \cdots \times \mathbb{Z}_{{p_n}^{m_n}}$, $p_i, i \in [n]$ are primes and not necessarily distinct. We are given query access to the truth table of $f$, that is, for any $x \in \mathcal{G}$ our algorithm can query $f(x)$. We build upon the Generalized Implicit Sieve algorithm (Algorithm~\ref{algo:Implicit_Sieve_generalized}) for testing if a known function $g: \mathcal{G} \to \zone$ with spectral norm at most $s$ is $\epsilon$-close to being isomorphic to an unknown function $f: \mathcal{G} \to \zone$ or $\epsilon+ \tau$-far from being isomorphic to $f$.

Also, let $\mathcal{L}= \LCM\{p_1^{m_1}, \ldots, p_n^{m_n}\}$, and $x*y$ denotes the following. 
\begin{align*}
    x*y = \sum_{i=1}^n \frac{\mathcal{L}}{p_i^{m_i}} x^{(i)} y^{(i)} \pmod{\mathcal{L}},
\end{align*}
where $x^{(i)},y^{(i)}$ are the $i^{th}$ component of $x$ and $y$ respectively.

The fractional Hamming distance between two functions has a nice connection with their Fourier coefficients, which we describe in the following well-known observation (see~\cite{ODonnellbook2014}).

\begin{observation}
    \label{obs: distance vs. Fourier coefficient}
   For any two Boolean valued functions  $f:\mathcal{G} \to \zone$ and $g:\mathcal{G} \to \zone$,  $\delta(f , g) \leq \epsilon$ if and only if $\sum_{r \in \mathcal{G}} \widehat{f}(\chi_r) \overline{\widehat{g}(\chi_r)} \geq (1 -2\epsilon)$.
\end{observation}
\begin{proof}
For any  $f:\mathcal{G} \to \R$ and $g:\mathcal{G} \to \R$ we have, 
\begin{align*}
    \E_x[f(x) g(x)] = \sum_{r \in \mathcal{G}} \widehat{f}(\chi_r) \overline{\widehat{g}(\chi_r)}. \tag*{see~\cite[Section 1.4]{ODonnellbook2014}}
\end{align*}
Now for $f:\mathcal{G} \to \zone$ and $g:\mathcal{G} \to \zone$ we have, 
\begin{align*}
    \E_x[f(x) g(x)] =  \Pr_{x} [f(x) = g(x)] - \Pr_{x} [f(x) \neq g(x)] = 1 - 2\delta(f,g).
\end{align*}
Hence, $\delta(f , g) \leq \epsilon$ if and only if $\sum_{r \in \mathcal{G}} \widehat{f}(\chi_r) \overline{\widehat{g}(\chi_r)} \geq (1 -2\epsilon)$.
\end{proof}

\begin{remark}
\label{remark: k small in implicit sieve algo}
    Algorithm~\ref{algo:Implicit_Sieve_generalized}, however, does not return $\mathcal{S}$. Also observe that by Parseval's identity,  $|\mathcal{S}| = O(1/\theta^2)$.
\end{remark}

We also need the following lemma. 
% \textcolor{red}{@Arijit da: Can we write a line for what the following lemma says.}

\begin{remark}
    Note that the functions are real-valued in our case. So, $f(x) \overline{g(x)} = f(x)g(x)$, and $\sum_{r \in \mathcal{G}} \widehat{f}(\chi_r) \overline{\widehat{g}(\chi_r)}$ is a real number, since $\sum_{r \in \mathcal{G}} \widehat{f}(\chi_r) \overline{\widehat{g}(\chi_r)} = \E_x[f(x) g(x)]$ by Parseval's inequality and $\E_x[f(x) g(x)]$ is a real number when $f,g$ are real-valued.
\end{remark}

\begin{algorithm}[]
% \SetAlgoLined
    \textbf{Input.}
    Given query access to $f: \mathcal{G} \to \zone$ and $g : \mathcal{G} \to \R$ such that $\|g\|_1 \leq s$. 
   % Let $G:\mathcal{G} \to \R$ be a function such that $\|G\| \leq t$ and $G$ $(\omega/4)$-approximates $g$. \\
    
    \textbf{Output.} \textsf{Accept} if $\dist_{\G}(f \circ A, g) \leq \epsilon$, and output \textsf{Reject} when $\dist_{\G}(f, g) \geq \epsilon+ \tau$.

\begin{enumerate}
    \item Let $\theta = (\tau/4) \cdot 1/(3s) \cdot 10\pi/\mathcal{L}$.
    Let $\mathcal{M}$ be a set of $\Tilde{m}$ elements chosen uniformly and independently from $\mathcal{G}$, where $\Tilde{m} = \Tilde{\Theta}(\frac{s^2}{\tau^2})$. Run the \textsf{Generalized Implicit Sieve Algorithm} for groups with $\mathcal{M}$ and $\theta$ as input to obtain a $(\Tilde{m} \times (\mathcal{N}+1))$ matrix $Q$, where $\mathcal{N} = O(1/\theta^2)$ (see Remark~\ref{remark: k small in implicit sieve algo}). \label{line: query algo: call Implicit Sieve}

    \item Label the rows of $Q$ with $1, \dots ,\Tilde{m}$ (which corresponds to the set $\mathcal{M}$) and columns with $1, \dots, \mathcal{N}$ (which corresponds to the set of large Fourier coefficients of $f$), leaving the last column. Let us denote the set of columns by $B$.

    % \item For a subset $S = \{s_1, \dots, s_{|S|}\} \subseteq [k]$, let $\Pi_{S,\lambda_1, \ldots, \lambda_{|S|}}$ denote an element satisfying $(\Pi_{S,\lambda_1, \ldots, \lambda_{|S|}})_i = \prod_{j = 1}^{|S|} Q[i,s_j]^{\lambda_j}$ for all $i \in [\Tilde{m}]$.

    \item \label{line: query algo: find B}
        Find the smallest $\Tilde{B} = \{r_1, \ldots, r_{|\Tilde{B}|}\} \subseteq B$ such that, 
        % $\Pi_{B,\lambda_1,\ldots,\lambda_{|\Tilde{B}|}}$ does not have all entries $=1$ for any $\lambda_1,\ldots,\lambda_{|\Tilde{B}|} \in \mathbb{N}\cup\{0\}$ with at least one $\lambda_i \neq 0$, and
        for any column $r \in B\setminus \Tilde{B}$, there exists $\lambda, \lambda_1, \ldots, \lambda_{|\Tilde{B}|} \in \mathbb{N}\cup\{0\}$, with $\lambda$ an unit in $\mathbb{Z}_\mathcal{L}$ (that is, $\lambda$ is invertible in $\mathbb{Z}_\mathcal{L}$) such that $\lambda r +\sum_{i\in [|\Tilde{B}|]} \lambda_i r_i = 0$.

        \begin{itemize}
            \item If such a $\Tilde{B}$ does not exist, then return \textsf{Fail}.
        \end{itemize}

    \item \label{line: query algo: relabel columns}
        Relabel the columns of $Q$ by $\{S_1, \dots ,S_{\mathcal{N}}\}$ where $S_1 = e_1, \dots S_{|\Tilde{B}|} = e_{|\Tilde{B}|}$ such that for all $j \in \{{|\Tilde{B}|}+1, \dots, \mathcal{N}\}$, $S_j$ can be written as $\frac{1}{\lambda} \biggl( \sum_{i\in [|\Tilde{B}|]} \lambda_i S_i \biggr)$, $\lambda, \lambda_1, \ldots, \lambda_{|\Tilde{B}|} \in \mathbb{N}\cup\{0\}$, and $\lambda$ is an unit in $\mathbb{Z}_\mathcal{L}$. Note that the columns $S_1, \dots, S_{|\Tilde{B}|}$ correspond to those in $\Tilde{B}$.

    \item \label{line: query algo: estimate ^f}
        Query $f(x^{(1)}), \dots, f(x^{(\Tilde{m})})$. 
        For all $j \in [\mathcal{N}]$, estimate $\widehat{f}(\chi_{S_i})$ to error $\pm \frac{\tau}{12 t}$ by $r_j$ where
        \begin{align*}
            r_j = \frac{\sum_{i=1}^{\Tilde{m}} f(x^{(i)})Q[i,j]}{\Tilde{m}}.
        \end{align*}
        Let $\tilde{f} : \mathcal{G} \to \R$ be the function such that for all $j \in [\mathcal{N}]$, $\widehat{\Tilde{f}}(\chi_{S_j}) = r_j$
        and $\widehat{\Tilde{f}}(\chi_S) = 0$ if $S \notin \{S_1, \dots, S_{\mathcal{N}}\}$.

    \item \label{line: query algo: accept if close}
        If there exists a group automorphism $A$ such that
        \begin{align*}
            \sum_{r \in \mathcal{G}} \widehat{\tilde{f \circ A}}(\chi_r) \overline{\widehat{g}(\chi_r)} \geq \left(1 - 2\epsilon -\frac{\tau}{2}\right),
        \end{align*}
        then return \textsf{Accept.}

    \item \label{line: query algo: reject if far}
        If for all group automorphism $A$
        \begin{align*}
            \sum_{r \in \mathcal{G}} \widehat{\tilde{f \circ A}}(\chi_r) \overline{\widehat{g}(\chi_r)} \leq \left(1 - 2\epsilon - \frac{3 \tau}{2}\right),
        \end{align*}
        then return \textsf{Reject.}

    \item Otherwise return \textsf{Fail}.
    \end{enumerate}
\caption{Query Algorithm}
\label{algo 1 testing}
\end{algorithm}

We now give a proof of the main theorem of this section, i.e. Theorem~\ref{theorem: testing algorithm}, which is stated below for convenience. The proof closely follows~\cite{WY13}.

Let us recall the theorem that we will be proving now.

\testingalgorithmquery*

\begin{proof}
Consider Algorithm~\ref{algo 1 testing}. Observe that the only queries made by this algorithm are in Step~\ref{line: query algo: call Implicit Sieve}. Thus by Theorem~\ref{lemma: Implicit Sieve}, the query complexity of the algorithm is upper bounded by $\tilde{O}\left(\frac{s^{8}}{\tau^{8}}\right)$. 

There are three bad events for this algorithm:
\begin{enumerate}
    \item The output of the \textsf{Implicit Sieve Algorithm} is in Step~\ref{line: query algo: call Implicit Sieve}.
    
    \item The algorithm fails to find $B$ in Step~\ref{line: query algo: find B}, in which case it returns \textsf{Fail}.

    \item The estimates in Step~\ref{line: query algo: estimate ^f} are incorrect.
\end{enumerate}

% \textcolor{red}{@Arijit da: Please check this paragraph for correctness.}
By Theorem~\ref{lemma: Implicit Sieve}, the error probability of the first bad event is upper bounded by $1/100$.
The probability of the third bad event can also be bounded by $1/100$ be a Chernoff bound for the choice of $\Tilde{m}$.

Next, we upper bound the probability of the second bad event. Consider the output $Q$ of the \textsf{Generalized Implicit Sieve Algorithm} in Step~\ref{line: query algo: call Implicit Sieve}. 
The rows are labelled by $1, \dots, \Tilde{m}$, where $\Tilde{m} = \Tilde{\Theta}(\frac{s^2}{\tau^2})$, which corresponds to $x^{(1)}, \dots, x^{(\Tilde{m})}$, each distributed uniformly and independently in $\mathcal{G}$. 
Also the columns are labelled by $1, \dots, \mathcal{N}$, where $\mathcal{N} = O(1/\theta^2) = O(\frac{s^2}{\tau^2})$ (were the first equality is by Remark~\ref{remark: k small in implicit sieve algo} and the second by the choice of $\theta$ in the algorithm), which correspond to some $T_1, \dots, T_{\mathcal{N}} \in \mathcal{G}$ (the $T_i$'s are not known).
By Theorem~\ref{lemma: Implicit Sieve}, the $(i,j)$th entry of $Q$ is $\chi_{s_j}(x_i) \in \{1, \omega_\mathcal{L}, \ldots, \omega_\mathcal{L}^{\mathcal{L}-1}\}$.

% \textcolor{red}{The purpose of identifying such a pseudo-independent set $\Tilde{B}= \{r_1, \ldots r_{|\Tilde{B}|}\}$ in Step~\ref{line: query algo: find B} is to reduce the work of choosing group automorphisms on $\G$.} 
Now, let us consider all the group automorphisms on $\G$ which are formed by only choosing the images of the elements in $\Tilde{B}$. Then, by the property of group automorphism, the images of any other element get defined automatically. That is, for an element $r \notin \Tilde{B}$, let $\lambda_1, \ldots, \lambda_{|\Tilde{B}|}, \lambda \in \mathbb{Z}_\mathcal{L}$, with $\lambda$ invertible in $\mathbb{Z}_\mathcal{L}$ be such that $\lambda r = \sum_{i=1}^\mathcal{L} \lambda_i r_i$. If $\psi$ is a group automorphism on $\G$, then $\psi(r) = \frac{1}{\lambda} \sum_{i=1}^\mathcal{L} \lambda_i \psi(r_i)$.

% Consider $S = \{s_1 \dots, s_{|S|}\} \subseteq [k]$ such that $\sum_{i \in [|S|]} \lambda_i s_i = 0^n$, for some $\lambda_i \in \mathbb{N}$ with $i \in [|S|]$, with at least one invertible $\lambda_j \in \mathbb{Z}_n$. For any $x \in \mathcal{G}$, 

% \begin{align*}
%     \prod_{i \in [|S|]} (\chi_{s_i}(x))^{\lambda_i} 
%     &=  \prod_{i \in [|S|]} (\omega_n)^{\lambda_i (s_i * x)} \\
%     &=  (\omega_n)^{(\sum_{i \in [|S|]} \lambda_i s_i) * x} \\
%     &= 1.
% \end{align*}
% Thus, for such a set $S$, $\Pi_{S, \lambda_1, \ldots, \lambda_{|S|}}$ has all entries $=1$. in Step~\ref{line: query algo: find B}. This implies that a dependent subset of $\{T_1, \dots, T_k\}$ is never chosen as $B$ in Step~\ref{line: query algo: find B}.

For a subset $S = \{s_1, \dots, s_{|S|}\} \subseteq [\mathcal{N}]$, let $\Pi_{S,\lambda_1, \ldots, \lambda_{|S|}}$ denote an element satisfying $(\Pi_{S,\lambda_1, \ldots, \lambda_{|S|}})_i = \prod_{j = 1}^{|S|} Q[i,s_j]^{\lambda_j}$ for all $i \in [\Tilde{m}]$. Observe that, if a dependent $\Tilde{B}$ is chosen in Step~\ref{line: query algo: find B}, then all the entries of $\Pi_{\Tilde{B}, \lambda_1, \ldots, \lambda_{|\Tilde{B}|}}$ are equal to $1$. 
% If we show that the probability with which the entries of $\Pi_{\Tilde{B}, \lambda_1, \ldots, \lambda_{|\Tilde{B}|}}$ are equal to $1$ is small, then we are done. Note that, due to the presence of zero divisors, in the case of an independent set, we can also have the entries equal to $1$, but that won't affect the probability, as we are choosing in such a way that not all entries of $\Pi_{\Tilde{B}, \lambda_1, \ldots, \lambda_{|\Tilde{B}|}}$ are equal to $1$.
Now we want to argue that for any pseudo-independent $R = \{r_1, \dots,r_{|R|}\} \subseteq \{T_1, \dots, T_{\mathcal{N}}\}$ and $\lambda_1,\ldots, \lambda_{|R|} \in \mathbb{N}\cup\{0\}$, the probability that $\Pi_{R,\lambda_1,\ldots, \lambda_{|R|}}$ has all entries $=1$ is at most $\mathcal{L}^{-\Tilde{m}} = \mathcal{L}^{-\tilde{O}(s^2/\tau^2)}$.
% Since, by Remark~\ref{remark: k small in implicit sieve algo}, the number of possible subsets of $[k]$ is upper bounded by $n^k = n^{O(t^2/\theta^2)}$, for a large enough constant in choice of $\Tilde{m}$, every such independent set $R$ satisfies $\Pi_{R, \lambda_1,\ldots, \lambda_{|R|}} \neq 1^{\Tilde{m}}$ with probability at least $99/100$. 
To see this, fix $R = \{r_1, \dots,r_{|R|}\}$ and observe that $r_j$'s are pseudo-independent. Since $x^{(1)}, \dots, x^{(\Tilde{m})}$ are uniformly and independently distributed in $\mathcal{G}$, $(\Pi_{R, \lambda_1,\ldots, \lambda_{|R|}})_i = 1$ with probability $1/\mathcal{L}$ for all $i \in [\Tilde{m}]$. This follows from the fact that each entry of $\Pi_{R, \lambda_1,\ldots, \lambda_{|R|}}$ can take values from $\{1, \omega_\mathcal{L}, \ldots, \omega_\mathcal{L}^{\mathcal{L}-1}\}$, hence the probability that it takes the value $1$ is $=\frac{1}{\mathcal{L}}$. Thus the probability that $\Pi_{R, \lambda_1,\ldots, \lambda_{|R|}} = 1^{\Tilde{m}}$ is upper bounded by $\mathcal{L}^{-\Tilde{m}}$, for each $\lambda_1, \ldots, \lambda_{|R|}$. Therefore, the probability that a dependent set is chosen in Step~\ref{line: query algo: find B} is at most $\mathcal{L}^{-\Tilde{m}} \times \frac{\binom{\mathcal{L}}{1} \varphi(\mathcal{L}) \mathcal{L}^{|R|-1}}{\varphi(\mathcal{L})!} = \mathcal{L}^{-\Tilde{m}} \times \frac{\mathcal{L}^{|R|}}{(\phi(\mathcal{L})-1)!}$, since it is upper bounded by the probability that $\Pi_{R, \lambda_1,\ldots, \lambda_{|R|}} = 1^{\Tilde{m}}$. Since, by Remark~\ref{remark: k small in implicit sieve algo}, the number of possible subsets of $[\mathcal{N}]$ is upper bounded by $\mathcal{L}^{\Tilde{m}} = \mathcal{L}^{O(s^2/\tau^2)}$, for a large enough constant in choice of $\Tilde{m}$, a pseudo-independent set is chosen with probability $\geq 1-\frac{1}{\mathcal{L}^{\Tilde{m}-|R|} (\varphi(\mathcal{L})-1)!}$.
%\sdc{This paragraph needs to be checked. Changes in the later part of the proof need to be made after confirmation.}

In particular, let $\Tilde{B} \subseteq \{T_1, \dots, T_k\}$ such that $\{T_1, \dots, T_k\} \subseteq S_R$, where $S_R$ is the subgroup generated by $R$, and elements of $\Tilde{B}$ are pseudo-independent. With probability at least $99/100$, $B$ will be considered in Step~\ref{line: query algo: find B} of the algorithm (if not such set has already been found). By choosing the elements of $\Tilde{B}$ to be $e_1, \dots, e_{|\Tilde{B}|}$, a brute-force algorithm (which makes no queries to the unknown function) can be used to check whether $\{T_1, \dots, T_{\mathcal{N}}\} \subseteq S_R$. Furthermore, the dependencies obtained can be used in the next step (Step~\ref{line: query algo: relabel columns}) of the algorithm.

Thus the probability of bad events happening for Algorithm~\ref{algo 1 testing} is at most $1/3$.

In the rest of this section we show that conditioned on bad events not happening, the algorithm always decides whether $ \dist_{\G}(f,g) \leq \epsilon$ or $ \dist_{\G}(f,g) \geq (\epsilon + \tau)$ correctly. Thus the error probability of the algorithm is upper bounded by $1/3$. We first consider the case when $ \dist_{\G}(f,g) \leq \epsilon$ and then the case when $ \dist_{\G}(f,g) \geq (\epsilon + \tau)$.

\begin{description}
    \item[Case 1:] We show that if $\dA(f,g) \leq \epsilon$, 
    then Algorithm~\ref{algo 1 testing} always returns \textsf{Accept}, conditioned on the bad events not happening.
    Since $\delta(f \circ A, g) \leq \epsilon$, by Observation~\ref{obs: distance vs. Fourier coefficient} we have
     \begin{align}
        \sum \widehat{f \circ A}(S) \widehat{g}(S) \geq \left(1 -2 \epsilon \right). \label{eqn: corr approx and bool 1}
    \end{align}
   However, since we do not know $f$, we would like to use ${\tilde{f}}$ instead. To do this, we first show that ${\tilde{f}}$ approximates all Fourier coefficients of $f$ to error at most $\frac{\tau}{4}\cdot \frac{1}{3s}$, up to some unknown group automorphism.

As Algorithm 1 uses $\theta = \frac{\tau}{4}\cdot \frac{1}{3s}$ while calling the \textsf{Generalized Implicit Sieve Algorithm} (see Theorem~\ref{lemma: Implicit Sieve}) we have the following cases
\begin{itemize}
    \item If $|\widehat{f}(\chi_\alpha)| \leq \theta/2$ then $\alpha \notin \mathcal{S}$. In this case the corresponding Fourier coefficient of ${\tilde{f}}$ is $0$.

    \item If  $|\widehat{f}(\chi_\alpha)| \geq \theta$ then $\alpha \in \mathcal{S}$. In this case the corresponding Fourier coefficient of ${\tilde{f}}$ is a $\theta$-error approximation of $\widehat{f}(\chi_\alpha)$.

    \item If $\theta /2 \leq |\widehat{f}(\chi_\alpha)| \leq \theta$ then $\alpha \in \mathcal{S}$. In this case also the corresponding Fourier coefficient of ${\tilde{f}}$ lies in $[0, 2\theta]$. 
\end{itemize}

In Step~\ref{line: query algo: accept if close} the algorithm checks the correlation between $g$ and $\Tilde{f}$ against all group automorphisms $A$. 
Since we assumed that the automorphism distance between $f$ and $g$ is at most $\epsilon$, analogous to Equation~\ref{eqn: corr approx and bool 1}, there exists group automorphisms $A,U$ (the existence of $U$ is guaranteed by Lemma~\ref{lemma_pontryagin} and Lemma~\ref{lemma_aut_perm_char}) such that 

\begin{align*}
&|\sum_{r \in \mathcal{G}} \widehat{\widetilde{f \circ UA}}(\chi_r) \overline{\widehat{g}(\chi_r)}| \\
&\geq |\sum_{r \in \mathcal{G}} \widehat{{f \circ A}}(\chi_r)\overline{\widehat{g}(\chi_r)}| - |\sum_{r \in \mathcal{G}} \bigl(  \widehat{{f \circ A}}(\chi_r) - \widehat{\widetilde{f \circ UA}}(\chi_r) \bigr) \overline{\widehat{g}(\chi_r)}| \\
 &\geq |\sum_{r \in \mathcal{G}} \widehat{{f \circ A}}(\chi_r) \widehat{g}(\chi_r)| - |\sum_{r \in \mathcal{G}}  \bigl(\frac{\tau}{4}\cdot \frac{1}{3s} + \frac{\theta^4}{32}\bigr)\overline{\widehat{g}(\chi_r)}| \\
 &\geq \left(1 - 2\epsilon \right) - \sum_{r \in \mathcal{G}} (\frac{\tau}{4}\cdot \frac{1}{3s} + \frac{\theta^4}{32}) |\overline{\widehat{g}(\chi_r)}| \tag*{from Observation~\ref{obs: distance vs. Fourier coefficient}} \\
 &\geq \left(1 - 2\epsilon -  \frac{\tau}{4} - \frac{\tau^4}{12^4\times 32s^3}\right) \\
 &\geq \left(1 - 2\epsilon -  \frac{\tau}{2} \right),
\end{align*}
where the first inequality assumed that $\delta(f\circ A, g) \leq \epsilon$
and the last inequality uses that fact that $|\sum_{r \in \mathcal{G}} \overline{\widehat{g}(\chi_r)}| \leq \sum_{r \in \mathcal{G}} |\widehat{g}(\chi_r)|$. Thus in this case the algorithm returns \textsf{Accept} in Step~\ref{line: query algo: accept if close}.
\end{description}

\begin{description}
\item[Case 2:] 
In this case we show that if $\dA(f,g) \geq (\epsilon + \tau)$, 
then Algorithm~\ref{algo 1 testing} always returns \textsf{Reject}, conditioned on the bad events not happening. To show this we again use the fact from the previous case that, up to some unknown group automorphism, the Fourier coefficients of ${\tilde{f}}$ approximate all Fourier coefficients of $f$ to error at most $\frac{\tau}{4}\cdot \frac{1}{3s}$. 
In Step~\ref{line: query algo: reject if far} the algorithm checks the correlation between $g$ and $\Tilde{f}$ against all group automorphisms $A$.
Since we have assumed that the automorphism distance between $f$ and $g$ is at least $(\epsilon + \tau)$, for all group automorphisms $U,A$ (the existence of $U$ is guaranteed by Lemma~\ref{lemma_pontryagin} and Lemma~\ref{lemma_aut_perm_char}) we have 

 \begin{align*}
 &|\sum_{r \in \mathcal{G}} \widehat{\widetilde{f \circ UA}}(\chi_r) \overline{\widehat{g}(\chi_r)}| \\
&\leq |\sum_{r \in \mathcal{G}} \widehat{{f \circ A}(\chi_r)}\overline{\widehat{g}(\chi_r)}| + |\sum_{r \in \mathcal{G}} \bigl(\widehat{\widetilde{f \circ UA}}(\chi_r) - \widehat{{f \circ A}(\chi_r)}\bigr)\overline{\widehat{g}(\chi_r)}| \\
 &\leq |\sum_{r \in \mathcal{G}} \widehat{{f \circ A}}(\chi_r) \overline{\widehat{g}(\chi_r)}| + |\sum_{r \in \mathcal{G}} \bigl( \frac{\tau}{4}\cdot \frac{1}{3s} + \frac{\theta^4}{32} \bigr) \overline{\widehat{g}(\chi_r)}| \\
 &\leq (1 - 2(\epsilon + \tau)) + \sum_{r \in \mathcal{G}} \bigl( \frac{\tau}{4}\cdot \frac{1}{3s} + \frac{\theta^4}{32} \bigr) |\overline{\widehat{g}(\chi_r)}| \tag*{from Observation~\ref{obs: distance vs. Fourier coefficient}} \\
 &\leq \left(1 - 2\epsilon - 2\tau + \frac{\tau}{4} + \frac{\tau^4}{12^4\times 32s^3}\right) \\
 &\leq (1 - 2\epsilon -  \frac{3\tau}{2}),
\end{align*}
where the last inequality uses that fact that $|\sum_{r \in \mathcal{G}} \overline{\widehat{g}(\chi_r)}| \leq \sum_{r \in \mathcal{G}} |\widehat{g}(\chi_r)|$. Thus in this case the algorithm returns \textsf{Reject} in Step~\ref{line: query algo: reject if far}.
\end{description}

\begin{lemma}
    The query complexity of Algorithm~\ref{algo 1 testing} is $\Tilde{O}\bigl( \frac{s^8}{\tau^8} \bigr)$, where $\Tilde{O}$ hides multiplicative factors in $\log s$ and $\log(1/\tau)$.
\end{lemma}

\begin{proof}
    The query complexity of Algorithm~\ref{algo 1 testing} is same as Generalized Implicit Algorithm~\ref{algo:Implicit_Sieve_generalized}. Putting $\theta = (\tau/4) \cdot 1/(3s) \cdot 10\pi/\mathcal{L}$ and $\Tilde{m} = \Tilde{\Theta}(\frac{s^2}{\tau^2})$, we have the query complexity to be $\tilde{O}\bigl( \frac{s^8}{\tau^8} \bigr)$, where $\tilde{O}(\cdot)$ hides multiplicative factors in $\log s$ and $\log(1/\tau)$ up to constants, and $\mathcal{L}$ is constant.
\end{proof}

%\textcolor{blue}{SD: Observe that, when we consider those group automorphism that are formed only by mapping the elements of the independent set $\Tilde{B}$ to its images, the images of the rest of the elements automatically gets defined since they are dependent on $\Tilde{B}$ by the property of group automorphism.}

\medskip

%\textcolor{blue}{SD: I could not find a lower bound for the query algorithm. The proof of Chakraborty et al. does not go through for finite Abelian groups.}

\end{proof}

\section{Isomorphism testing for sparse functions}

\subsection{Generalized Implicit Seive for sparse functions}

\begin{theorem}\label{thm_implicit_seive_sparse}
    Let $\G$ be a finite Abelian group, hence $\G =\mathbb{Z}_{p_1^{m_1}}\times \cdots \times \mathbb{Z}_{p_n^{m_n}}$ where $p_i$ are primes for all $i \in [n]$ and not necessarily distinct.
    Also, let $\theta \geq 0$ be a threshold and $\mathcal{M} = \left\{x_i\right\}_{i=1}^{\Tilde{m}} \subseteq \mathcal{G}$ be a set of independently and uniformly chosen $\Tilde{m}$ random points in $\mathcal{G}$, where $\Tilde{m} = \Tilde{\Theta}(\frac{s^2}{\tau^2})$.
    There exists an algorithm, \textsf{Generalized Implicit Sieve Algorithm for sparse functions} (\cref{algo:Implicit_Sieve_generalized_sparse}), that takes $\theta$ and $\mathcal{M}$ as input, makes $\Tilde{O}\left( \bigl(\frac{s^2}{\theta^4} + \Tilde{m} s^2\bigr) \right)$ queries to the truth table of a $s$-sparse Boolean valued function $f: \mathcal{G} \to \zone$, 
    % $c_\mathcal{L}$ is a function of $\mathcal{L}$, $\mathcal{L}= \LCM\{p_1^{m_1}, \ldots, p_n^{m_n}\}$, 
    and returns, with probability at least $\frac{96}{100}$, a labeled set of $\Tilde{m}$ examples, and the value of the function $f$ at $x_i$ for each $i \in [\Tilde{m}]$, of the form $\left\{ \chi_{\alpha_1}\left(x_i\right), \dots, \chi_{\alpha_\mathcal{N}}\left(x_i\right), f\left(x_i\right) \mid x_i \in \mathcal{M} \right\}$, where $\mathcal{S} = \left\{\alpha_1, \dots, \alpha_{\mathcal{N}} \right\}$ is some set that satisfies the following two properties:
    \begin{enumerate}
        \item
            $\forall \alpha \in \mathcal{G}$ with $\left|\widehat{f}(\chi_\alpha)\right| \geq \theta$ then $\alpha \in \mathcal{S}$.
        \item 
            $\forall \alpha \in \mathcal{S}$, $\left| \widehat{f}(\chi_\alpha) \right| > {\theta}/{2}$.
    \end{enumerate}
    The output of this algorithm can be seen as a $\Tilde{m} \times (\mathcal{N}+1)$ matrix $Q$ with entries $\omega_\mathcal{L}$, where $\mathcal{L}= \LCM\{p_1^{m_1}, \ldots, p_n^{m_n}\}$.
\end{theorem}

\begin{algorithm}[H]
\caption{Generalized Implicit Sieve for sparse functions}\label{algo:Implicit_Sieve_generalized_sparse}
\begin{algorithmic}[1]
    \Statex \textbf{Input:} Let $\G = \mathbb{Z}_{p_1^{m_1}}\times \cdots \times \mathbb{Z}_{p_n^{m_n}}$ be a finite Abelian group where $p_i$ are primes for all $i \in [n]$ and not necessarily distinct, its set of all characters, a threshold $\theta$, a set $M = \{x_1, \ldots, x_{\Tilde{m}}\}$ of characters, which is sufficiently large, and a Boolean valued function $f:\G \to \{-1,+1\}$ which has sparsity $s_f =s$.
    \Statex \textbf{Output:} A matrix $Q$, whose columns are given by $\chi_\alpha(x_i), \ i \in [\Tilde{m}]$ for some character $\chi_\alpha$. Also output $\mathcal{F}$, the $\Tilde{m} \times 1$ column matrix whose entries are $f(x_1), \ldots, f(x_{\Tilde{m}})$.

    \bigskip

    \State \textbf{Initialize} $t\geq \log_{\mathcal{L}} \bigl( 100s^2 \bigr)$. 

    \State Consider a randomly permuted coset structure $(H,\mathcal{C})$ such that the subgroup $H$ has codimension $\leq t$ (see \cref{defn_random_coset_structure}).

    \For{for each $C \in \mathcal{C}$}
    
    \State Estimate $\wt_2$ for each bucket with error $= \pm \frac{\theta^2}{4}$ and with confidence $1- \frac{1}{100 s^2}$. Let $\Tilde{\wt_2}(C)$ be the estimated weight of the bucket $C$.

    \If{$\Tilde{\wt_2}(C) > \frac{\theta^2}{2}$} 
    
    \State Keep the bucket $C$.
    \Else { discard.}

    \EndIf

    \EndFor

    \State Pick $\{y_1, \ldots,y_{\Tilde{m}}\}$ randomly from $\mathcal{G}$.

    \For{for each $i \in [\Tilde{m}]$}

    \State Estimate $P_Cf(y_i)$ and $P_Cf(y_i-x_i)$, and calculate $P_Cf(y_i) \overline{P_Cf(y_i-x_i)}$.
    \EndFor

    \State $Q_{ij} = \frac{P_{C_j}f(y_i) \overline{P_{C_j}f(y_i-x_i)}}{|P_{C_j}f(y_i) \overline{P_{C_j}f(y_i-x_i)}|}$, where $C_j, \ j \in [\mathcal{N}]$ are the survived buckets, and $Q_{ij}$ is the element in the $i^{th}$ row and $j^{th}$ column of the matrix $Q$.
\end{algorithmic}
\end{algorithm}

\begin{theorem}\label{thm:random-coset}
Let $S\subseteq \G$ be such that $|S| \leq s+1$. Then, if $t \geq 2\log_\mathcal{L} s + \log_\mathcal{L} 100$, all $\chi \in S$ belong to different cosets except with probability at most $\frac{1}{100}$, where $\mathcal{L} = \LCM\{p_1^{m_1}, \ldots, p_n^{m_n}\}$. 
\end{theorem}

\begin{algorithm}[H]
% \SetAlgoLined
    \textbf{Input.}
    Given query access to $f: \mathcal{G} \to \zone$, with sparsity $s_f =s$, and $g : \mathcal{G} \to \R$ with sparsity $s$. \\
    
    \textbf{Output.} \textsf{Accept} if there is a group automorphism $A : \mathcal{G} \to \mathcal{G}$ (let us think of $A$ in its matrix representation) such that $\delta(f \circ A, g) \leq \epsilon$, output \textsf{Reject} when $\dist_{\G}(f, g) \geq \epsilon+ \tau$.

    \bigskip
    
\begin{enumerate}
    \item Let $\theta = (\tau/4) \cdot 1/(3s)$.
    Let $\mathcal{M}$ be a set of $\Tilde{m}$ vectors chosen uniformly and independently from $\mathcal{G}$, where $\Tilde{m} = \Tilde{\Theta}(\frac{s^2}{\tau^2})$. Run the \textsf{Implicit Sieve Algorithm} for groups with $\mathcal{M}$ and $\theta$ as input to obtain a $(\Tilde{m} \times (\mathcal{N}+1))$ matrix $Q$, where $\mathcal{N} = O(1/\theta^2)$ (see Remark~\ref{remark: k small in implicit sieve algo}). 

    \item Label the rows of $Q$ with $1, \dots ,\Tilde{m}$ (which corresponds to the set $\mathcal{M}$) and columns with $1, \dots, \mathcal{N}$ (which corresponds to the set of large Fourier coefficients of $f$), leaving the last column. Let us denote the set of columns by $B$.

    % \item For a subset $S = \{s_1, \dots, s_{|S|}\} \subseteq [k]$, let $\Pi_{S,\lambda_1, \ldots, \lambda_{|S|}}$ denote an element satisfying $(\Pi_{S,\lambda_1, \ldots, \lambda_{|S|}})_i = \prod_{j = 1}^{|S|} Q[i,s_j]^{\lambda_j}$ for all $i \in [\Tilde{m}]$.

    \item
        Find the smallest $\Tilde{B} = \{r_1, \ldots, r_{|\Tilde{B}|}\} \subseteq B$ such that, 
        % $\Pi_{B,\lambda_1,\ldots,\lambda_{|\Tilde{B}|}}$ does not have all entries $=1$ for any $\lambda_1,\ldots,\lambda_{|\Tilde{B}|} \in \mathbb{N}\cup\{0\}$ with at least one $\lambda_i \neq 0$, and
        for any column $r \in B\setminus \Tilde{B}$, there exists $\lambda, \lambda_1, \ldots, \lambda_{|\Tilde{B}|} \in \mathbb{N}\cup\{0\}$, with $\lambda$ an unit in $\mathbb{Z}_\mathcal{L}$ (that is, $\lambda$ is invertible in $\mathbb{Z}_\mathcal{L}$) such that $\lambda r +\sum_{i\in [|\Tilde{B}|]} \lambda_i r_i = 0$.

        \begin{itemize}
            \item If such a $\Tilde{B}$ does not exist, then return \textsf{Fail}.
        \end{itemize}

    \item 
        Relabel the columns of $Q$ by $\{S_1, \dots ,S_{\mathcal{N}}\}$ where $S_1 = e_1, \dots S_{|\Tilde{B}|} = e_{|\Tilde{B}|}$ such that for all $j \in \{{|\Tilde{B}|}+1, \dots, \mathcal{N}\}$, $S_j$ can be written as $\frac{1}{\lambda} \biggl( \sum_{i\in [|\Tilde{B}|]} \lambda_i S_i \biggr)$, $\lambda, \lambda_1, \ldots, \lambda_{|\Tilde{B}|} \in \mathbb{N}\cup\{0\}$, and $\lambda$ is an unit in $\mathbb{Z}_\mathcal{L}$. Note that the columns $S_1, \dots, S_{|\Tilde{B}|}$ correspond to those in $\Tilde{B}$.

    \item 
        Query $f(x^{(1)}), \dots, f(x^{(\Tilde{m})})$. 
        For all $i \in [\mathcal{N}]$, estimate $\widehat{f}(\chi_{S_i})$ to error $\pm \frac{\omega}{12 t}$ by $r_j$ where
        \begin{align*}
            r_j = \frac{\sum_{i=1}^{\Tilde{m}} f(x^{(i)})Q[i,j]}{\Tilde{m}}.
        \end{align*}
        Let $\tilde{f} : \mathcal{G} \to \R$ be the function such that for all $j \in [\mathcal{N}]$, $\widehat{\Tilde{f}}(\chi_{S_j}) = r_j$
        and $\widehat{\Tilde{f}}(\chi_S) = 0$ if $S \notin \{S_1, \dots, S_{\mathcal{N}}\}$.

    \item 
        If there exists a group automorphism $A$ such that
        \begin{align*}
            \sum_{r \in \mathcal{G}} \widehat{\tilde{f \circ A}}(\chi_r) \overline{\widehat{g}(\chi_r)} \geq \left(1 - 2\epsilon -\frac{\tau}{4}\right),
        \end{align*}
        then return \textsf{Accept.}

    \item 
        If for all group automorphism $A$
        \begin{align*}
            \sum_{r \in \mathcal{G}} \widehat{\tilde{f \circ A}}(\chi_r) \overline{\widehat{g}(\chi_r)} \leq \left(1 - 2\epsilon - \frac{7 \tau}{4}\right),
        \end{align*}
        then return \textsf{Reject.}

    \item Otherwise return \textsf{Fail}.
    \end{enumerate}
\caption{Query Algorithm}
\label{algo 1 testing sparse}
\end{algorithm}

\begin{proof}
    Consider the $t$-dimensional coset structure $(H, \mathcal{C})$ (\cref{defn_random_coset_structure}). Let $\chi_{S_1}$ and $\chi_{S_2}$ be two distinct characters in $\widehat{\G}$. We have that 
    \begin{align*}
        \Pr \left[ \chi_{S_1}, \chi_{S_2} \text{ belong to the same coset} \right]
        =\Pr \left[ \forall i \in [t], r_i * (S_1-S_2) =0 \right]
        =\frac{1}{\mathcal{L}^t},
    \end{align*}
    where the last equality holds because $\chi_{S_1}$ and $\chi_{S_2}$ are distinct, and $r_1, \ldots, r_t$ are independent, and $\mathcal{L} = \LCM\{p_1^{m_1}, \ldots, p_n^{m_n}\}$.
    
     Since $|S|\leq s+1$, therefore the number of ways in which two distinct $\chi_{S_1}, \chi_{S_2}$ can be chosen from $S$ is $\binom{s+1}{2} \leq s^2$. Therefore, probability that all $\chi\in S$ belong to different cosets is given by
        \begin{align*}
            \Pr[\text{all } \chi\in S \text{ belong to different cosets}] &\geq 1- \binom{s+1}{2} \frac{1}{\mathcal{L}^t} \\
            &\geq 1- s^2 \frac{1}{\mathcal{L}^t} \\
            &\geq 1- s^2 \frac{1}{\mathcal{L}^{2\log_\mathcal{L} s + \log_\mathcal{L} 100}} \\
            &= 1- s^2 \frac{1}{100s^2} \\
            &= 1- \frac{1}{100} = \frac{99}{100}.
        \end{align*}
\end{proof}

\begin{lemma}
Let 
$\chi_{\alpha(C)}$ be the character in the bucket $C$ (with high probability). Algorithm~\ref{algo:Implicit_Sieve_generalized_sparse} does the following with probability $\geq 1-\frac{1}{100}$:
    \begin{enumerate}
        \item It does not discard any bucket $C$ with $|\widehat{f}(\chi_{\alpha(C)})| \geq \theta$.
        \item It discards any bucket $C$ with $|\widehat{f}(\chi_{\alpha(C)})| \leq \frac{\theta}{2}$.
    \end{enumerate}
\end{lemma}

\begin{proof}
\begin{description}
    \item[\textbf{Proof of part (1).}] $|\widehat{f}(\chi_{\alpha(C)})| \geq \theta \Rightarrow \Tilde{\wt_2}(C) \geq \theta^2 - \frac{\theta^2}{4} = \frac{3\theta^2}{4}$, which is $> \frac{\theta^2}{2}$. Hence the bucket $C$ is not discarded.

    \item[\textbf{Proof of part (2).}] When $|\widehat{f}(\chi_{\alpha(C)})| \leq \frac{\theta}{2}$, the estimated weight is either $\Tilde{\wt_2}(C) \leq \frac{\theta^2}{4}+\frac{\theta^2}{4} = \frac{\theta^2}{2}$, or $\Tilde{\wt_2}(C) = \frac{\theta^2}{4}-\frac{\theta^2}{4} =0$. Therefore, the bucket is discarded.
\end{description}
\end{proof}

\begin{lemma}
    \begin{align*}
        \Pr[\frac{P_{C_\alpha}f(y_i) \overline{P_{C_\alpha}f(y_i-x_i)}}{|\widehat{f}(\chi_{\alpha(C)})|^2} = \chi_{\alpha(C)}(x_i)] \geq 0.98.
    \end{align*}
\end{lemma}

\begin{proof}
    We know all the characters belong to distinct buckets with probability $\frac{99}{100}$. Therefore, $P_{C_\alpha}(y_i) = \widehat{f}(\chi_{\alpha(C)}) \chi_{\alpha(C)}(y_i)$ and $P_{C_\alpha}(y_i-x_i) = \widehat{f}(\chi_{\alpha(C)}) \chi_{\alpha(C)}(y_i-x_i)$ with probability $\frac{99}{100}$ each. Therefore, $\frac{P_{C_\alpha}f(y_i) \overline{P_{C_\alpha}f(y_i-x_i)}}{|\widehat{f}(\chi_{\alpha(C)})|^2} = \chi_{\alpha(C)}(x_i)$ with probability $\geq 1- \frac{2}{100} = \frac{98}{100}$.
\end{proof}

Therefore, with probability $\geq 1- (\frac{1}{100} + \frac{1}{100} + \frac{2}{100}) = \frac{96}{100}$, Algorithm~\ref{algo:Implicit_Sieve_generalized_sparse} works.

Let $\mathcal{N}$ denotes the number of survived buckets. Then, the matrix $Q$ is given by
\begin{align*}
Q=
\begin{bmatrix}
    \chi_{\alpha(C_1)}(x_1) & \chi_{\alpha(C_2)}(x_1) & \cdots & \chi_{\alpha(C_\mathcal{N})}(x_1) & \mid & f(x_1) \\
    \chi_{\alpha(C_1)}(x_2) & \chi_{\alpha(C_2)}(x_2) & \cdots & \chi_{\alpha(C_\mathcal{N})}(x_2) & \mid & f(x_2) \\
    . & . & . & . & \mid & . \\
    . & . & . & . & \mid & . \\
    . & . & . & . & \mid & . \\
    \chi_{\alpha(C_1)}(x_{\Tilde{m}}) & \chi_{\alpha(C_2)}(x_{\Tilde{m}}) & \cdots & \chi_{\alpha(C_\mathcal{N})}(x_{\Tilde{m}}) & \mid &  f(x_{\Tilde{m}})
\end{bmatrix}
\end{align*}

\begin{theorem}
    The query complexity of Algorithm~\ref{algo:Implicit_Sieve_generalized_sparse} is $\Tilde{O}((\frac{1}{\theta^4} + \Tilde{m}))$.
    % $c_\mathcal{L}$ is a function of $\mathcal{L}$, $\mathcal{L} = \LCM\{p_1^{m_1},\ldots, p_n^{m_n}\}$.
\end{theorem}

\begin{proof}
    We know that the number of buckets $=s^2$. For each bucket, the $\wt_2$ estimate makes $O(\frac{1}{\theta^4})$ (from \cref{lemma_wt__2_estimate}). Also, each $P_{C_\alpha}$ needs to be estimated for $y_i, y_i-x_i, \ i \in [\Tilde{m}]$. Since queries can be reused, therefore, the total number of queries made by the algorithm is $\Tilde{O}(\frac{1}{\theta^4} + \Tilde{m})$.
\end{proof}

\subsection{Testing isomorphism for sparse functions}
%\sdc{This won't work unless $g$ is $s$-sparse. Then it becomes a generalization of Manmatha da's algorithm for sparse functions.}
\begin{theorem}\label{thm_isomorphism_test_sparse}
    Let $\G$ be a finite Abelian group, hence $\G =\mathbb{Z}_{p_1^{m_1}}\times \cdots \times \mathbb{Z}_{p_n^{m_n}}$ where $p_i$ are primes for all $i \in [n]$ and not necessarily distinct. 
    Given a known function $g: \mathcal{G} \to \zone$ with sparsity $s$ and query access to a unknown $s$-sparse Boolean valued function  $f: \mathcal{G} \to \zone$, the query complexity of deciding whether $\dist_{G}(f,g) \leq \epsilon$ or $\dist_{G}(f,g) \geq \epsilon + \tau$ is upper bounded by $\Tilde{O}\left(\frac{s^{4}}{\tau^{4}}\right)$, where $\Tilde{O}(\cdot)$ hides multiplicative factors in $\log s$ and $\log\log(1/\tau)$ up to constants and $\epsilon, \tau \in (0,1/2]$.
    % $c_\mathcal{L}$ is a function of $\mathcal{L}$, $\mathcal{L} = \LCM\{p_1^{m_1},\ldots, p_n^{m_n}\}$.
\end{theorem}

The analysis of Algorithm~\ref{algo 1 testing sparse} follows exactly along the lines of Algorithm~\ref{algo 1 testing}. The query complexity is same as Algorithm~\ref{algo:Implicit_Sieve_generalized_sparse}, that is equal to $\Tilde{O}(\frac{s^4}{\tau^4})$, taking $\theta = (\frac{\tau}{4}) (\frac{1}{3s})$ and $\Tilde{m} = \Tilde{\Theta}(\frac{s^2}{\tau^2})$.

\bibliography{ref}

\end{document}